\documentclass[aps,preprint]{revtex4}
\usepackage{graphics,graphicx}
\usepackage{subfigure}
\usepackage{bm}
\usepackage{subcaption}
\usepackage{amsmath,amssymb}
\usepackage{epsfig}
\usepackage{txfonts}
\usepackage{subfigure}
\usepackage{xcolor}
\usepackage{epstopdf}
\usepackage{hyperref}
\usepackage{booktabs}
\usepackage{multirow}
\usepackage{amsmath}
\usepackage{array}
\usepackage{booktabs}
\usepackage{caption}

\begin{document}
\title{Chains of rotating boson stars with quartic or sextic self-interaction}
\author{Hao-Ran Sun $^{1,2}$\footnote{Email:sunhr2024@lzu.edu.cn}}
\author{Jing-Kang Bin $^{1,2}$\footnote{Email:220220939311@lzu.edu.cn}}
\author{Li Zhao $^{1,2}$\footnote{Email:lizhao@lzu.edu.cn, corresponding author}}

\affiliation{\small{$^{1}$ Institute of Theoretical Physics, Lanzhou University,  Lanzhou 730000, China\\
                  $^{2}$ School of Information Science and Engineering, Lanzhou University, Lanzhou 730000, China\\\\}}

\begin{abstract}
This paper investigates chains of rotating boson stars (BSs) within Einstein gravity coupled to a complex scalar field. The model incorporates quartic or sextic self-interactions in the scalar Lagrangian, which support the existence of stationary, solitonic, gravitationally bound solutions. We numerically construct these multi-component systems and investigate how the self-interactions alter their global properties—specifically the Arnowitt-Deser-Misner (ADM) mass $M$, the angular momentum $J$ and the morphology of their ergospheres. A central result is the distinct dependence of the $(\omega, M)$ and $(\omega, J)$ relations on the parity of the chains. Specifically, systems with an even number of constituents display spiraling curves, while those with an odd number exhibit loop structures. Moreover, we observe that two initially distinct ergospheres merge into a single one as the frequency $\omega$ increases. Our analysis also indicates that the quartic interaction imposes more restrictive existence bounds, particularly for odd-numbered chains, thereby restricting stable configurations to the weak-coupling regime. In contrast, the sextic interaction has a weaker effect and enables stable solutions at substantially stronger couplings.


\end{abstract}

\maketitle

\section{Introduction}

The ubiquitous rotation of compact objects and the interaction of their gravitational fields with matter are crucial to understanding stellar evolution and gravitational wave emission \cite{LIGOScientific:2016aoc, LIGOScientific:2017vwq, Bondi:1962px, LIGOScientific:2017ync, Kapadia:2019uut}. Boson stars (BSs), which are stationary soliton configurations arising from the coupling of complex scalar fields to Einstein's theory of gravity  \cite{Rosen:1968mfz,Friedberg:1976me,Liebling:2012fv,Kaup:1968zz,Wheeler:1955zz,Ruffini:1969qy}, are promising candidates for dark matter and black hole mimickers \cite{DellaMonica:2022kow,Olivares:2018abq}. Their lack of a singularity and their capacity to modulate their mass through self-interactions provide critical testing grounds for general relativity and dark matter theories \cite{Hui:2016ltb,Colpi:1986ye,Guzman:2009zz,Guzman:2009xre}. These distinctive features offer a viable alternative to black holes for explaining anomalous observations, such as the remnant emissions from gravitational-wave events \cite{Palenzuela:2006wp,Cardoso:2016oxy,CalderonBustillo:2020fyi,Bezares:2018qwa}. Investigating the formation, distribution, and evolution of BSs provides key insights into the structure of dark matter halos \cite{Feng:2010gw, Lee:2008jp}. Such an investigation also contributes to explaining various cosmological observations \cite{Palenzuela:2017kcg}.

The study of BSs can be divided into several key directions, on the basis of their intrinsic physical mechanisms and macroscopic properties. The most fundamental model, that of mini-BSs, consists of a complex scalar field without self-interaction \cite{Lee:1988av,Friedberg:1986tq}. Its maximum mass is inversely proportional to the particle mass, specifically scaling as $M_{PL}^2/m$ \cite{Kaup:1968zz,Herdeiro:2022gzp}, where $M_{PL}$ is the Planck mass and $m$ is the mass of the scalar particle. This property implies that the resulting mass would be too small for astrophysical relevance if the scalar particle is a Standard Model particle, such as the Higgs boson \cite{Colpi:1986ye}. To overcome the mass limitation of mini-BSs and explore richer phenomena, the self-interacting BSs model has been introduced \cite{Mielke:1980sa}. Incorporating a quartic self-interaction into the Lagrangian scalar field introduces a repulsive force that significantly enhances the maximum stable mass of BSs, which then scales as $\lambda^{1/2}M_{PL}^3/m^2$, where $\lambda$ is the coupling strength of self-interaction \cite{Colpi:1986ye}. In addition, the effect of repulsive and attractive self-interaction is studied in \cite{Eby:2015hsq}. Furthermore, a strong quartic self-interaction can stabilize BSs in excited states \cite{Sanchis-Gual:2021phr}.

Recently, a novel class of static BSs chains, bound by gravity and scalar repulsion, is discovered \cite{Herdeiro:2020kvf,Jaramillo:2024cus}. Static BSs chains are bound by both gravity and scalar repulsion \cite{Zhang:2023rwc,Liang:2023ywv,Sun:2022duv}, yet the rotation inherent to real astrophysical systems requires their rotating counterparts. While rotating single BSs and binaries have been thoroughly investigated \cite{Herdeiro:2016gxs,Ryan:1996nk,Zhang:2023rwc,Vaglio:2022flq}, their rotating multi-star configurations remain far less explored. In addition, the regime of sextic self-interaction remains an unexplored area of research. The sextic self-interaction has a negligible effect in BSs with weak coupling strength due to its higher-order nature. In contrast, for systems with strong coupling strength, it can yield stable solutions for BSs. These self-interactions modify the frequency dependence of the mass $M$ and angular momentum $J$, often leading to characteristic spiraling or looping behaviors in the $(\omega, M)$ and $(\omega,J)$ diagrams \cite{Sun:2022duv,Sun:2023ord,Gervalle:2022fze,Liang:2025myf,Herdeiro:2021mol}. Rotating spacetime exhibits the ergospheres, where matter cannot remain stationary \cite{Wang:2018xhw,Herdeiro:2014jaa,Herdeiro:2014goa,Kunz:2019sgn}, and in rotating BSs chains, two ergospheres can merge into one \cite{Sun:2023ord}.

This work provides the first systematic comparison of quartic and sextic self-interactions in rotating BSs chains.  Through numerical construction of multi-component configurations, we demonstrate that the interaction type critically influences both the ADM mass and the total momentum. A main result is that the system exhibits a parity-dependent response. And the quartic interactions impose stricter existence bounds, particularly in odd-numbered chains, while sextic interactions allow persistence under stronger coupling. 

The structure of this paper is as follows. Section \ref{sec2} presents the theoretical model for chains of rotating BSs with both quartic and sextic self-interactions. Section \ref{sec3} details the boundary conditions and describes the numerical methods employed to compute the physical quantities of interest. Our main numerical results are reported and discussed in Section \ref{sec4}. Finally, Section \ref{sec5} provides a summary of our work and outlines prospects for future research. Throughout this paper, we adopt the metric signature $(-,+,+,+)$ and use natural units where $\hbar=c=1$. Consequently, the Planck mass is given by $M_{Planck}=G^{-1/2}$.

\section{Setup and field equations}\label{sec2}

We consider the action describing a complex scalar field $\psi$ minimally coupled to Einstein's theory of gravity
\begin{equation}\label{action}
S = \int \left(\frac{1}{16\pi G}R+\mathcal{L}_M\right)\sqrt{-g}\,d^4x ,
\end{equation}
where $R$ is the Ricci scalar of the spacetime represented by the metric $g_{ab}$, $G$ is the gravitational constant and $\mathcal{L}_M$ is the Lagrangian density of the matter fields, which is expressed as
\begin{eqnarray}\label{Lagrangian1}
  {\cal L_M} =
   -\frac{1}{2}g^{ab}\nabla_a\psi^{*}\nabla_b\psi-V(|\psi^2|).
\end{eqnarray}

The quartic and sextic interactions are characterized by distinct functional forms of $V(|\psi|^2)$
\begin{equation}\label{V of phi4}
	 V(|\psi|^{2})=\mu^{2}\psi^*\psi+\lambda(\psi^*\psi)^{2},
\end{equation}
\begin{equation}\label{V of phi6}
	V(|\psi|^{2})=\mu^{2}\psi^*\psi+\lambda(\psi^*\psi)^{3},
\end{equation}
where Eqs. (\ref{V of phi4}) and (\ref{V of phi6}) give the explicit forms of the potential $V(|\psi|^2)$  for the  quartic and sextic self-interaction, respectively. Here  $\psi^*$ is the complex conjugate of the scalar field $\psi$ and $V$ is the potential depending only on the magnitude of the field. The parameters 
$\mu$ and $\lambda$ denote the field mass and self-interaction coupling strength.

By varying the action (\ref{action}) with respect to the complex conjugate $\psi^*$ of the complex scalar field and the metric $g_{ab}$, we obtain the following Einstein-Klein-Gordon (EKG) equations 
\begin{equation}
\Box\psi=\mu^2\psi\ +2\lambda(\psi^{*}\psi)\psi,
\label{eq:EKG2}
\end{equation}
\begin{equation}
	\Box\psi=\mu^2\psi\ +3\lambda(\psi^{*}\psi)^2\psi,
	\label{eq:EKG3}
\end{equation}
\begin{equation}
	E_{ab}=R_{ab}-\frac{1}{2}g_{ab}R-8\pi GT_{ab}.
	\label{eq:EKG1}
\end{equation}
Here
\begin{equation}
   \Box = g^{\mu\nu}\nabla_\mu \nabla_\nu=-\frac{\partial^2}{\partial t^2}+\nabla^2,
\end{equation}
is the d'Alembert operator. Eq. (\ref{eq:EKG2}) is the KG equation in the quartic self-interaction case, Eq. (\ref{eq:EKG3}) is the KG equation in the sextic self-interaction case and Eq. (\ref{eq:EKG1}) represents the Einstein field equations.

Additionally, the energy-momentum tensors for the quartic and sextic self-interaction cases are given by
\begin{equation}
T_{ab}=\nabla_{a}\psi^*\nabla_{b}\psi+\nabla_{b}\psi^*\nabla_{a}\psi-g_{ab}\left\{\frac{1}{2}(\nabla_{c}\psi^*\nabla^{c}\psi+\nabla_{c}\psi\nabla^{c}\psi^*)+\mu^2\psi^*\psi+\lambda(\psi^*\psi)^2\right\},
\label{eq:T1}
\end{equation}
\begin{equation}
	T_{ab}=\nabla_{a}\psi^*\nabla_{b}\psi+\nabla_{b}\psi^*\nabla_{a}\psi-g_{ab}\left\{\frac{1}{2}(\nabla_{c}\psi^*\nabla^{c}\psi+\nabla_{c}\psi\nabla^{c}\psi^*)+\mu^2\psi^*\psi+\lambda(\psi^*\psi)^3\right\},
	\label{eq:T2}
\end{equation}
where Eqs. (\ref{eq:T1}) and (\ref{eq:T2}) define the energy-momentum tensor for the case of quartic and sextic self-interaction, respectively.

To construct stationary solutions for rotating BS chains with quartic or sextic self-interactions, we employ an axisymmetric metric under the following ansatz
\begin{eqnarray}
\label{ansatz1}
ds^2=e^{2F_1(r,\theta)}\left(dr^2+r^2 d\theta^2\right)+e^{2F_2(r,\theta)}r^2 \sin^2\theta (d\varphi-W(r,\theta)dt)^2-e^{2F_0(r,\theta)}  dt^2.
\end{eqnarray}
Here the functions $F_{i}~(i=0,1,2)$ and $W$ depend on the radial distance $r$ and the polar angle $\theta$. Additionally, the ansatz for the complex scalar field is given by
\begin{eqnarray}\label{ansatz2}
\psi=\phi(r,\theta) e^{i(m\varphi-\omega t)},\;\;\; m=\pm1,\pm2,  \cdots ,
\label{scalar_ansatz}
\end{eqnarray}
where $\varphi$ is the azimuthal angle, the constant $\omega$ is the frequency of the complex scalar field and $m$ is the azimuthal harmonic index.

Considering the Einstein field Eq. (\ref{eq:EKG1}), the non-zero component equations are $E_{t}^{t}$, $E_{r}^{r}$, $E_{\theta}^{\theta}$, $E_{\varphi}^{\varphi}$, $E_{\varphi}^{t}$ and $E_{r}^{\theta}$. Through the combination of these six equations, the derivation yields four independent partial differential equations (PDEs).
\begin{equation}
	\begin{aligned}
		& E_{r}^{r}+E_{\theta}^{\theta}-E_{\varphi}^{\varphi}-E_{t}^{t}=0, \\
		& E_{r}^{r}+E_{\theta}^{\theta}-E_{\varphi}^{\varphi}+E_{t}^{t}+2WE_{\varphi}^{t}=0, \\
		& E_{r}^{r}+E_{\theta}^{\theta}+E_{\varphi}^{\varphi}-E_{t}^{t}-2WE_{\varphi}^{t}=0, \\
		& E_{\varphi}^{t}=0.
    \end{aligned}
    \label{PDE}
\end{equation}

Substituting the ansatz Eqs. (\ref{ansatz1}) and (\ref{ansatz2}) into Eqs. (\ref{PDE}) allows us to derive the second-order partial differential equations governing $F_{i}$ ($i=0,1,2$) and $W$ in systems with quartic and sextic self-interactions.

For the quartic self-interaction case, the second-order PDEs for $F_{i}~(i=0,1,2)$ and $W$ are given by
\begin{equation}
	\begin{aligned}
		&-32e^{2F_{1}}G\pi \phi^2(e^{2F_{2}}r^2\omega^2-e^{2F_{0}}m^2\csc^2\theta-2e^{2F_{2}}mr^2\omega W+e^{2F_{2}}m^2r^2W^2)\\&+e^{2F_{2}}(4e^{2F_{0}}\partial_\theta F_{0}(\cot\theta+\partial_\theta F_2)-32e^{2F_0}G\pi\partial_\theta\phi^2+e^{2F_2}r^2\sin^2\theta\partial_\theta W^2-4e^{2F_0}\partial^2_\theta F_1\\&+4e^{2F_0}r\partial_{r}F_{0}-4e^{2F_{0}}r\partial_{r}F_{1}+4e^{2F_{0}}r^2\partial_{r}F_{0}\partial_{r}F_{2}-32e^{2F_{0}}G\pi r^2\partial_{r}\phi^2\\&+e^{2F_{2}}r^4\sin^2\theta\partial_{r}W^2-4e^{2F_{0}}r^2\partial^2_{r}F_1)=0,
	\end{aligned}
	\label{PDE1}
\end{equation}
\begin{equation}
	\begin{aligned}
		&16e^{2(F_0+F_1)}G\pi r^2\lambda\phi^4+16e^{2F_1}G\pi r^2\phi^2(e^{2F_{0}}\mu^2-2\omega^2+4m\omega W-2m^2W^2)\\&+2e^{2F_{0}}\partial_\theta F_{0}^2+2e^{2F_{0}}\partial_\theta F_{0}(\cot\theta+\partial_\theta F_2)-e^{2F_2}r^2\sin^2\theta\partial_\theta W^2\\&+2e^{2F_{0}}\partial^2_{\theta}F_0+4e^{2F_0}r\partial_{r}F_{0}+2e^{2F_0}r^2\partial_{r}F^2_0+2e^{2F_{0}}r^2\partial_{r}F_{0}\partial_{r}F_{2}\\&-e^{2F_2}r^4\sin^2\theta\partial_{r}W^2+2e^{2F_0}r^2\partial^2_{r}F_0=0,
	\end{aligned}
\end{equation} 
\begin{equation}
	\begin{aligned}
		&16e^{2(F_{0}+F_{1})}G\pi(e^{2F_{2}}r^2\mu^2+2m^2\csc^2\theta)\phi^2+16e^{2(F_0+F_1+F_2)}G\pi r^2\lambda\phi^4\\&+e^{2F_2}(4e^{2F_0}\cot\theta\partial_\theta F_2+2e^{2F_0}\partial_\theta F^2_2+2e^{2F_0}\partial_\theta F_{0}(\cot\theta+\partial_\theta F_2)\\&+e^{2F_2}r^2\sin^2\theta\partial_\theta W^2+2e^{2F_0}\partial^2_\theta F_2+2e^{2F_0}r\partial_{r}F_{0}+6e^{2F_0}r\partial_{r}F_{2}\\&+2e^{2F_0}r^2\partial_{r}F_{0}\partial_{r}F_{2}+2e^{2F_0}r^2\partial_{r}F^2_2+e^{2F_2}r^4\sin^2\theta\partial_{r}W^2+2e^{2F_0}r^2\partial^2_{r}F_2)=0,
	\end{aligned}
\end{equation} 
\begin{equation}
	\begin{aligned}
		&32e^{2F_1}Gm\pi\phi^2(-\omega+mW)-e^{2F_2}\sin\theta((-\sin\theta\partial_\theta F_0+3(\cos\theta+\sin\theta\partial_\theta F_2))\partial_\theta W\\&+\sin\theta(\partial^2_\theta W+r((4-r\partial_{r}F_{0}+3r\partial_{r}F_{2})\partial_{r}W+r\partial^2_{r}W)))=0.
	\end{aligned}
	\label{PDE4}
\end{equation} 

In contrast, the corresponding equations for the sextic self-interaction case are given by
\begin{equation}
	\begin{aligned}
		&-32e^{2F_{1}}G\pi \phi^2(e^{2F_{2}}r^2\omega^2-e^{2F_{0}}m^2\csc^2\theta-2e^{2F_{2}}mr^2\omega W+e^{2F_{2}}m^2r^2W^2)\\&+e^{2F_{2}}(4e^{2F_{0}}\partial_\theta F_{0}(\cot\theta+\partial_\theta F_2)-32e^{2F_0}G\pi\partial_\theta\phi^2+e^{2F_2}r^2\sin^2\theta\partial_\theta W^2-4e^{2F_0}\partial^2_\theta F_1\\&+4e^{2F_0}r\partial_{r}F_{0}-4e^{2F_{0}}r\partial_{r}F_{1}+4e^{2F_{0}}r^2\partial_{r}F_{0}\partial_{r}F_{2}-32e^{2F_{0}}G\pi r^2\partial_{r}\phi^2\\&+e^{2F_{2}}r^4\sin^2\theta\partial_{r}W^2-4e^{2F_{0}}r^2\partial^2_{r}F_1)=0,
	\end{aligned}
	\label{PDE11}
\end{equation}
\begin{equation}
	\begin{aligned}
		&16e^{2(F_0+F_1)}G\pi r^2\lambda\phi^6+16e^{2F_1}G\pi r^2\phi^2(e^{2F_{0}}\mu^2-2\omega^2+4m\omega W-2m^2W^2)\\&+2e^{2F_{0}}\partial_\theta F_{0}^2+2e^{2F_{0}}\partial_\theta F_{0}(\cot\theta+\partial_\theta F_2)-e^{2F_2}r^2\sin^2\theta\partial_\theta W^2\\&+2e^{2F_{0}}\partial^2_{\theta}F_0+4e^{2F_0}r\partial_{r}F_{0}+2e^{2F_0}r^2\partial_{r}F^2_0+2e^{2F_{0}}r^2\partial_{r}F_{0}\partial_{r}F_{2}\\&-e^{2F_2}r^4\sin^2\theta\partial_{r}W^2+2e^{2F_0}r^2\partial^2_{r}F_0=0,
	\end{aligned}
\end{equation} 
\begin{equation}
	\begin{aligned}
		&16e^{2(F_{0}+F_{1})}G\pi(e^{2F_{2}}r^2\mu^2+2m^2\csc^2\theta)\phi^2+16e^{2(F_0+F_1+F_2)}G\pi r^2\lambda\phi^6\\&+e^{2F_2}(4e^{2F_0}\cot\theta\partial_\theta F_2+2e^{2F_0}\partial_\theta F^2_2+2e^{2F_0}\partial_\theta F_{0}(\cot\theta+\partial_\theta F_2)\\&+e^{2F_2}r^2\sin^2\theta\partial_\theta W^2+2e^{2F_0}\partial^2_\theta F_2+2e^{2F_0}r\partial_{r}F_{0}+6e^{2F_0}r\partial_{r}F_{2}\\&+2e^{2F_0}r^2\partial_{r}F_{0}\partial_{r}F_{2}+2e^{2F_0}r^2\partial_{r}F^2_2+e^{2F_2}r^4\sin^2\theta\partial_{r}W^2+2e^{2F_0}r^2\partial^2_{r}F_2)=0,
	\end{aligned}
\end{equation} 
\begin{equation}
	\begin{aligned}
		&32e^{2F_1}Gm\pi\phi^2(-\omega+mW)-e^{2F_2}\sin\theta((-\sin\theta\partial_\theta F_0+3(\cos\theta+\sin\theta\partial_\theta F_2))\partial_\theta W\\&+\sin\theta(\partial^2_\theta W+r((4-r\partial_{r}F_{0}+3r\partial_{r}F_{2})\partial_{r}W+r\partial^2_{r}W)))=0.
	\end{aligned}
	\label{PDE44}
\end{equation} 

The explicit form of the Klein-Gordon (KG) equation for the quartic self-interaction case is 
obtained by substituting the Eqs. (\ref{ansatz1} and \ref{ansatz2}) into Eq. (\ref{eq:EKG2}).
\begin{equation}
	\begin{aligned}
		&-2e^{2(F_0+F_1+F_2)}r^2\lambda\phi^3-e^{2F_1}\phi(e^{2F_2}r^2(e^{2F_0}\mu^2-\omega^2)+e^{2F_0}m^2\csc^2\theta+2e^{2F_2}mr^2\omega W-e^{2F_2}m^2r^2W^2)\\&+e^{2(F_0+F_2)}((\cot\theta+\partial_\theta F_0+\partial_\theta F_2)\partial_\theta\phi+\partial^2_\theta\phi+r((2+r\partial_{r}F_0+r\partial_{r}F_2)\partial_{r}\phi+r\partial^2_{r}\phi)=0.
	\end{aligned}
	\label{PDE5}
\end{equation} 

Similarly, substituting the Eqs. (\ref{ansatz1} and \ref{ansatz2}) into Eq. (\ref{eq:EKG3}) leads to the explicit PDE governing the sextic self-interaction case

\begin{equation}
	\begin{aligned}
		&-3e^{2(F_0+F_1+F_2)}r^2\lambda\phi^5-e^{2F_1}\phi(e^{2F_2}r^2(e^{2F_0}\mu^2-\omega^2)+e^{2F_0}m^2\csc^2\theta+2e^{2F_2}mr^2\omega W-e^{2F_2}m^2r^2W^2)\\&+e^{2(F_0+F_2)}((\cot\theta+\partial_\theta F_0+\partial_\theta F_2)\partial_\theta\phi+\partial^2_\theta\phi+r((2+r\partial_{r}F_0+r\partial_{r}F_2)\partial_{r}\phi+r\partial^2_{r}\phi)=0.
	\end{aligned}
	\label{PDE55}
\end{equation} 

In addition, the two constraint equations can be derived from the following combination
\begin{equation}
	\begin{aligned}
		& E_{r}^{r}-E_{\theta}^{\theta}=0, \\
		& E_{\theta}^{r}=0.
	\end{aligned}
\end{equation}

By solving the five PDEs. ({\ref{PDE1}}) - ({\ref{PDE4}}) and PDE. (\ref{PDE5}) for the quartic self-interaction case or PDEs. ({\ref{PDE11}}) - ({\ref{PDE44}}) and PDE. (\ref{PDE55}) for the sextic self-interaction case, respectively. We can determine the unknown functions $F_{i}~(i=0,1,2)$, $W$ and $\phi$.

\section{Boundary conditions}\label{sec3}
This section details the boundary conditions for solving the PDE system in Eqs. (\ref{PDE1}) to (\ref{PDE55}). The requirement of asymptotic flatness at spatial infinity dictates the following boundary conditions:
\begin{equation}
	F_{i}|_{r\rightarrow\infty}=W|_{r\rightarrow\infty}=\phi|_{r\rightarrow\infty}=0.
\end{equation}

Due to the requirements of non-singularity and regularity at the origin, we impose the following conditions
\begin{equation}
	\partial_{r}F_{i}(0, \theta)=\partial_{r}W(0, \theta)=\phi(0, \theta)=0.
\end{equation}

In addition, because of the polar angle reflection symmetry $\theta\rightarrow\pi-\theta$ in the equatorial plane, we only need to consider the range $\theta\in [0, \pi/2]$. The boundary conditions at $\theta = \pi/2$ depend on the parity of the scalar field under equatorial reflection. For even and odd parity, they are respectively given by
\begin{equation}
	\begin{cases}
		 \partial_\theta F_{i}(r,\pi/2)=\partial_\theta W(r,\pi/2)=\partial_\theta\phi(r,\pi/2)=0, \quad \text{even parity}, \\
		 \partial_\theta F_{i}(r,\pi/2)=\partial_\theta W(r,\pi/2)=\phi(r,\pi/2)=0, \quad\quad \text{odd parity}.
    \end{cases}
\end{equation}

For axial symmetry, the boundary conditions at $\theta=0$ are
\begin{equation}
	\partial_\theta F_{i}(r, 0)=\partial_\theta W(r, 0)=\phi(r, 0)=0.
\end{equation}

The ADM mass $M$ and the total angular momentum $J$ are determined by the asymptotic behavior of the metric functions as $r$ approaches infinity
\begin{eqnarray}
	g_{tt}=-1+\frac{2GM}{r}+\cdots,
\end{eqnarray}
\begin{eqnarray}
	g_{\varphi t}=-\frac{2GJ}{r}\sin^2\theta+\cdots.
\end{eqnarray}

\section{Solving for Boson Stars}\label{sec4}
\subsection{The numerical method}
This section presents the numerical solutions of the EKG equations, showing the ADM mass $M$ and the angular momentum $J$ as functions of frequency $\omega$. We also present the ergospheres for chains of rotating BSs with both quartic and sextic self-interactions. The system is characterized by five input parameters: the gravitational constant $G$, the field mass $\mu$, the azimuthal harmonic index $m$, the coupling strength $\lambda$ and the frequency $\omega$. Furthermore, we employ the following redefined and rescaled quantities in our numerical analysis
\begin{eqnarray}
	\phi\rightarrow\frac{\phi}{\sqrt{8\pi G}},\:\: W\rightarrow\mu W,\:\: r\rightarrow \frac{r}{\mu},\:\: \omega\rightarrow\mu\omega,
\end{eqnarray}
by setting $G=\mu=m=1$, we reduce the input parameters to $\omega$ and $\lambda$. We introduce a new coordinate $x=\frac{r}{r+1}$, which maps the radial domain from $r\in[0, \infty)$ to $x\in[0,1)$. This transformation maps the asymptotic region $r\rightarrow\infty$ to the finite limit $x\rightarrow1$, which allows for the visualization of field behavior at spatial infinity.

All numerical computations are based on the finite element method, with a grid size of 150 $\times$ 150 for the integration regions \(0\le x\leq1\) and \(0\leq\theta\leq\pi/2\). The iteration process employs the Newton-Raphson method, and the relative error for the numerical solutions in this paper is below $10^{-4}$. Throughout this work, the results are presented using the coordinates $\rho=r\sin\theta$ and $z=r\cos\theta$.
\subsection{The result of quartic self-interaction}
\begin{figure}[htbp]
	\centering
	{
		\begin{minipage}[b]{.3\linewidth}
			\centering
			\includegraphics[scale=0.2]{"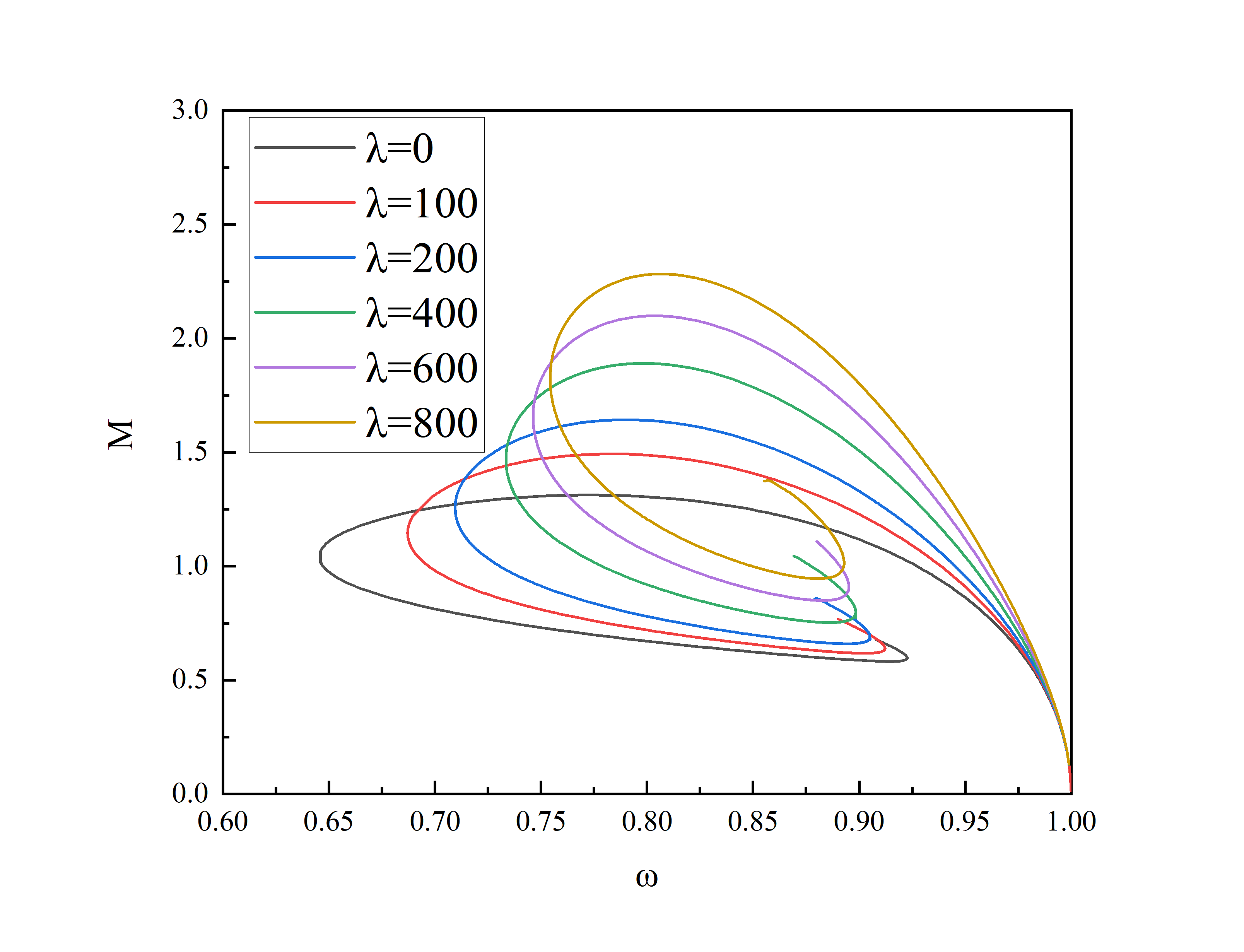"}
		\end{minipage}
		\begin{minipage}[b]{.3\linewidth}
			\centering
			\includegraphics[scale=0.2]{"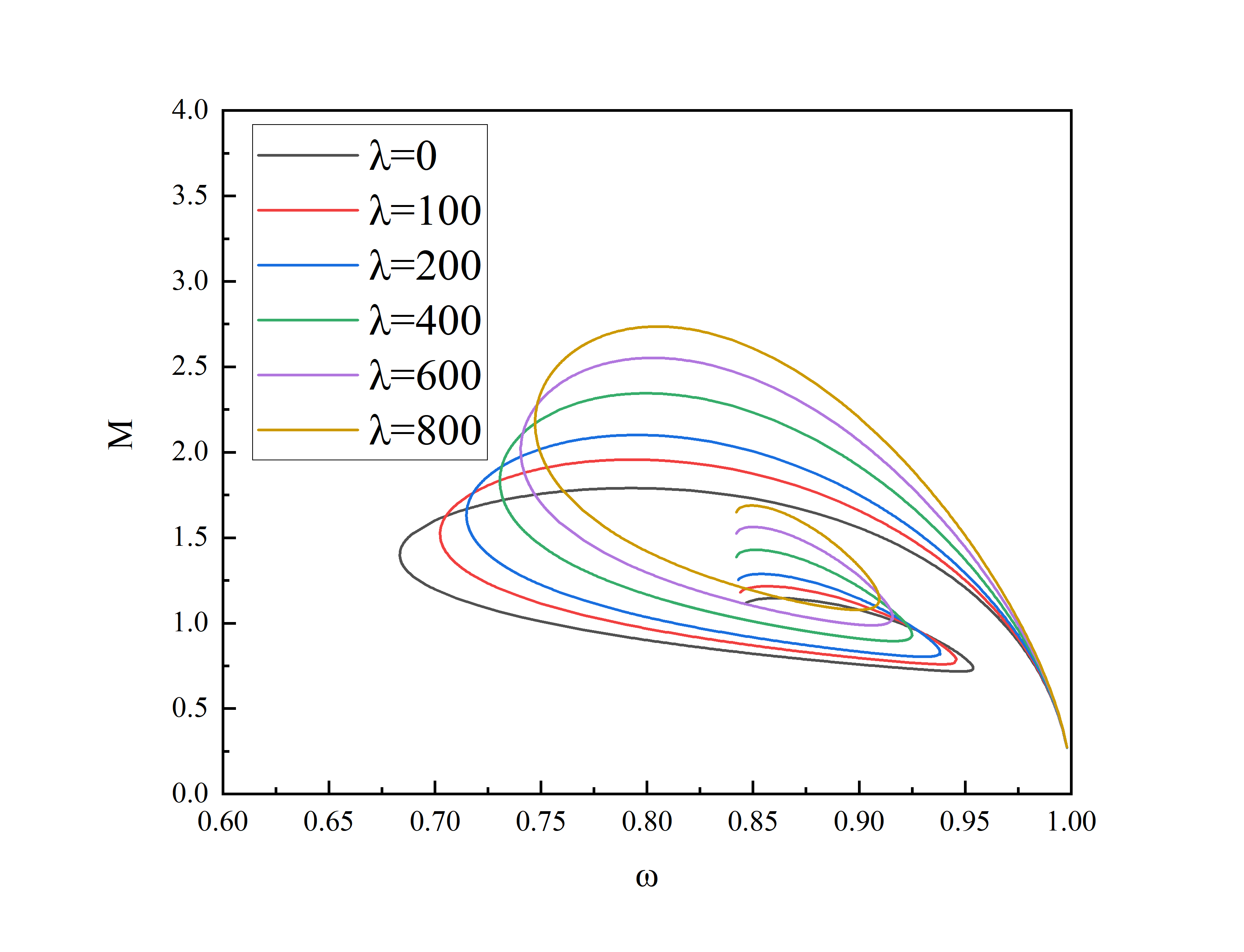"}
		\end{minipage}
		\begin{minipage}[b]{.3\linewidth}
			\centering
			\includegraphics[scale=0.2]{"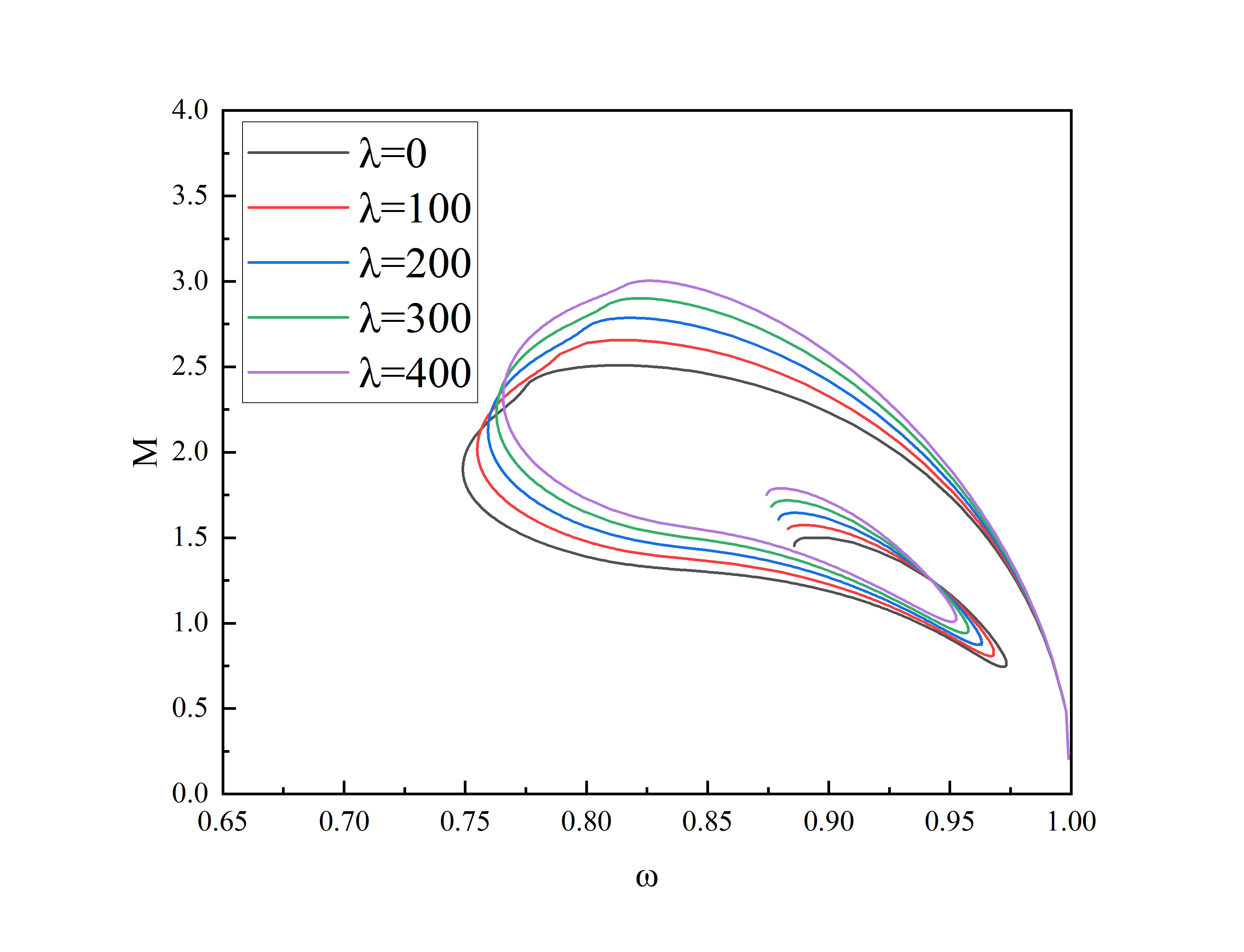"}
		\end{minipage}
	}
	\caption{The ADM mass $M$ as a function of the frequency $\omega$. Left: The rotating BSs with quartic self-interaction of one constituent. Middle: The rotating BSs with quartic self-interaction of two constituents. Right: The rotating BSs with quartic self-interaction of four constituents.}
\label{Fig.1}
\end{figure}

\begin{figure}[htbp]
	{
		\begin{minipage}[b]{.3\linewidth}
			\centering
			\includegraphics[scale=0.2]{"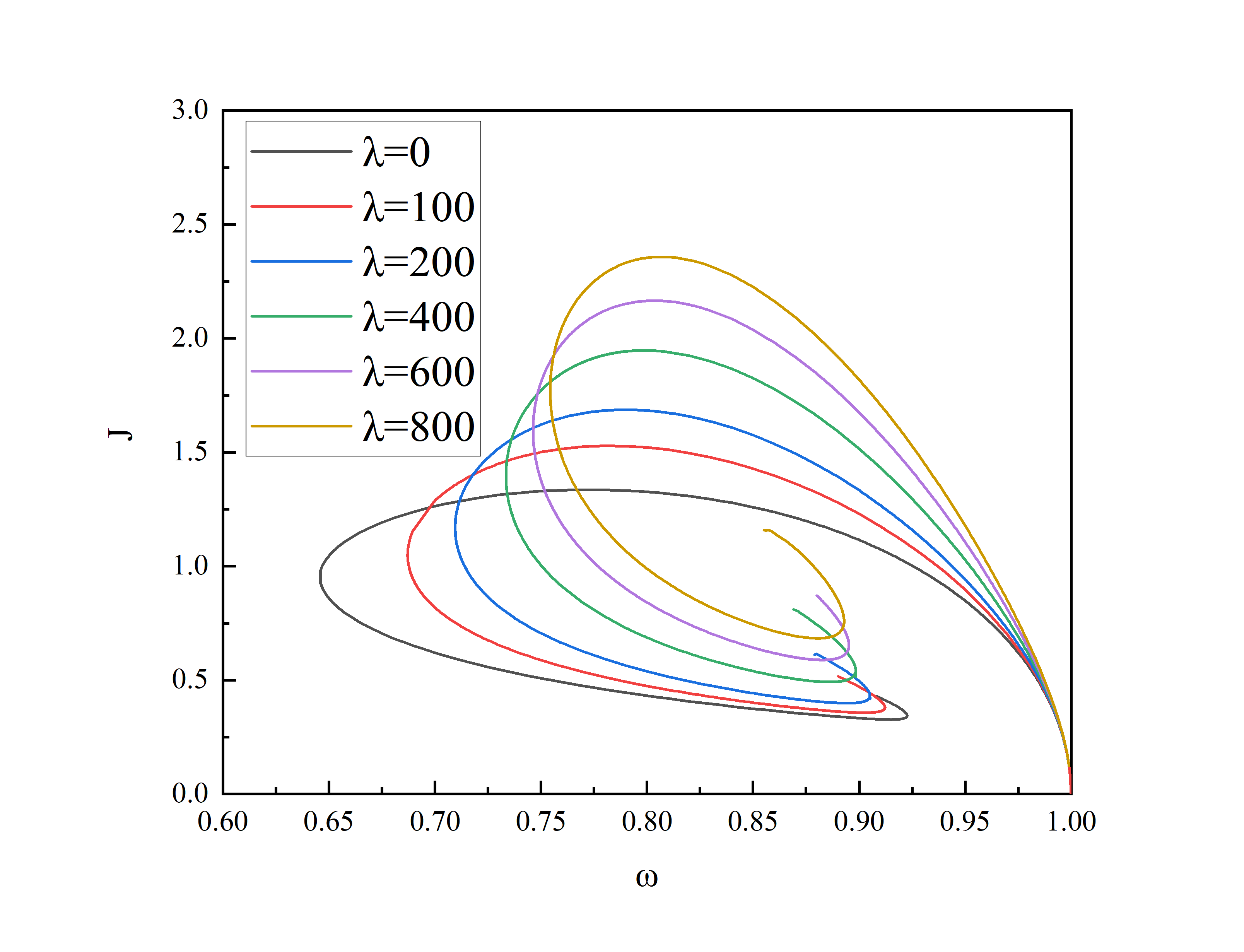"}
		\end{minipage}
		\begin{minipage}[b]{.3\linewidth}
			\centering
			\includegraphics[scale=0.2]{"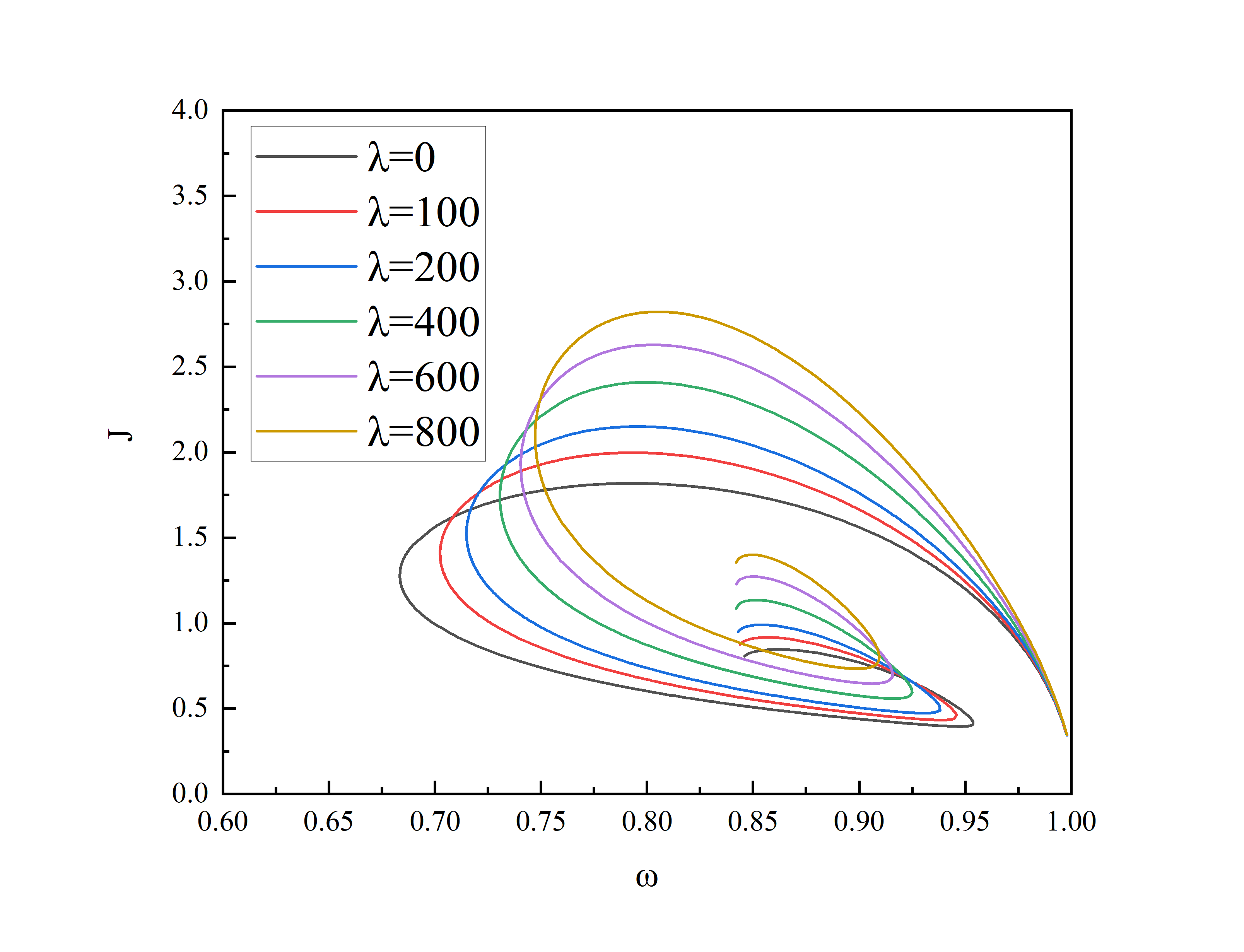"}
		\end{minipage}
		\begin{minipage}[b]{.3\linewidth}
			\centering
			\includegraphics[scale=0.2]{"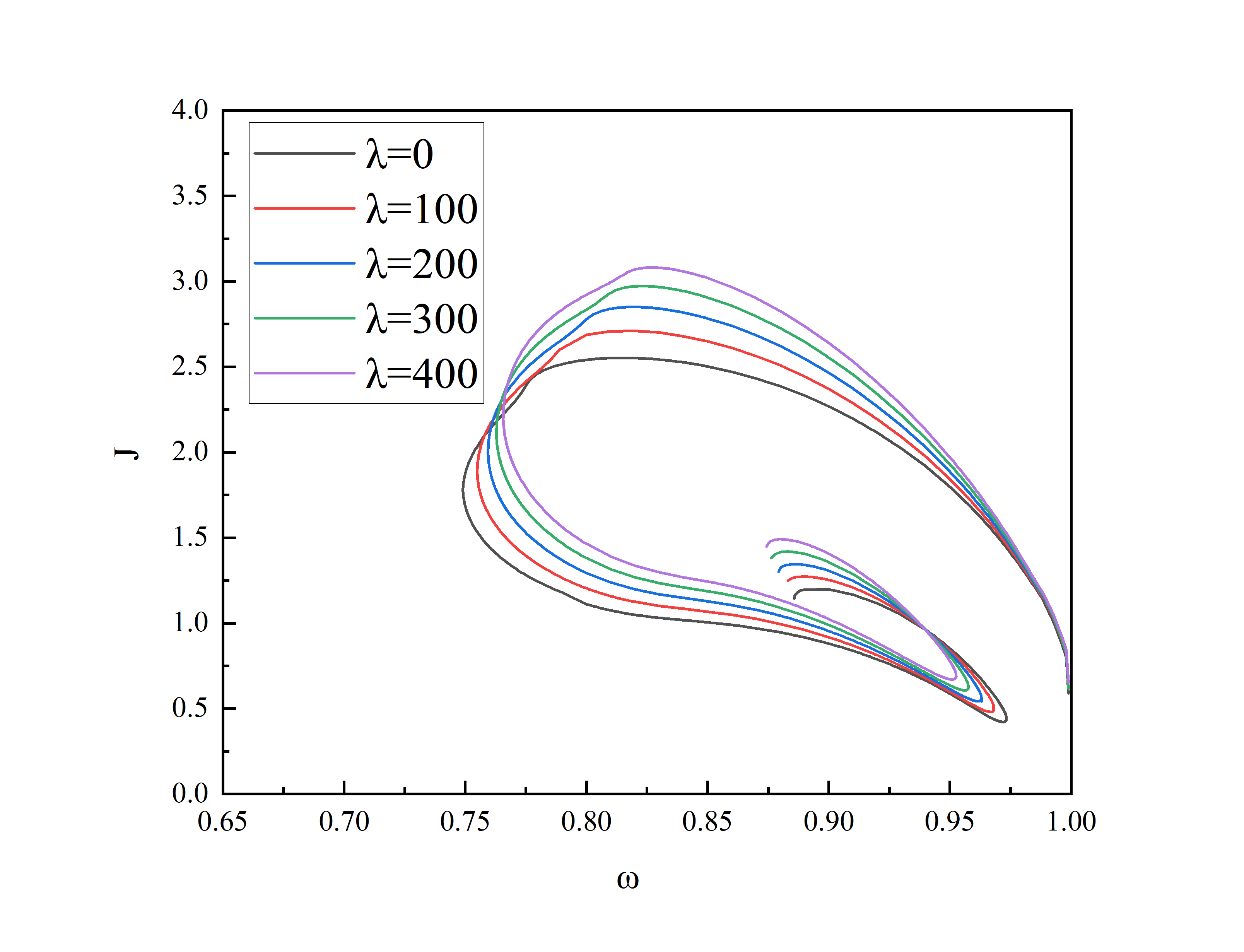"}
		\end{minipage}
	}
	\caption{The total angular momentum $J$ as a function of the frequency $\omega$. Left: The rotating BSs with quartic self-interaction of one constituent. Middle: The rotating BSs with quartic self-interaction of two constituents. Right: The rotating BSs with quartic self-interaction of four constituents.}
	\label{Fig.2}
\end{figure}

\begin{figure}[htbp]
	\centering
	\subfigure
	{
		\begin{minipage}[b]{0.45\textwidth}
			\centering
			\includegraphics[scale=0.3]{"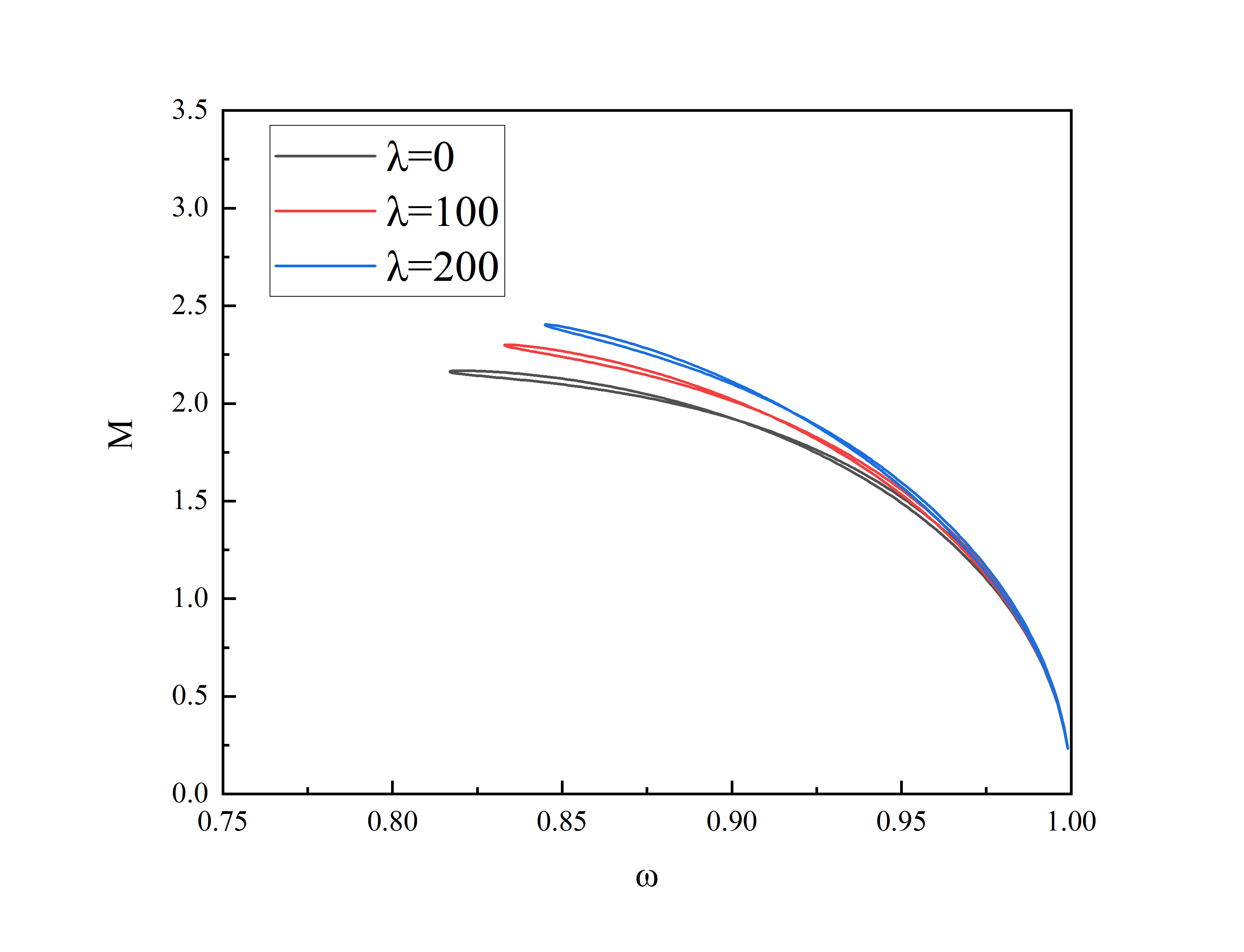"}
		\end{minipage}
	}
	\subfigure
	{
		\begin{minipage}[b]{0.45\textwidth}
			\centering
			\includegraphics[scale=0.3]{"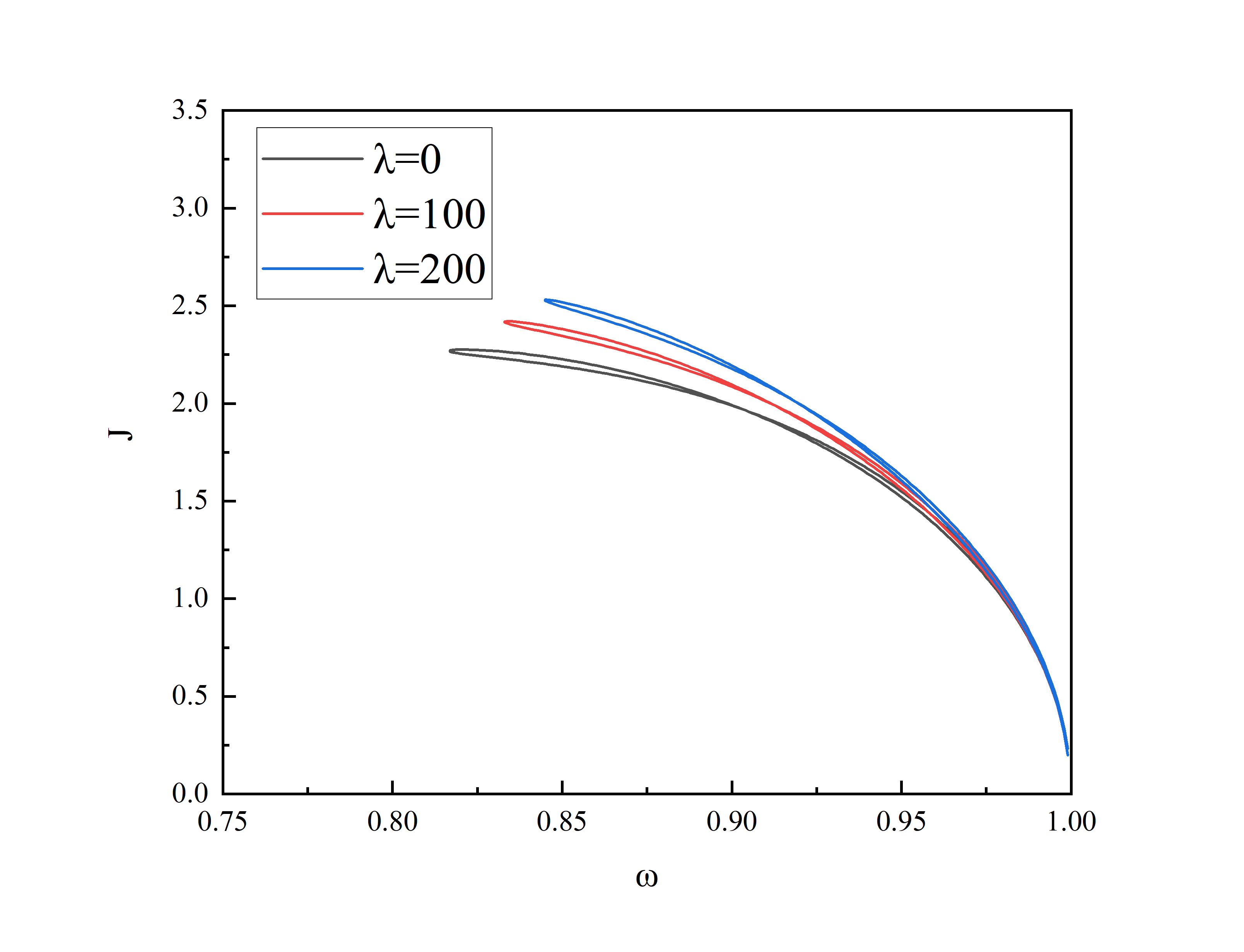"}
		\end{minipage}
	}
	\caption{The curves of $(\omega, M)$ and $(\omega, J)$ for three-constituent case. Left: The ADM mass $M$ with the different frequency $\omega$ of the rotating BSs with quartic self-interaction of three constituents. Right: The total angular momentum $J$ with different frequency $\omega$ of the rotating BSs with quartic self-interaction of three constituents.}
	\label{Fig.3}
\end{figure}

\begin{figure}[htbp]
	\centering
	{
		\begin{minipage}[b]{.3\linewidth}
			\centering
			\includegraphics[scale=0.133]{"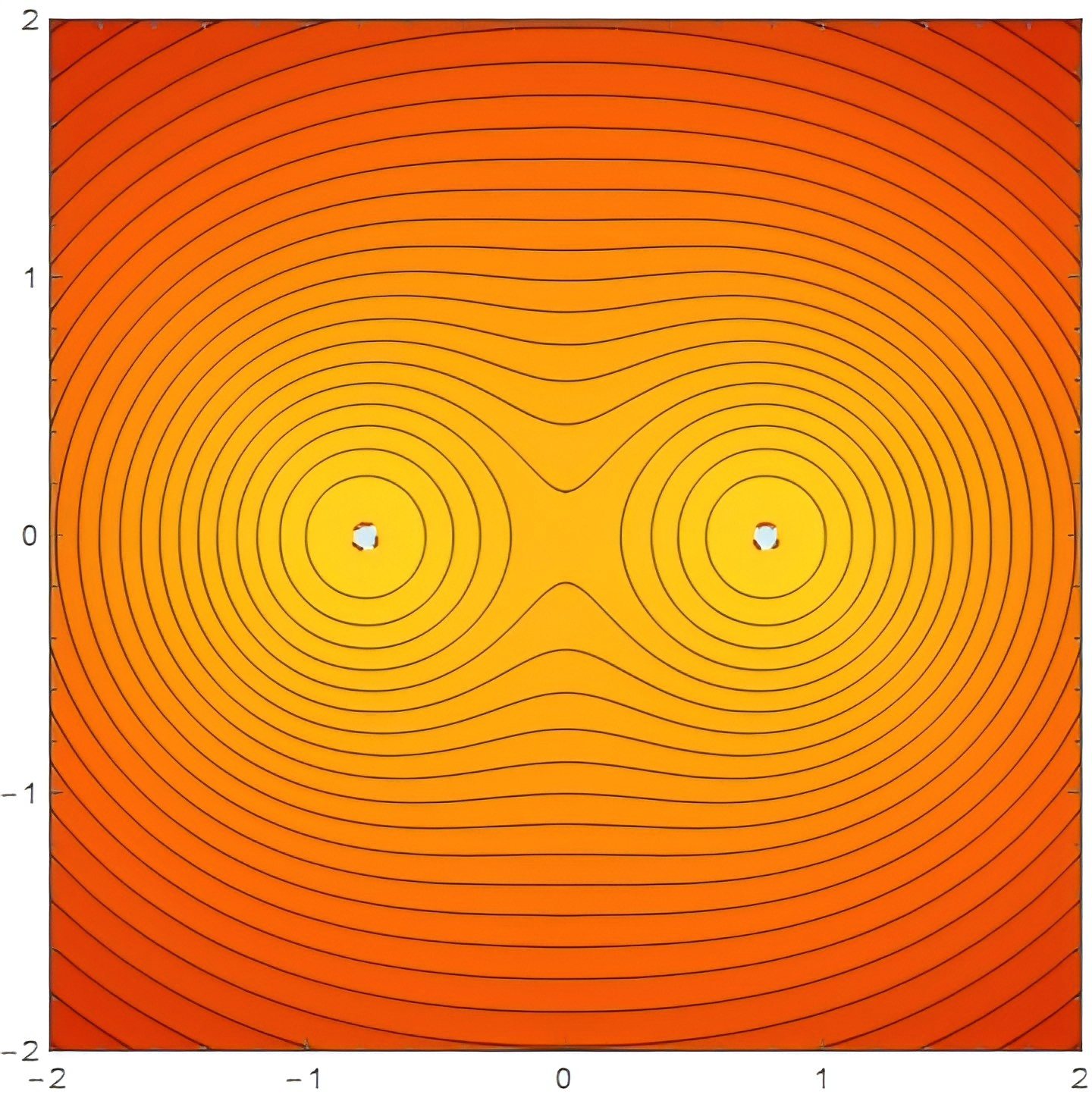"}
		\end{minipage}
		\begin{minipage}[b]{.3\linewidth}
			\centering
			\includegraphics[scale=0.133]{"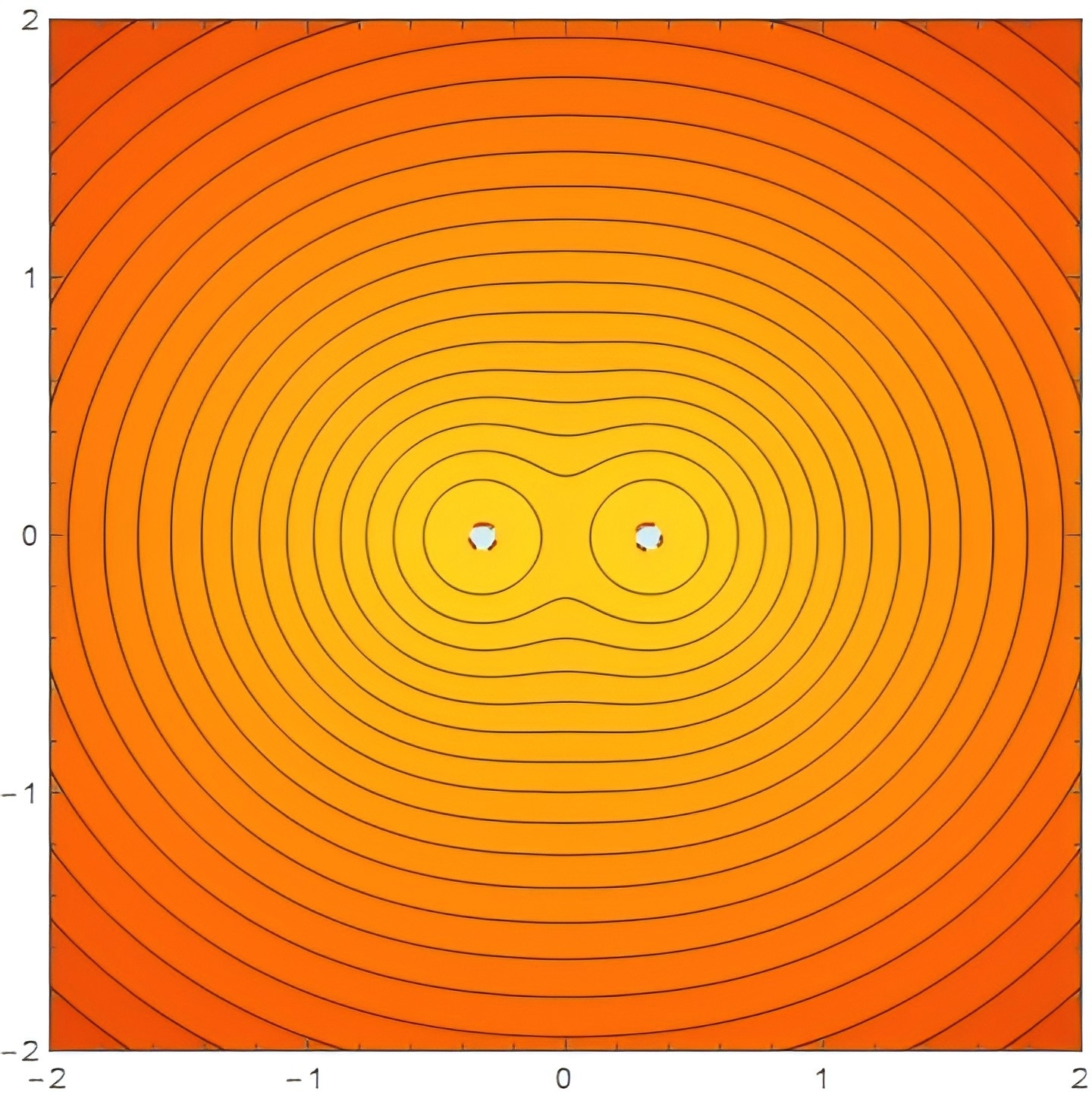"}
		\end{minipage}
		\begin{minipage}[b]{.3\linewidth}
			\centering
			\includegraphics[scale=0.133]{"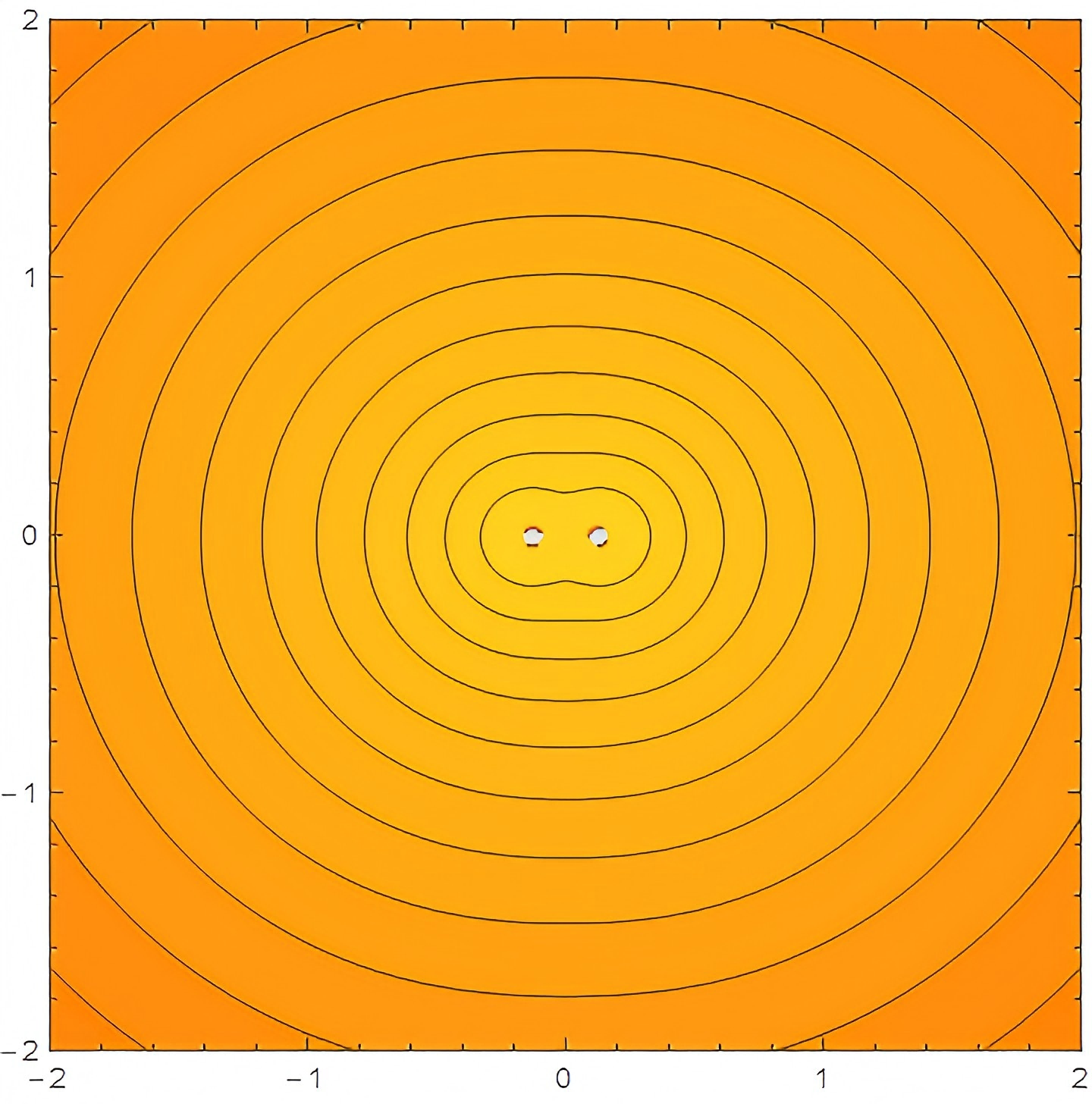"}
		\end{minipage}
	}
	\caption{Contours of $g_{tt}$ for chains of rotating BSs with one constituent. Left: When $\lambda = 0$, the ergosphere emerges at $\omega = 0.658$ on the first branch. Middle: When $\lambda = 400$, the ergosphere emerges at $\omega = 0.874$ on the second branch. Right: When $\lambda = 800$, the ergosphere emerges at $\omega = 0.851$ on the third branch. Warm and cold color schemes denote negative and positive $g_{tt}$ values, respectively. The red dashed line represents ergospheres ($g_{tt}$ = 0), and the black solid lines represent the contours of $g_{tt}$.}
	\label{Fig.4}
\end{figure}

We numerically solve the EKG system, Eqs. (\ref{PDE1}) - (\ref{PDE4}) and (\ref{PDE5}), to construct chains of rotating BSs with quartic self-interaction. Figs. \ref{Fig.1} and \ref{Fig.2} show the ADM mass $M$ and the angular momentum $J$ as functions of the frequency $\omega$ for rotating BSs chains with quartic self-interaction, comprising one, two, and four constituents. The results indicate that the $\omega$-dependence of $M$ and $J$ for the two- and four-constituent chains is qualitatively similar to that of the single-constituent case. Both the ADM mass $M$ and the angular momentum $J$ exhibit a consistent pattern of evolution with the frequency $\omega$. The first branch terminates at the minimum frequency. As $\omega$ decreases beyond this point, $M$ and $J$ initially increase to their maximum before declining. Starting from the frequency minimum on the second branch, $M$ and $J$ exhibit a non-monotonic dependence on $\omega$, initially decreasing, then increasing until a turning point marks the transition to the third branch. The third branch is qualitatively similar to the first, with $M$ and $J$ exhibiting a similar evolution. The configuration of three branches results in an overall spiral pattern. Moreover, the curves show a monotonic upward trend with both an increasing number of constituents and higher values of $\lambda$.

	\begin{figure}[htbp]
	\centering
	{
		\begin{minipage}[b]{.3\linewidth}
			\centering
			\includegraphics[scale=0.133]{"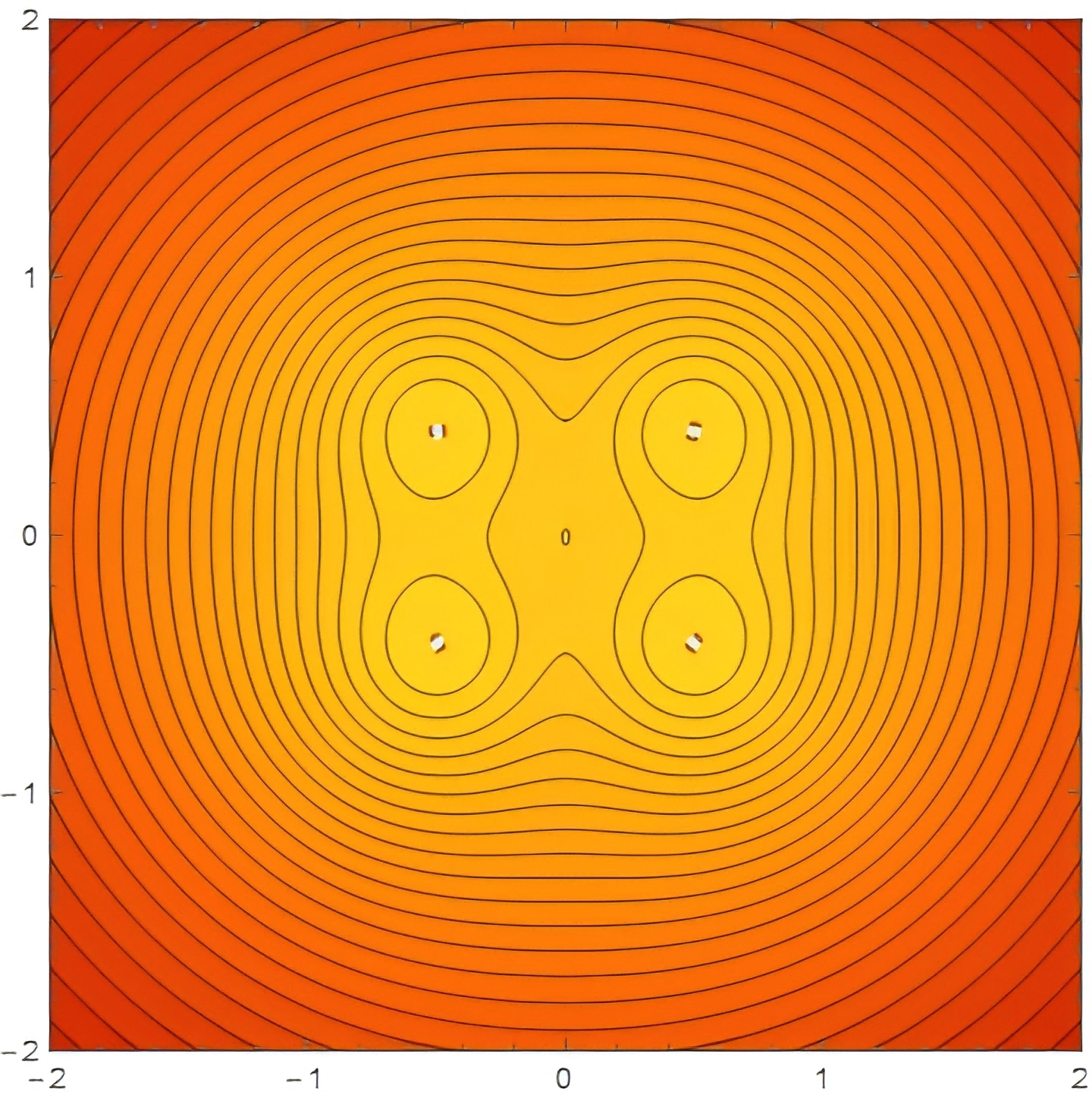"}
		\end{minipage}
		\begin{minipage}[b]{.3\linewidth}
			\centering
			\includegraphics[scale=0.133]{"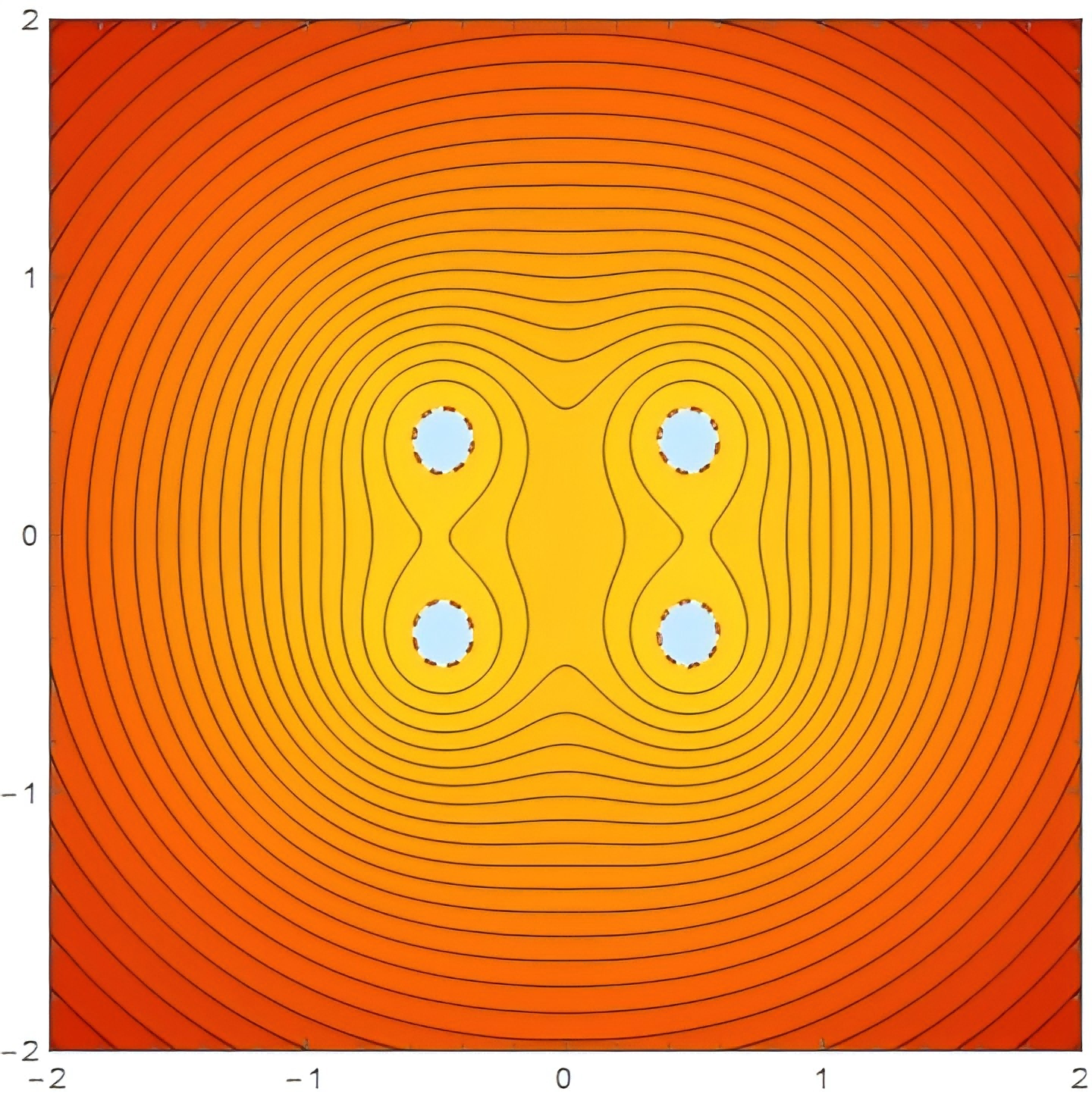"}
		\end{minipage}
		\begin{minipage}[b]{.3\linewidth}
			\centering
			\includegraphics[scale=0.133]{"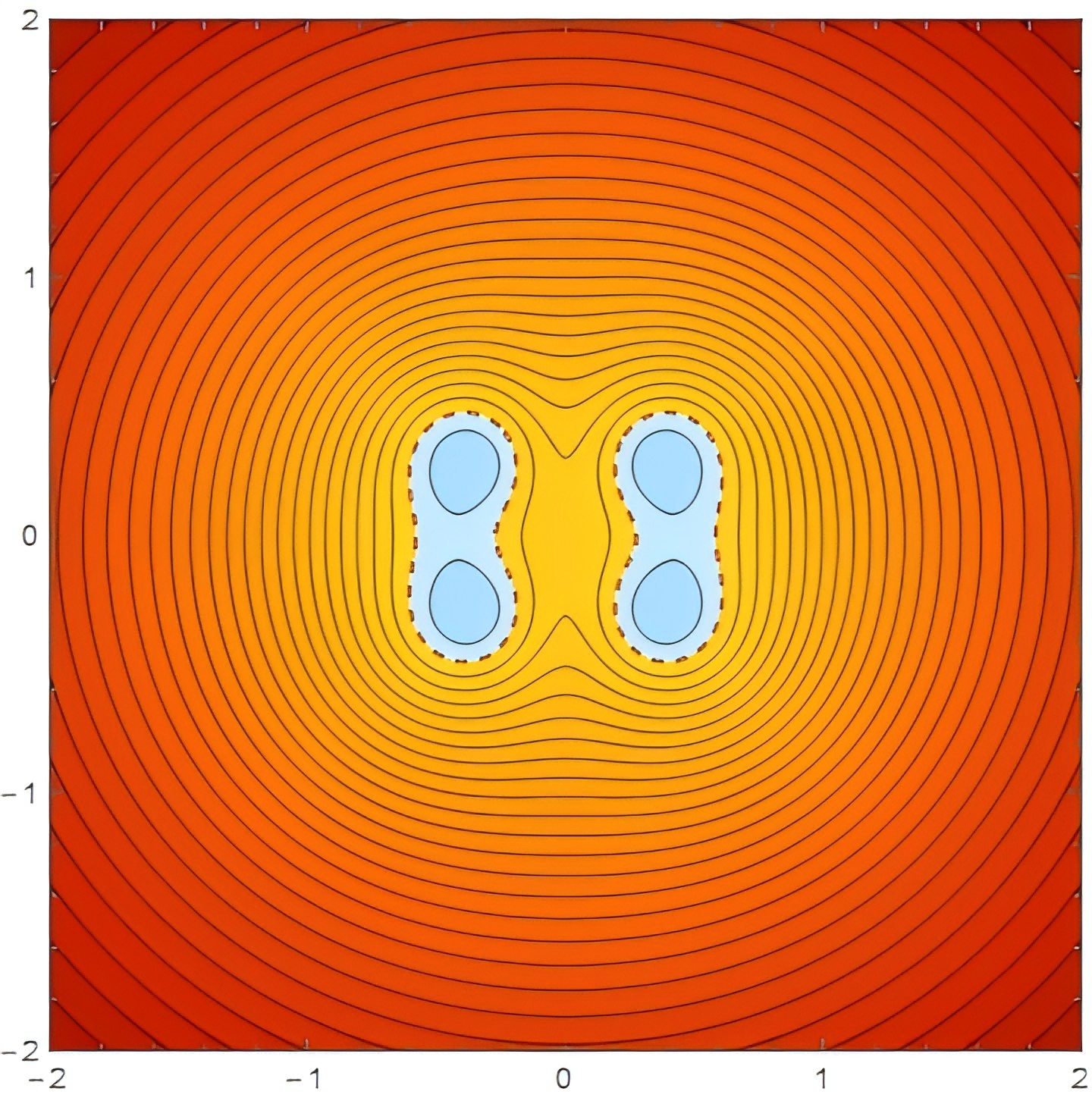"}
		\end{minipage}
	}
	\caption{Contours of $g_{tt}$ for chains of rotating BSs with two constituents. Left: The ergospheres emerge at $\omega=0.719$ on the second branch. Middle: With the increase of $\omega$, the ergospheres begin to merge at $\omega=0.730$ on the second branch. Right: The two ergospheres merge into one at $\omega=0.790$ on the second branch. Warm and cold color schemes denote negative and positive $g_{tt}$ values, respectively. The red dashed line represents ergospheres ($g_{tt}$ = 0), and the black solid lines represent the contours of $g_{tt}$.}
	\label{Fig.5} 
\end{figure}

In Fig. \ref{Fig.3}, the relationships between $M-\omega$ and $J-\omega$ are depicted for chains composed of three-constituent case. The figures show a loop structure in the dependence of both $M$ and $J$ on $\omega$. For the three-constituent case, the $(\omega, M)$ and $(\omega, J)$ diagram forms a two-branched loop, indicating a fragile balance between gravitational and repulsive interactions. The first branch is defined as the trajectory from the maximum frequency, where $M$ and $J$ initially rise to their peaks with decreasing $\omega$ before undergoing a monotonic decrease. On the second branch, the rotating case traces a monotonic decrease in both $M$ and $J$ with increasing $\omega$, ultimately closing the loop. Stable solutions for this case are only found for coupling strengths $\lambda \leq 250$. This contrasts with other constituents, where solutions also exist in the stronger-coupling regime.

 
\begin{figure}[htbp]
	\centering
	{
		\begin{minipage}[b]{.3\linewidth}
			\centering
			\includegraphics[scale=0.133]{"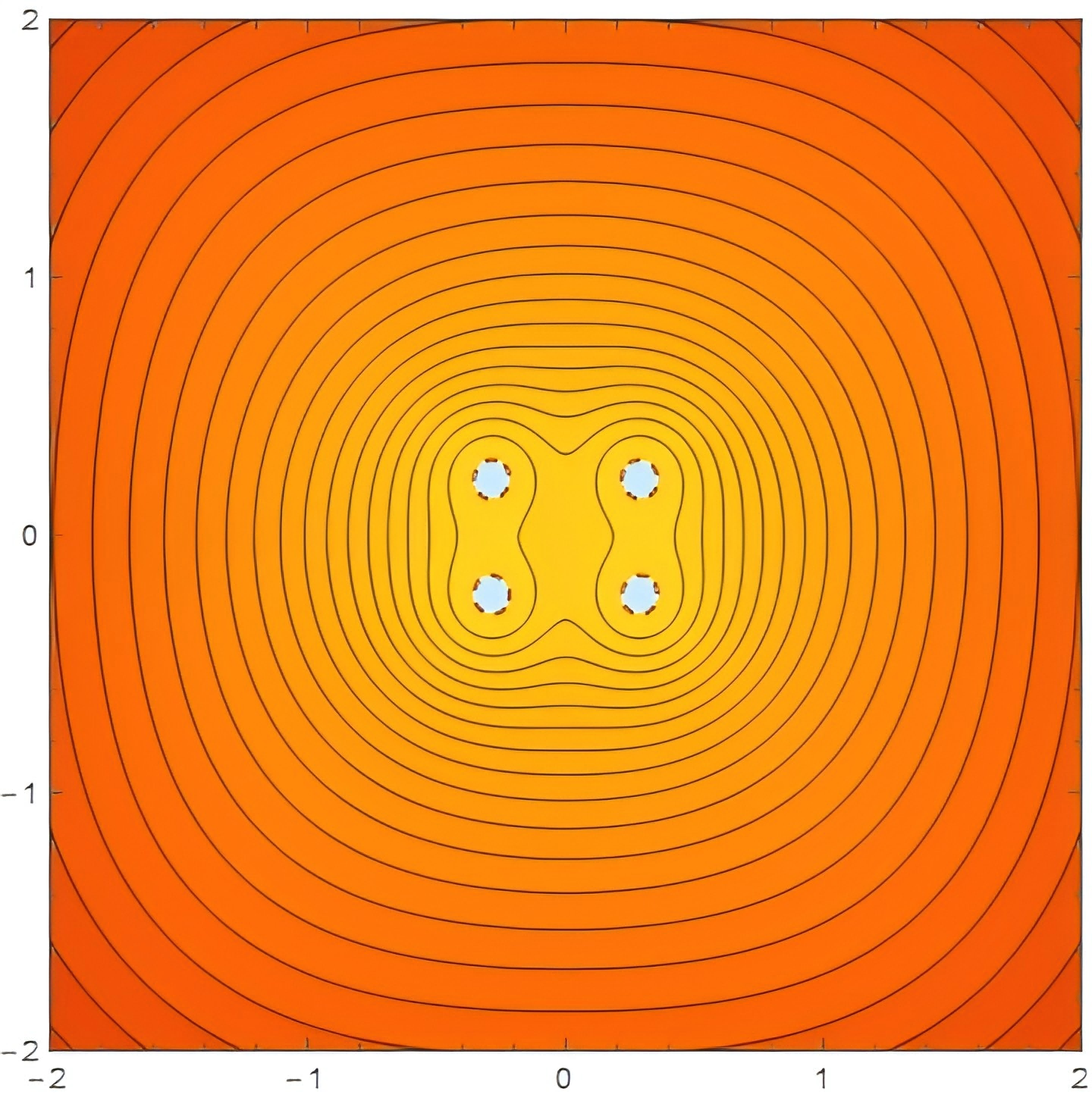"}
		\end{minipage}
		\begin{minipage}[b]{.3\linewidth}
			\centering
			\includegraphics[scale=0.133]{"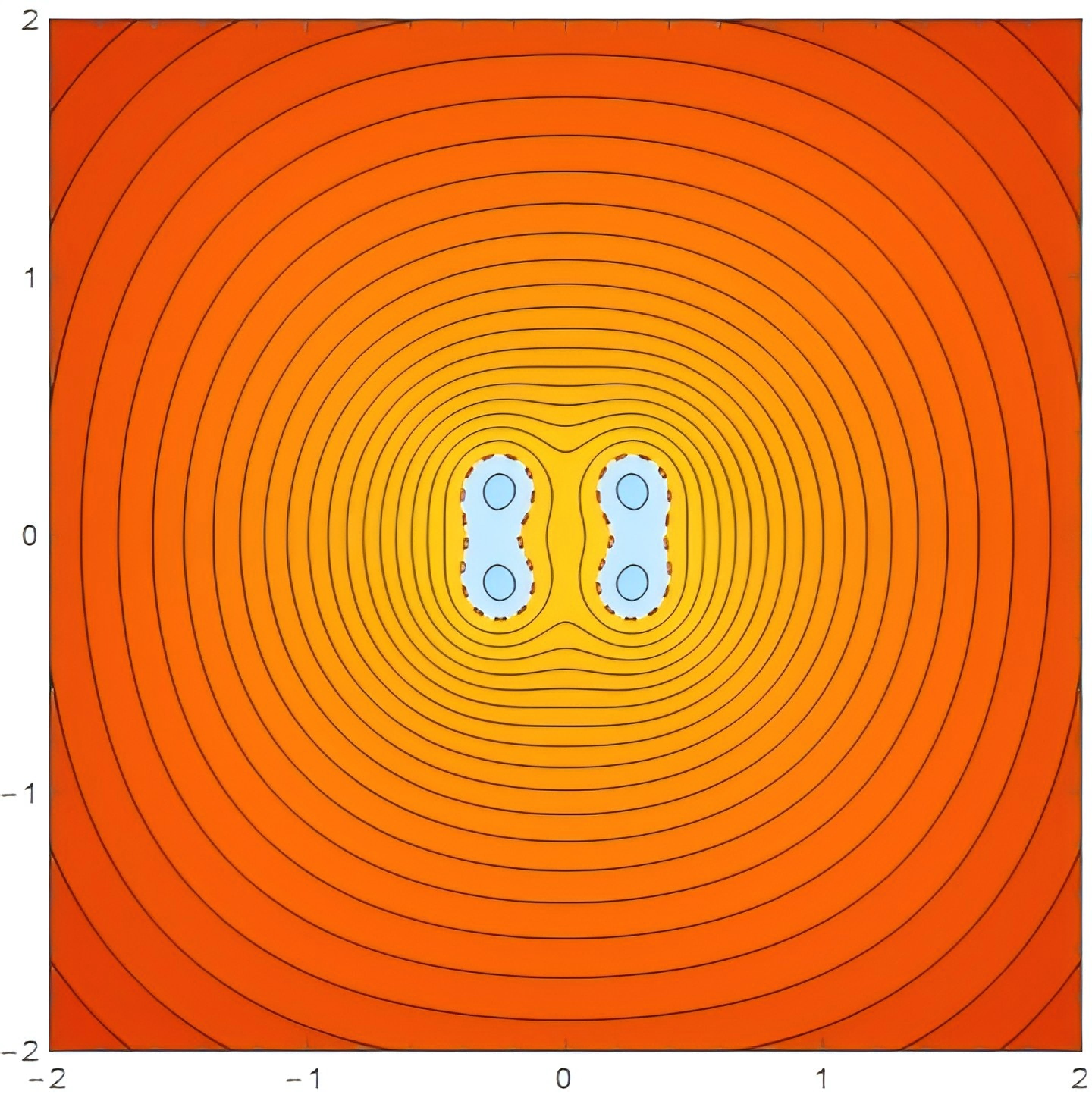"}
		\end{minipage}
		\begin{minipage}[b]{.3\linewidth}
			\centering
			\includegraphics[scale=0.133]{"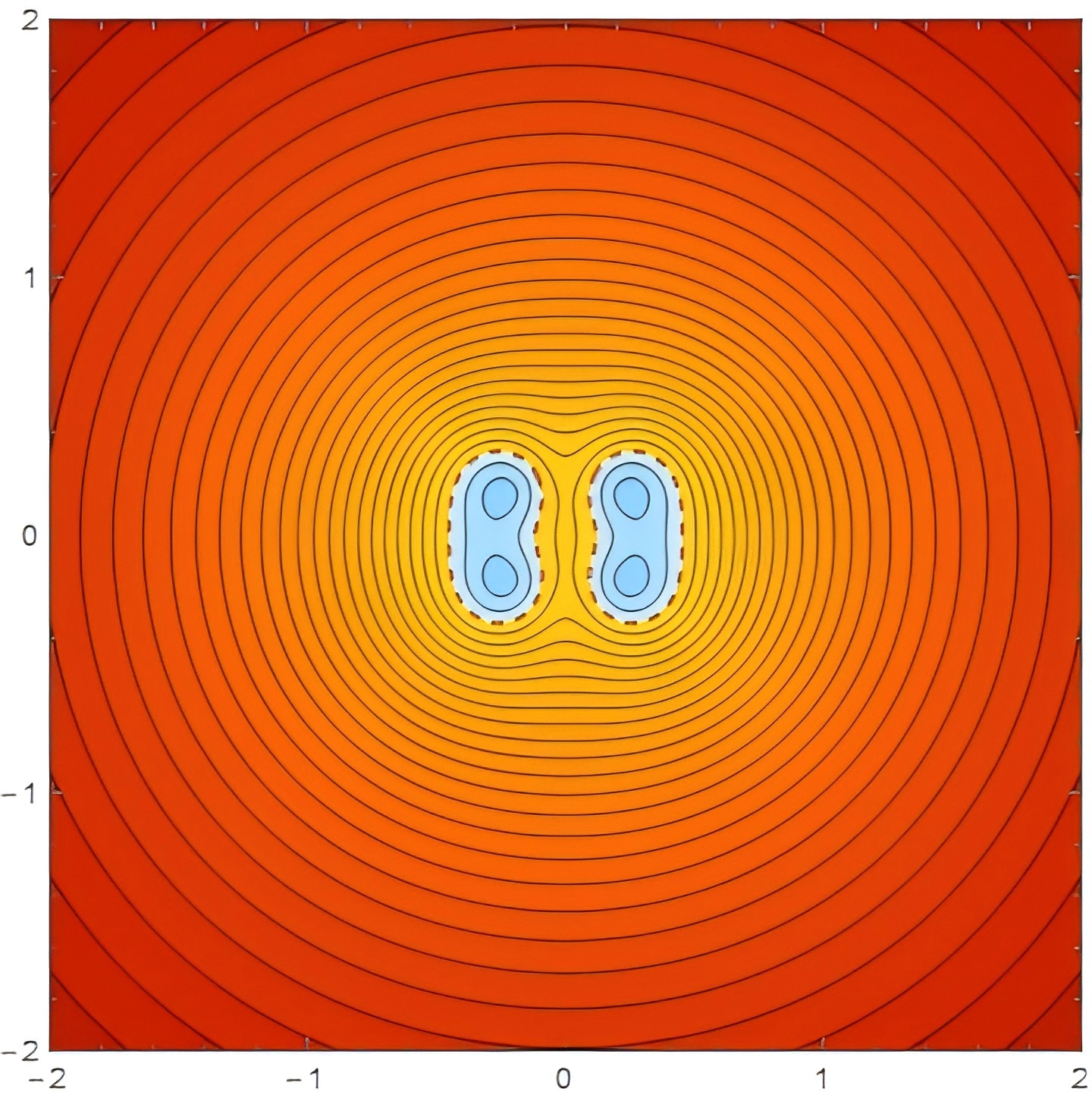"}
		\end{minipage}
	}
	\caption{Contours of $g_{tt}$ for chains of rotating BSs with four constituents. Left: The ergospheres emerge at $\omega=0.810$ on the second branch. Middle: With the increase of $\omega$, the ergospheres begin to merge at $\omega=0.840$ on the second branch. Right: The two ergospheres merge into one at $\omega=0.890$ on the second branch. Warm and cold color schemes denote negative and positive $g_{tt}$ values, respectively. The red dashed line represents ergospheres ($g_{tt}$ = 0), and the black solid lines represent the contours of $g_{tt}$.}
	\label{Fig.6}
\end{figure}
	
Ergospheres exist in certain chains of rotating BSs with quartic self-interaction, specifically those with one, two, and four constituents. The three-constituent case does not exhibit ergospheres due to insufficient compactness of the system. Fig. \ref{Fig.4} shows the ergosphere of a single-constituent rotating BS with quartic self-interaction. Ergospheres first form when the frequency drops to $\omega = 0.658$ on the first branch. With increasing coupling strength, the frequency at which ergospheres emerge also rises. For instance, at $\lambda = 400$ the emergence frequency is $\omega = 0.874$ on the second branch, while at $\lambda = 800$ it is $\omega = 0.851$ on the third branch. This trend also holds for the two- and four-constituent cases.


\begin{figure}[htbp]
	\centering
	{
		\begin{minipage}[b]{.3\linewidth}
			\centering
			\includegraphics[scale=0.133]{"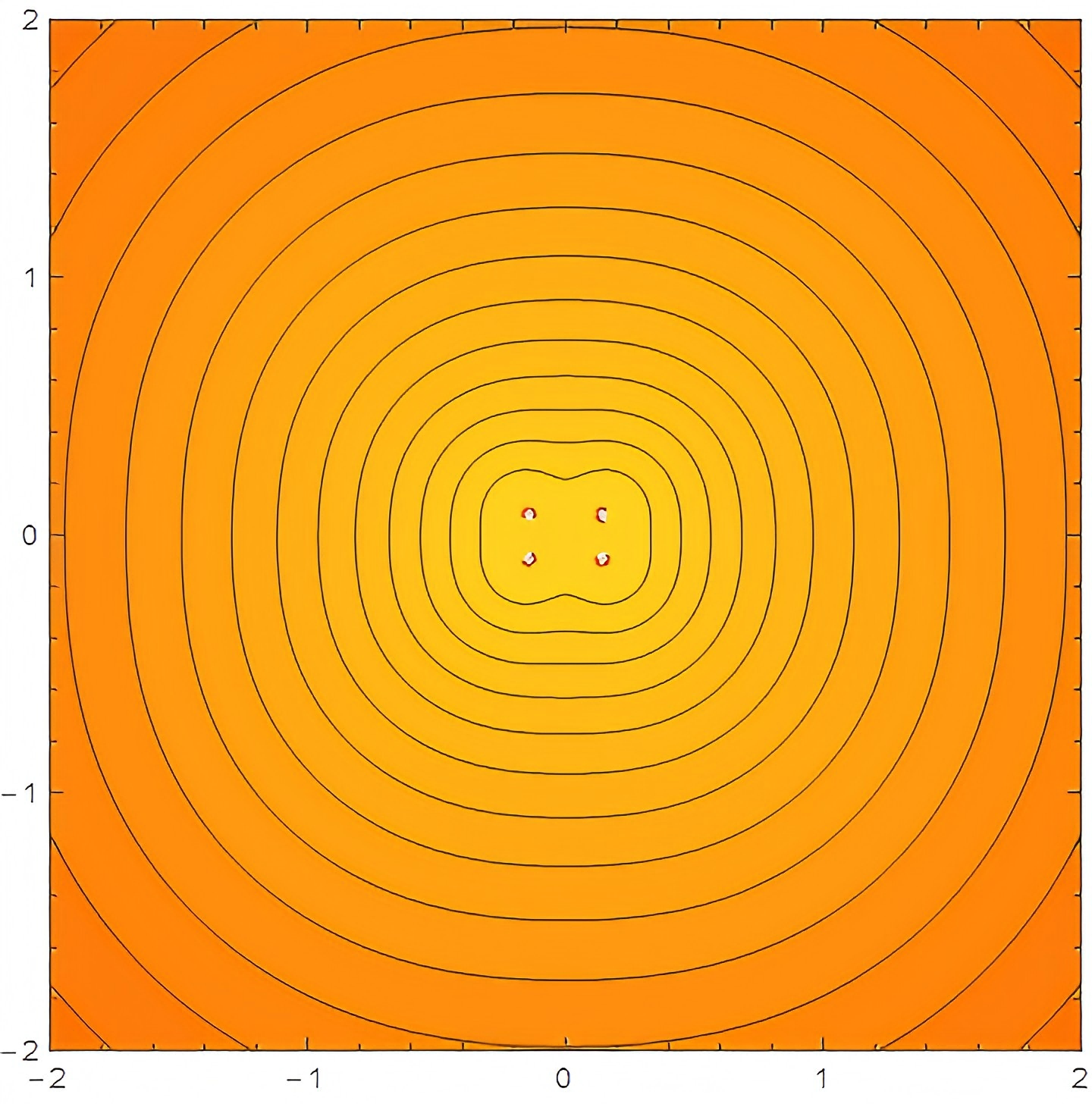"}
		\end{minipage}
		\begin{minipage}[b]{.3\linewidth}
			\centering
			\includegraphics[scale=0.133]{"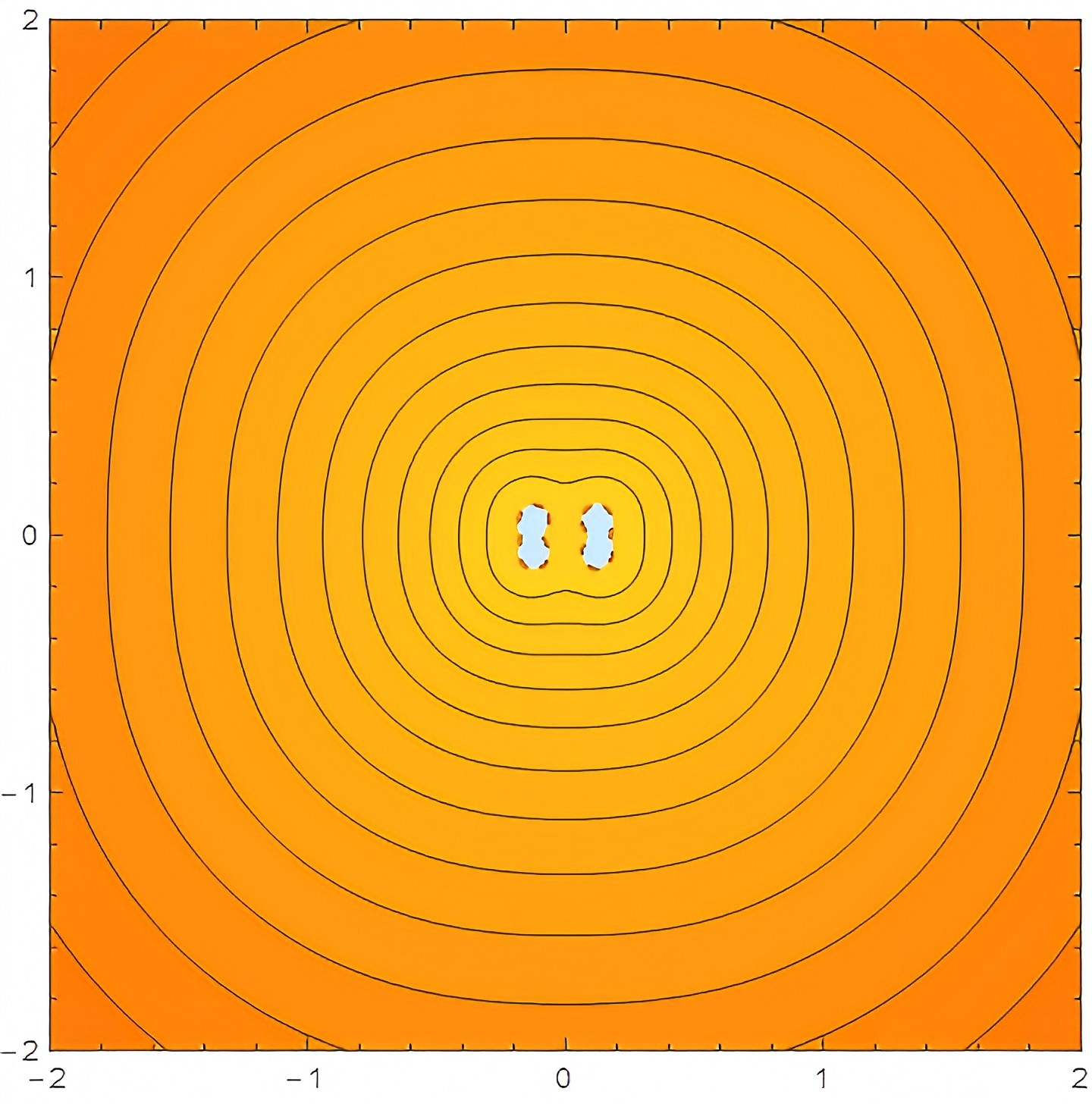"}
		\end{minipage}
		\begin{minipage}[b]{.3\linewidth}
			\centering
			\includegraphics[scale=0.133]{"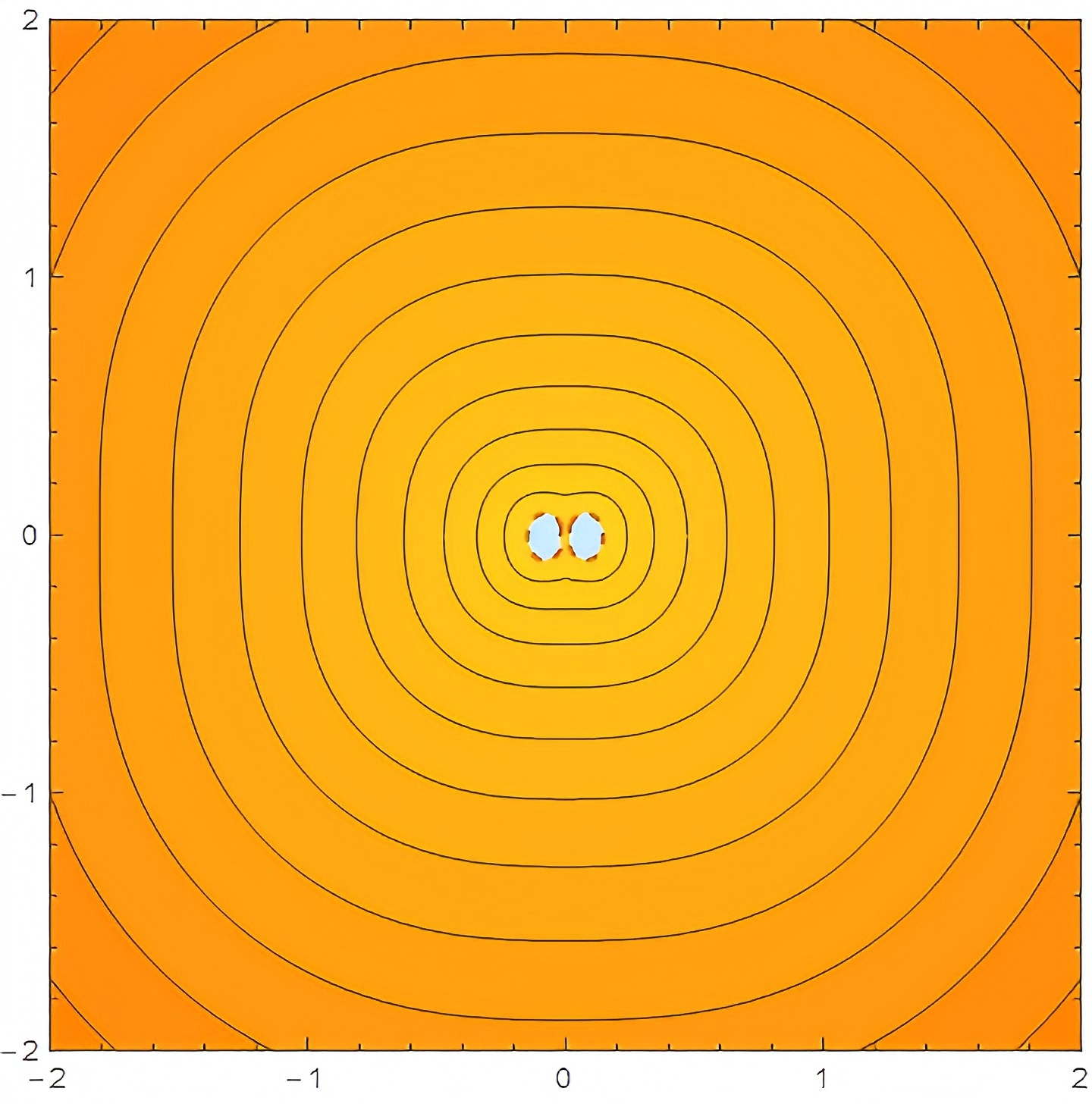"}
		\end{minipage}
	}
	\caption{Contours of $g_{tt}$ for chains of rotating BSs with two constituents with $\lambda = 200$. Left: The ergospheres emerge at $\omega=0.858$ on the third branch. Middle: With the increase of $\omega$, the ergospheres begin to merge at $\omega = 0.850$ on the third branch. Right: The two ergospheres merge into one at $\omega = 0.840$ on the third branch. Warm and cold color schemes denote negative and positive $g_{tt}$ values, respectively. The red dashed line represents ergospheres ($g_{tt}$ = 0), and the black solid lines represent the contours of $g_{tt}$.}
	\label{Fig.7}
\end{figure}

\begin{figure}[htbp]
	\centering	
	{
		\begin{minipage}[b]{.3\linewidth}
			\centering
			\includegraphics[scale=0.133]{"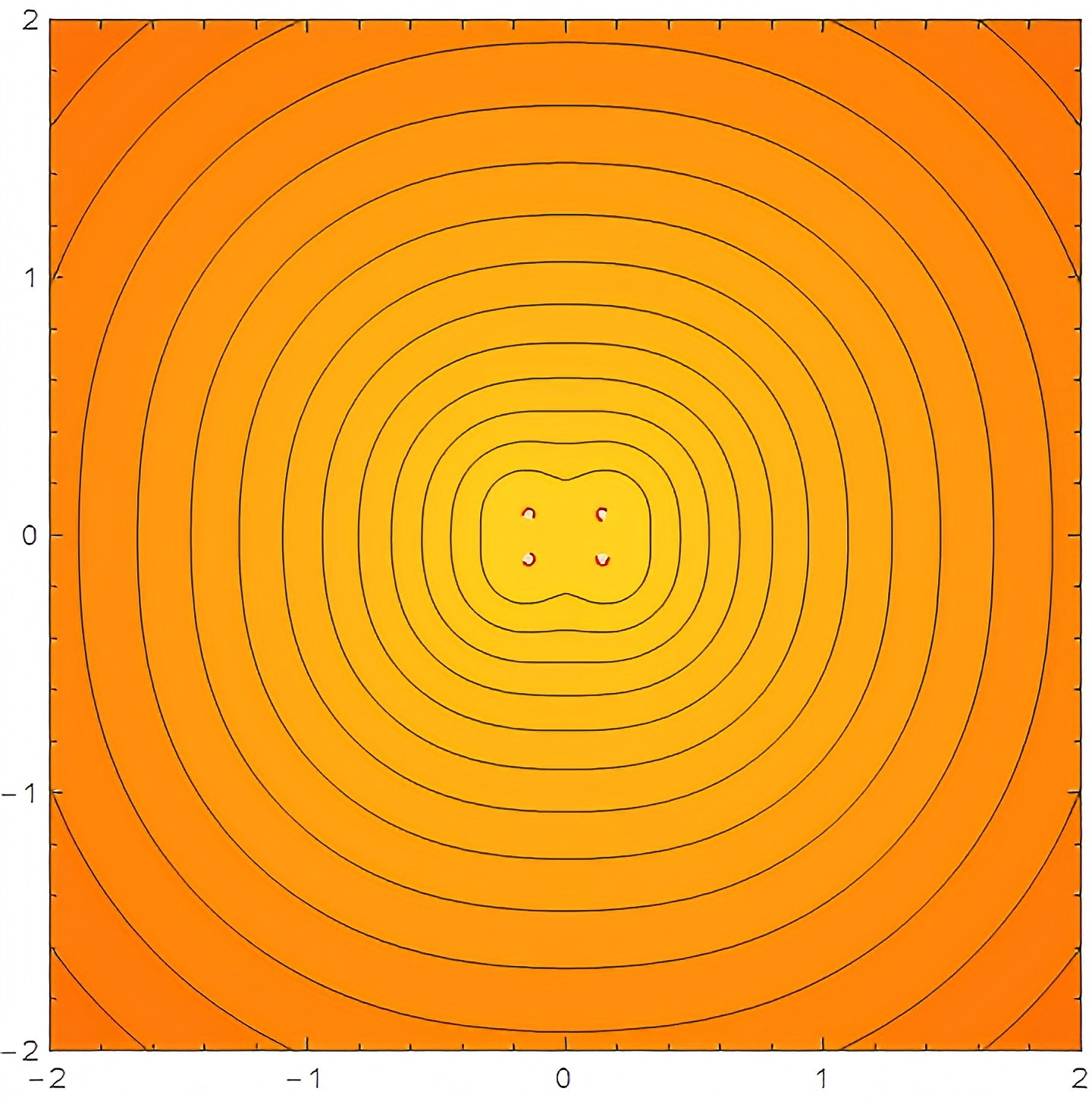"}
		\end{minipage}
		\begin{minipage}[b]{.3\linewidth}
			\centering
			\includegraphics[scale=0.133]{"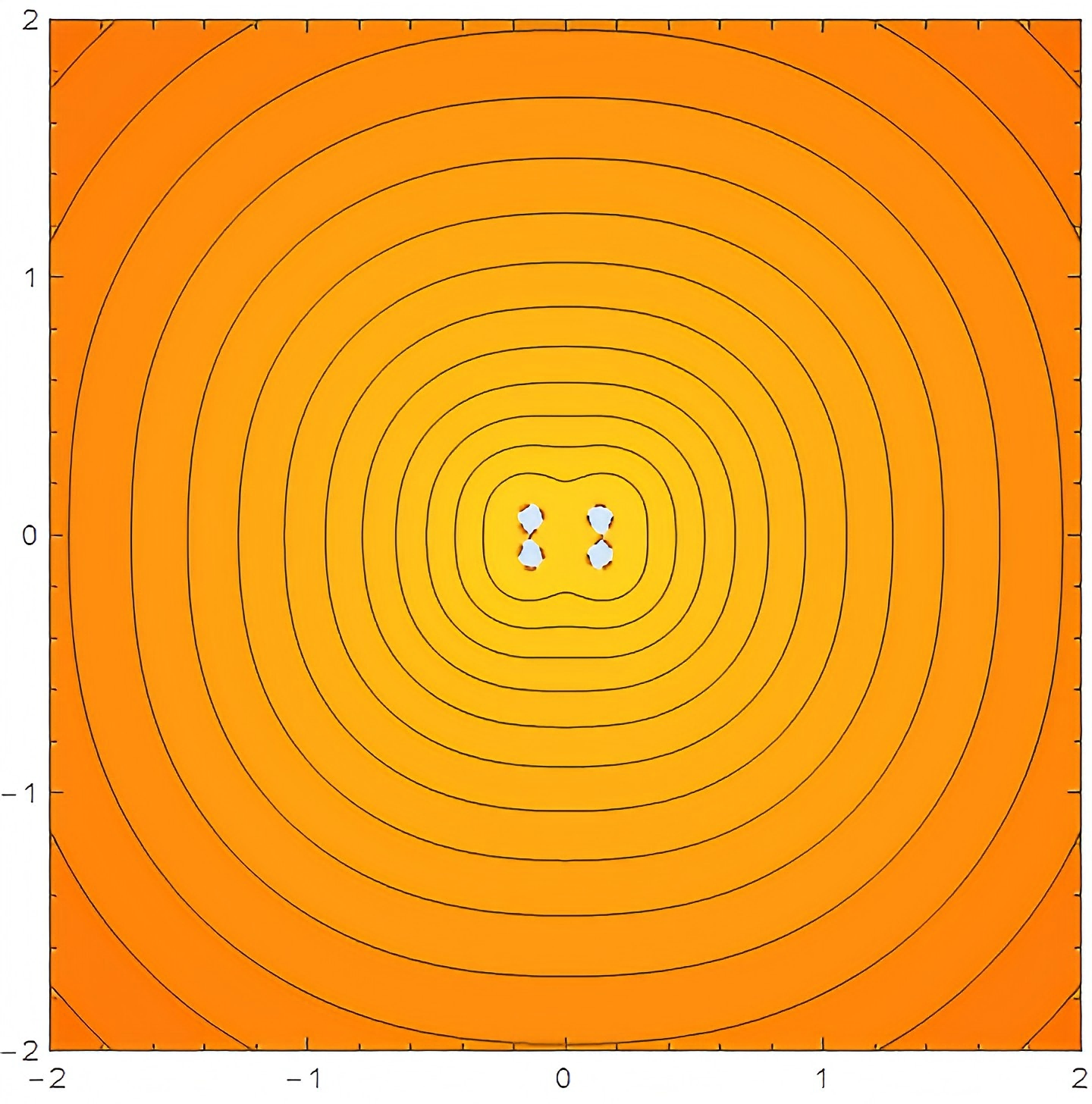"}
		\end{minipage}
		\begin{minipage}[b]{.3\linewidth}
			\centering
			\includegraphics[scale=0.133]{"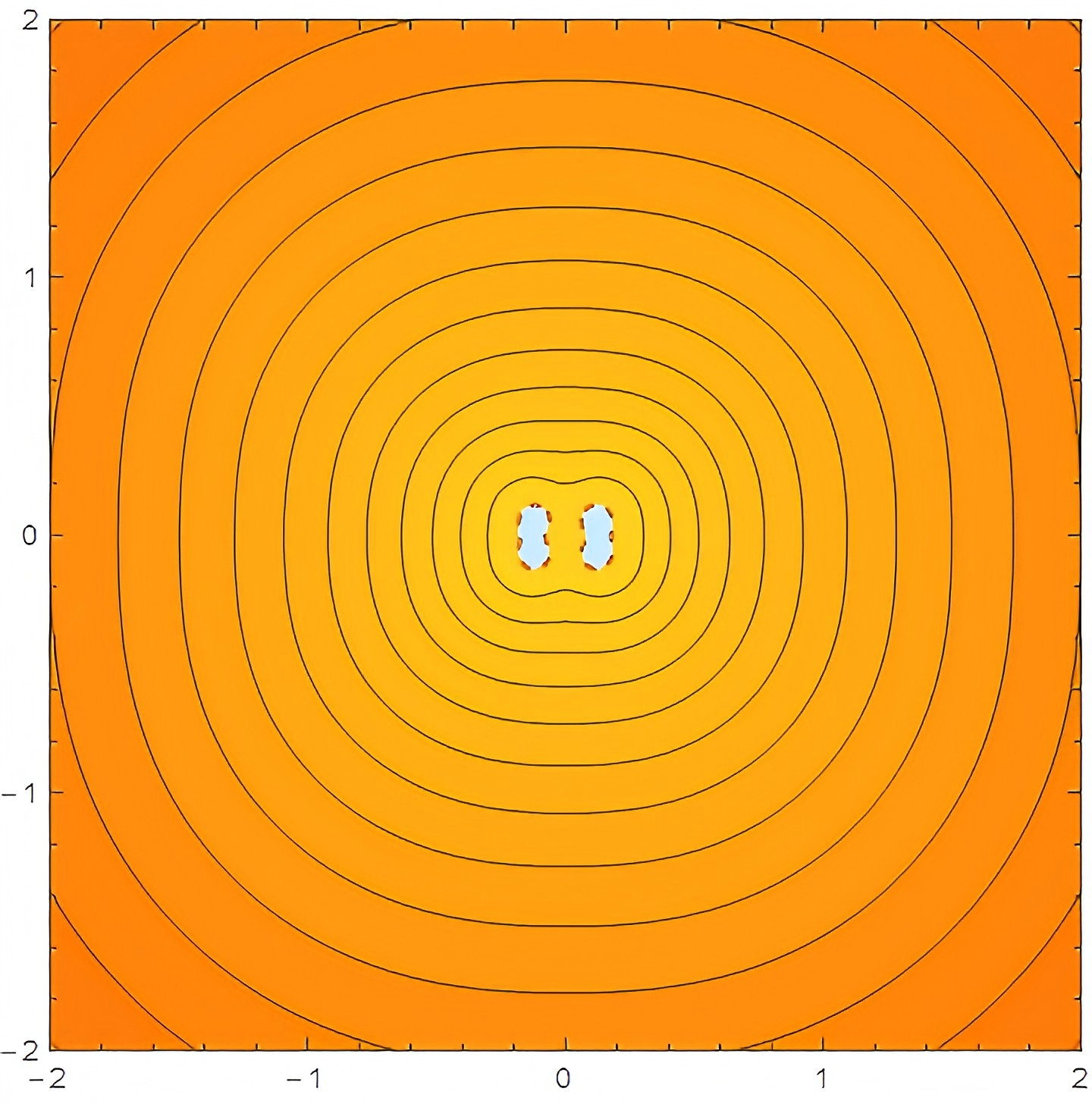"}
		\end{minipage}
	}
	\caption{Contours of $g_{tt}$ for chains of rotating BSs with four constituents with $\lambda = 200$. Left: The ergospheres emerge at $\omega = 0.869$ on the third branch. Middle: With the increase of $\omega$, the ergospheres begin to merge at $\omega = 0.865$ on the third branch. Right: The two ergospheres merge into one at $\omega = 0.860$ on the third branch. Warm and cold color schemes denote negative and positive $g_{tt}$ values, respectively. The red dashed line represents ergospheres ($g_{tt}$ = 0), and the black solid lines represent the contours of $g_{tt}$.}
	\label{Fig.8}
\end{figure}

The emergence of two ergospheres in chains with two and four constituents is shown in Figs. \ref{Fig.5} and \ref{Fig.6}, respectively. On the second branch, the ergospheres emerge at $\omega = 0.719$ for the two-constituent case. For the four-constituent case, the ergospheres emerge at $\omega = 0.810$. In addition, the frequency at which ergospheres emerge increases with the number of constituents. As the frequency $\omega$ increases, the reduced spatial separation between adjacent scalar fields causes their originally distinct ergospheres to converge and coalesce into a single connected region.

\begin{figure}[htbp]
	\centering
	\subfigure
	{
		\begin{minipage}[b]{0.45\textwidth}
			\centering
			\includegraphics[scale=0.15]{"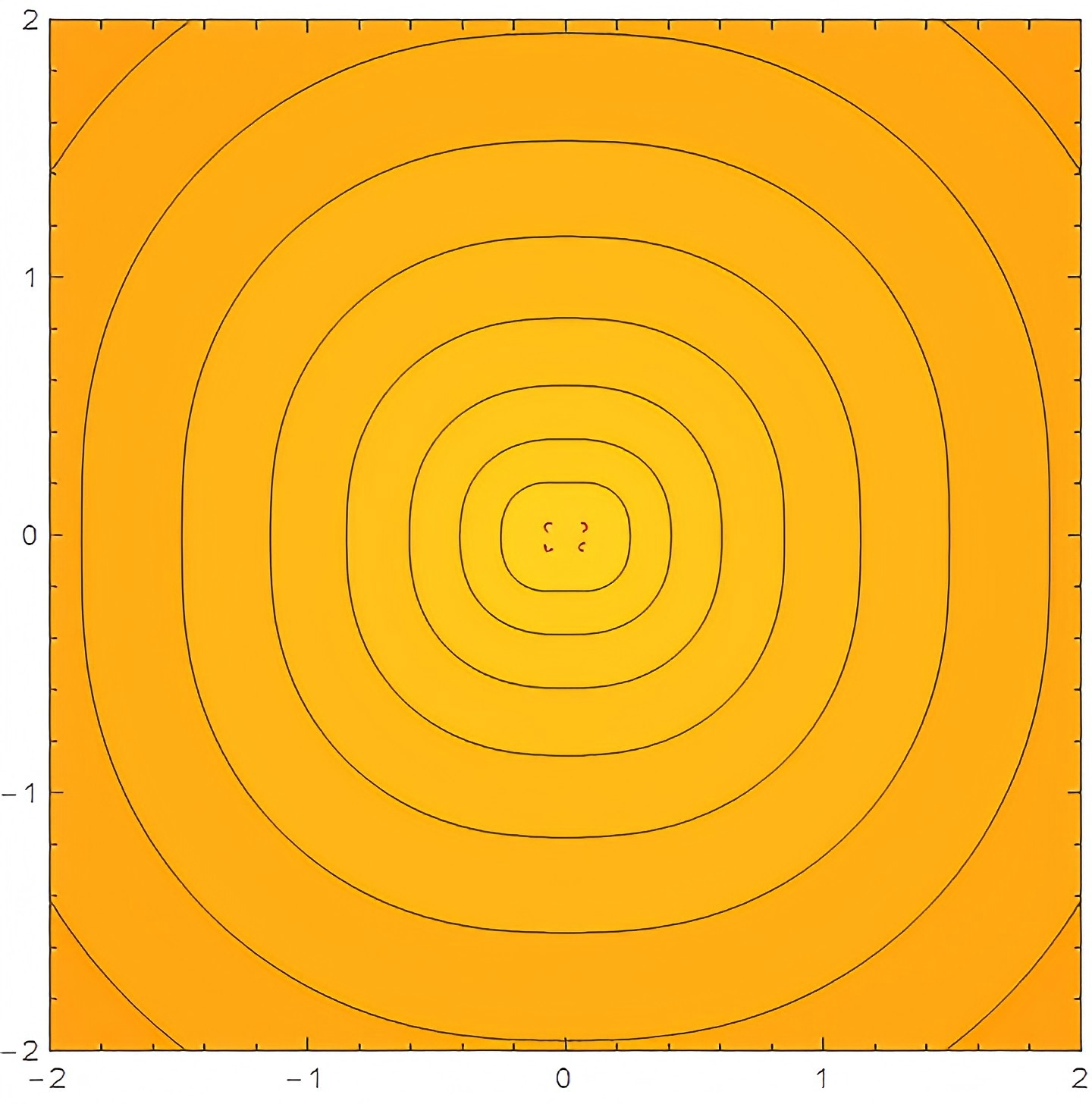"}
		\end{minipage}
	}
	\subfigure
	{
		\begin{minipage}[b]{0.45\textwidth}
			\centering
			\includegraphics[scale=0.15]{"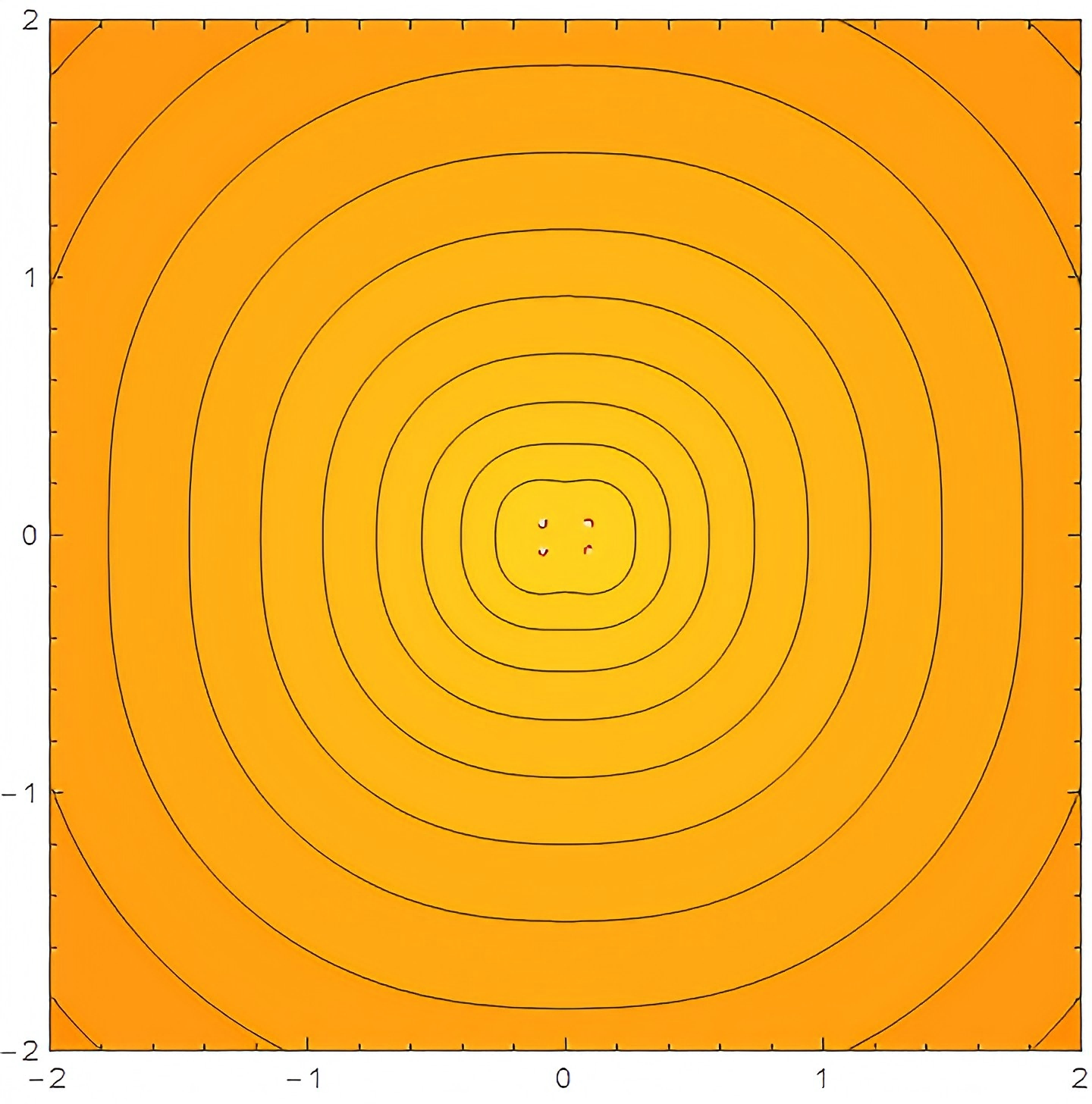"}
		\end{minipage}
	}
	\caption{Contours of $g_{tt}$ for chains of rotating BSs. Left: The ergospheres of two constituents with $\lambda = 600$, the ergospheres emerge at $\omega = 0.841$ on the third branch. Right: The ergospheres of four constituents with $\lambda = 400$, the ergospheres emerge at $\omega = 0.849$ on the third branch. Warm and cold color schemes denote negative and positive $g_{tt}$ values, respectively. The red dashed line represents ergospheres ($g_{tt}$ = 0), and the black solid lines represent the contours of $g_{tt}$.}
	\label{Fig.9}
\end{figure}

Figs. \ref{Fig.7} and \ref{Fig.8} show the ergospheres of the two-constituent and four-constituent chains at the coupling strength of $\lambda = 200$. The results of this study demonstrate that the merging of ergospheres is a commonly observed phenomenon. Our calculations across a range of coupling strengths systematically confirm that ergospheres invariably form and subsequently undergo this unification. Fig. \ref{Fig.9} shows the ergospheres emergence at the end of the third branch for the two-constituent system with $\lambda=600$ and the four-constituent system with $\lambda=400$, respectively. Furthermore, for the strongest couplings studied, the ergosphere emergence frequency exceeds the range investigated in this work.

\begin{figure}[htbp]
	\centering	
	{
		\begin{minipage}[b]{.3\linewidth}
			\centering
			\includegraphics[scale=0.4]{"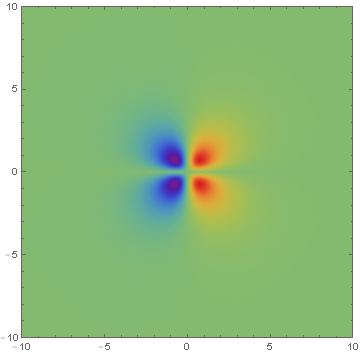"}
		\end{minipage}
		\begin{minipage}[b]{.3\linewidth}
			\centering
			\includegraphics[scale=0.4]{"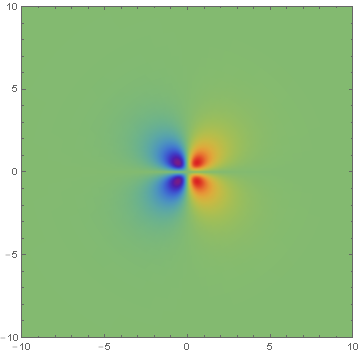"}
		\end{minipage}
		\begin{minipage}[b]{.3\linewidth}
			\centering
			\includegraphics[scale=0.4]{"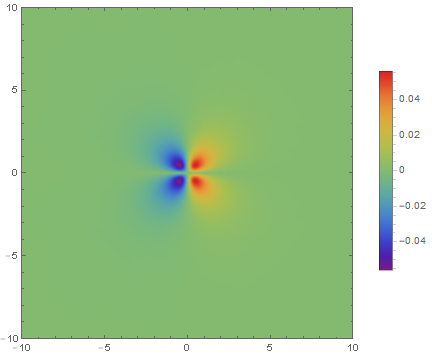"}
		\end{minipage}
	}
	\caption{Distributions of the scalar field function $\phi$ for chains of rotating BSs with two constituents with $\lambda = 400$ case in the $z-\rho$ plane. From left to right, the panels show solutions on the second branch with frequencies $\omega = 0.74$, $0.76$, and $0.78$, respectively.}
	\label{Fig.10}
\end{figure}

\begin{figure}[htbp]
	\centering	
	{
		\begin{minipage}[b]{.3\linewidth}
			\centering
			\includegraphics[scale=0.4]{"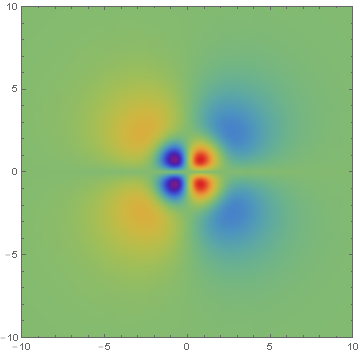"}
		\end{minipage}
		\begin{minipage}[b]{.3\linewidth}
			\centering
			\includegraphics[scale=0.4]{"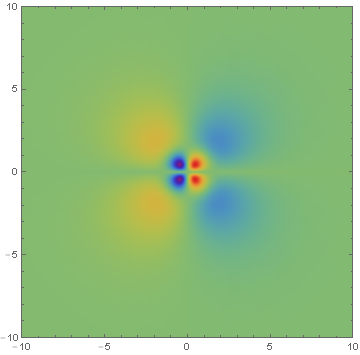"}
		\end{minipage}
		\begin{minipage}[b]{.3\linewidth}
			\centering
			\includegraphics[scale=0.4]{"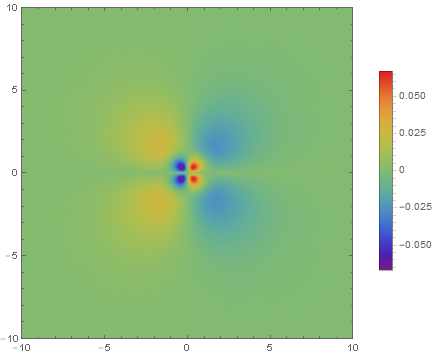"}
		\end{minipage}
	}
	\caption{Distributions of the scalar field function $\phi$ for chains of rotating BSs with four constituents with $\lambda = 200$ case in the $z-\rho$ plane. The three panels from left to right correspond to solutions with  $\omega=0.76$, $\omega=0.78$, and $\omega=0.8$ in the second branch.}
	\label{Fig.11}
\end{figure}

To illustrate the ergosphere merger, the $\phi$-field distribution for the quartic self-interaction case is presented in Figs. \ref{Fig.10} and \ref{Fig.11}. As the frequency increases, the maximum values of the $\phi$ field grow and the distance between the two extrema contracts. This suggests that the scalar field becomes more concentrated toward the center with increasing frequency. A pronounced asymmetry is seen in the four-constituent case in Fig. \ref{Fig.11}, where the off-center extremum has a substantially lower amplitude than the central extremum. The formation of only two ergospheres in the four-constituent system is a consequence of this asymmetry in the field distribution. Furthermore, the two ergospheres ultimately coalesce into one as they expand and converge, due to the central extremum evolving with frequency in the same manner as in the two-constituent case.

\subsection{The result of sextic self-interaction}

\begin{figure}[htbp]
	\centering
	{
		\begin{minipage}[b]{.3\linewidth}
			\centering
			\includegraphics[scale=0.2]{"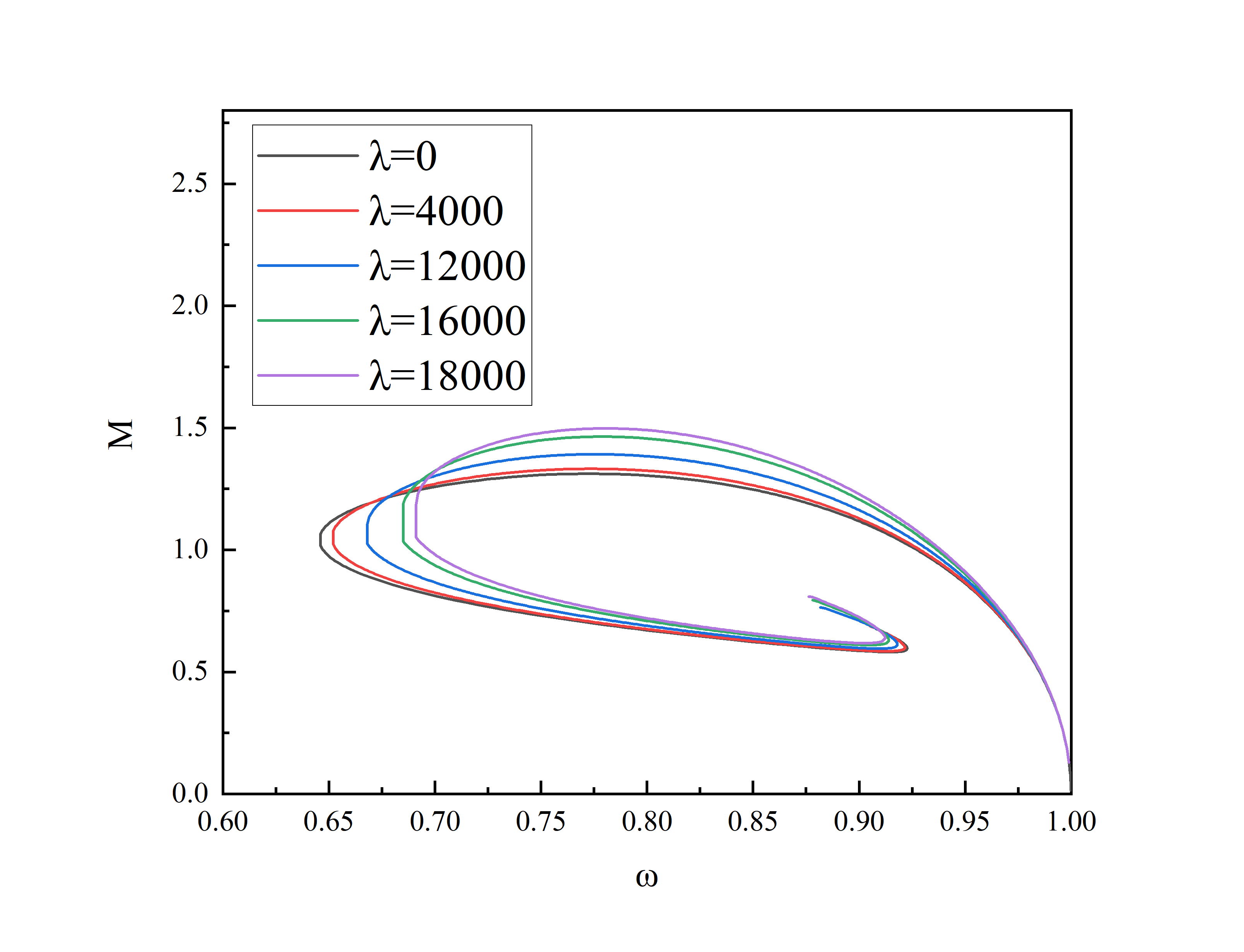"}
		\end{minipage}
		\begin{minipage}[b]{.3\linewidth}
			\centering
			\includegraphics[scale=0.2]{"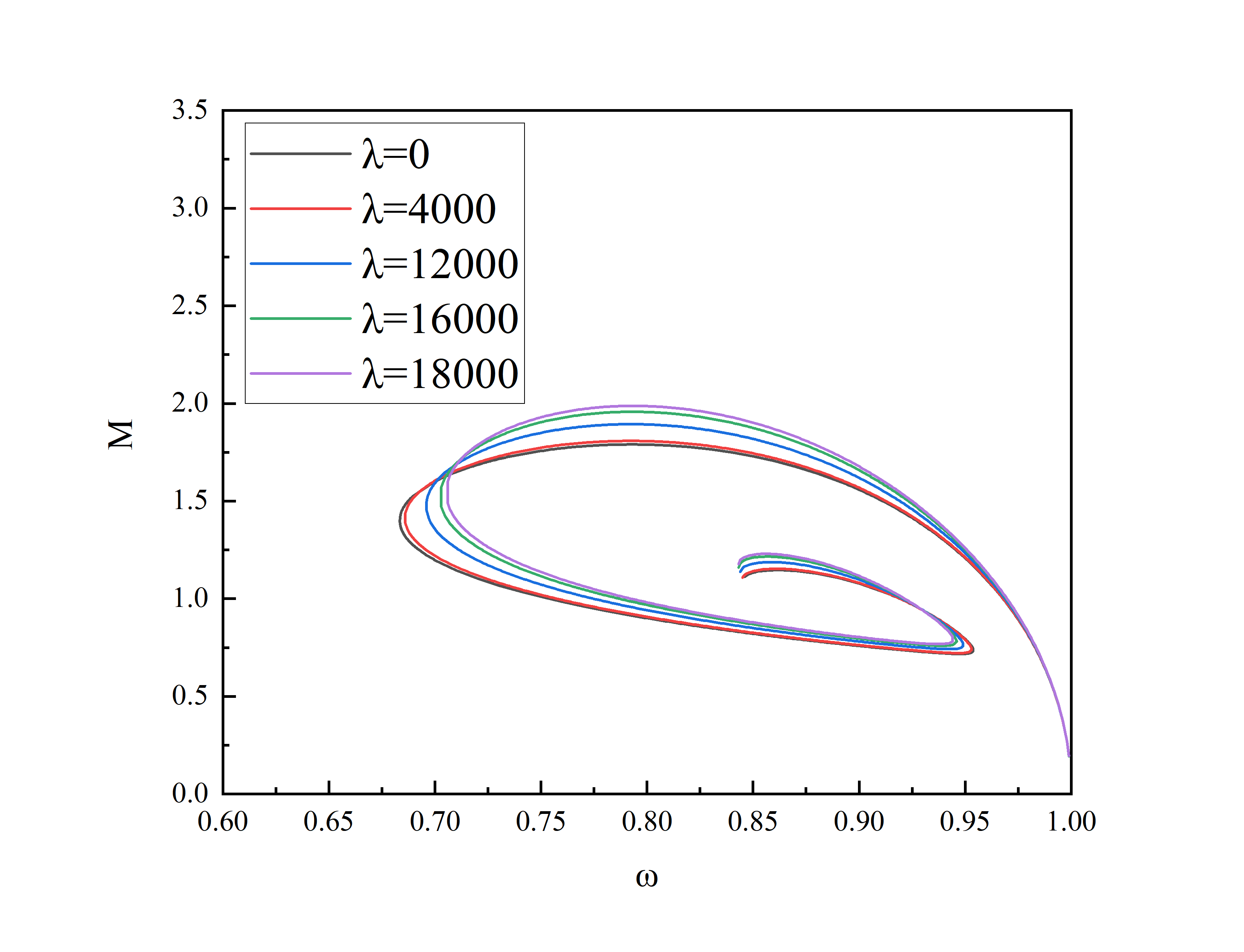"}
		\end{minipage}
		\begin{minipage}[b]{.3\linewidth}
			\centering
			\includegraphics[scale=0.2]{"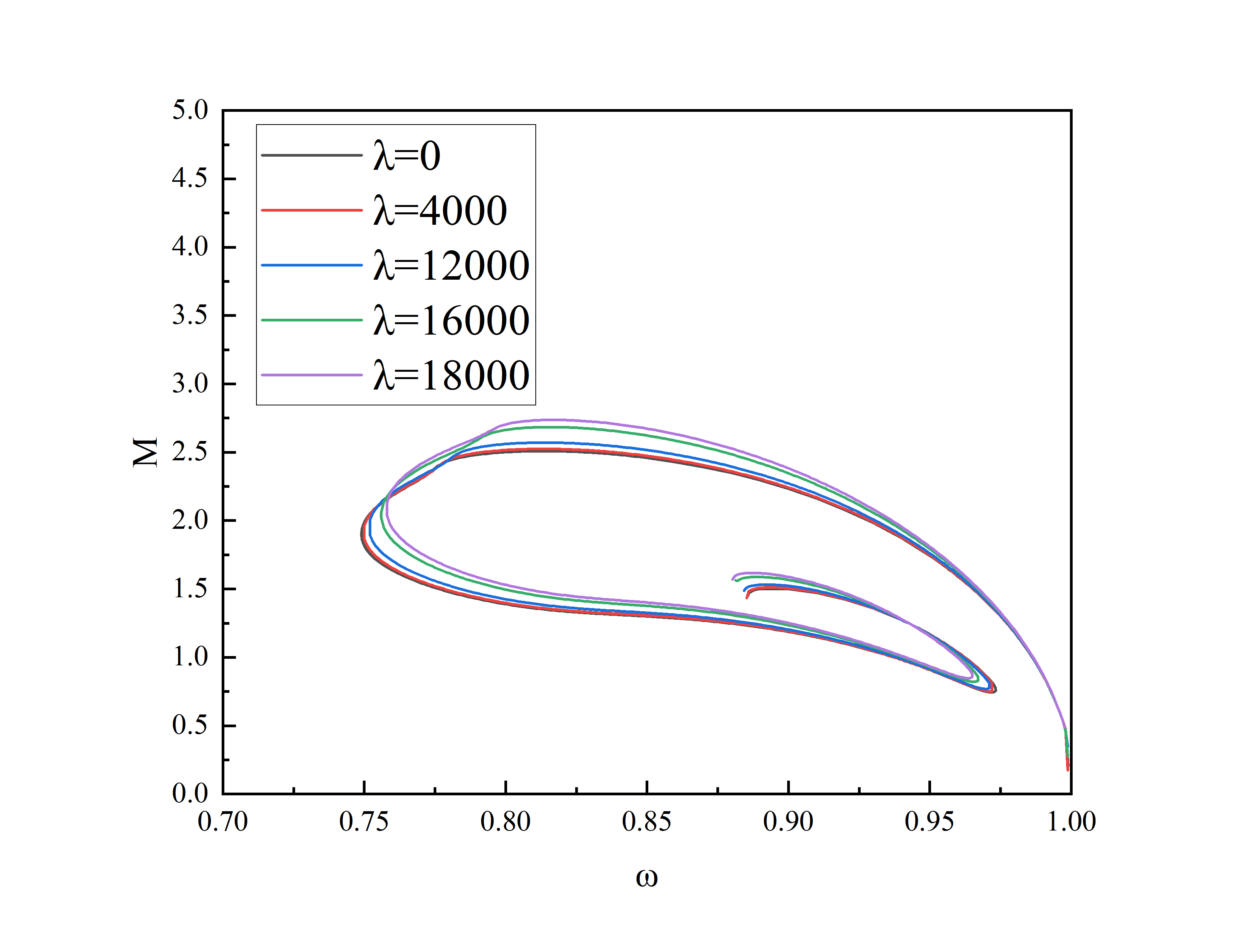"}
		\end{minipage}
	}
	\caption{The ADM mass $M$ as a function of the frequency $\omega$. Left: The rotating BSs with sextic self-interaction of one constituent. Middle: The rotating BSs with sextic self-interaction of two constituents. Right: The rotating BSs with sextic self-interaction of four constituents.}
	\label{Fig.12}
\end{figure}

The chain solutions for rotating BSs with sextic self-interactions, obtained from Eqs. (\ref{PDE11}) to (\ref{PDE44}) and (\ref{PDE55}), exhibit a significantly weaker dependence on the coupling strength $\lambda$ than the quartic case. This weak influence occurs because the small field amplitude renders the $|\phi|^6$ term a higher-order correction, contributing less to the energy density than the $|\phi|^4$ term. We perform separate computations for coupling strengths of $\lambda = 4000$, $12000$, $16000$, and $18000$, and the results are presented below.

Figs. \ref{Fig.12} and \ref{Fig.22} show the variation of the ADM mass $M$ and angular momentum $J$ with frequency $\omega$ for chains of rotating BSs with sextic self-interaction, comprising one, two and four constituents. The figures indicate that the curves are qualitatively similar to those in the quartic self-interaction case. The $M-\omega$ and $J-\omega$ curves trace a spiral, which can be categorized into three distinct branches. The three branches exhibit distinct non-monotonic behavior. On the first branch, $M$ and $J$ increase with decreasing $\omega$ until a peak, then decrease. On the second branch, they decrease to a minimum with $\omega$, then increase. On the third branch, they similarly reach a peak and then decline as $\omega$ decreases.

\begin{figure}[htbp]
	{
		\begin{minipage}[b]{.3\linewidth}
			\centering
			\includegraphics[scale=0.2]{"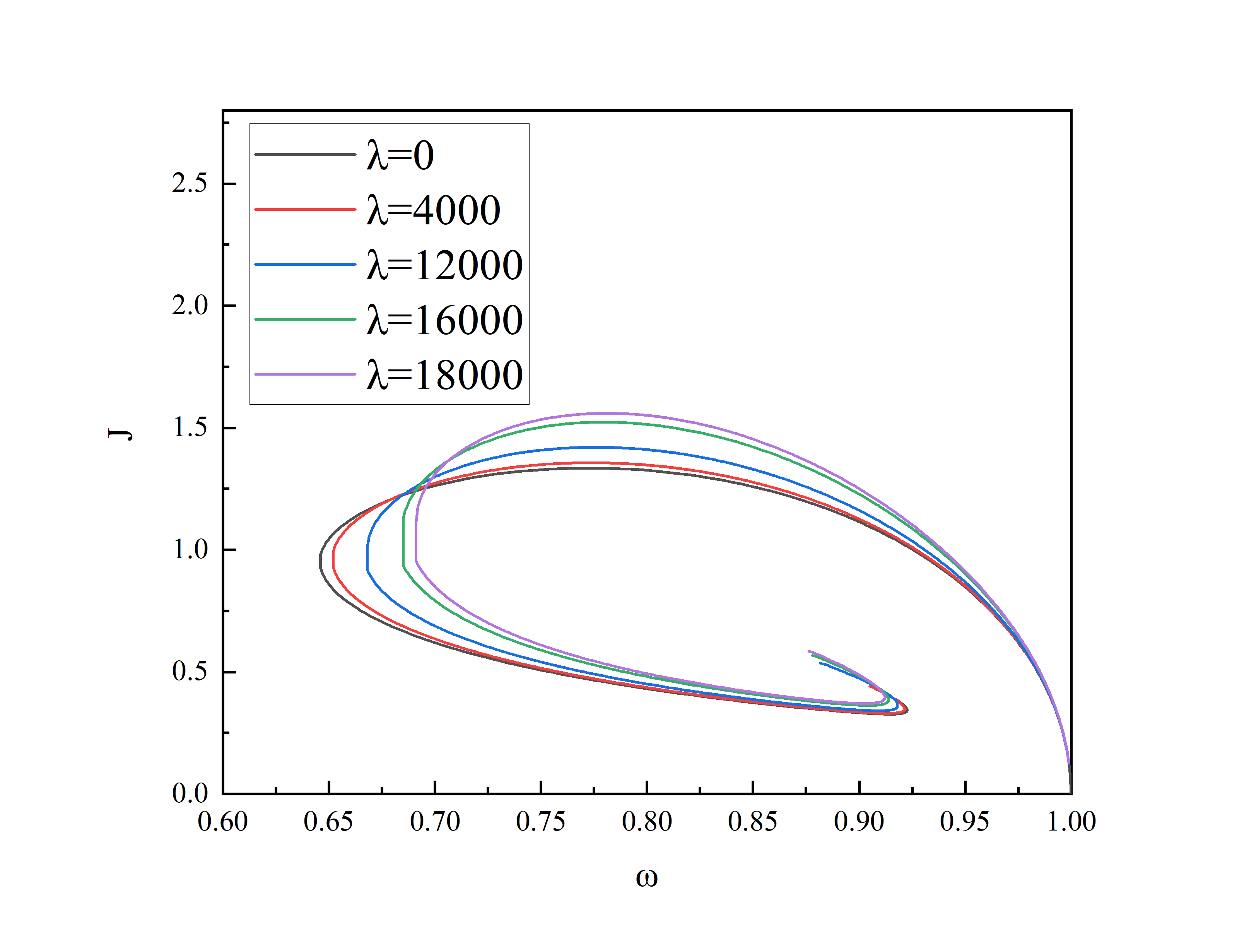"}
		\end{minipage}
		\begin{minipage}[b]{.3\linewidth}
			\centering
			\includegraphics[scale=0.2]{"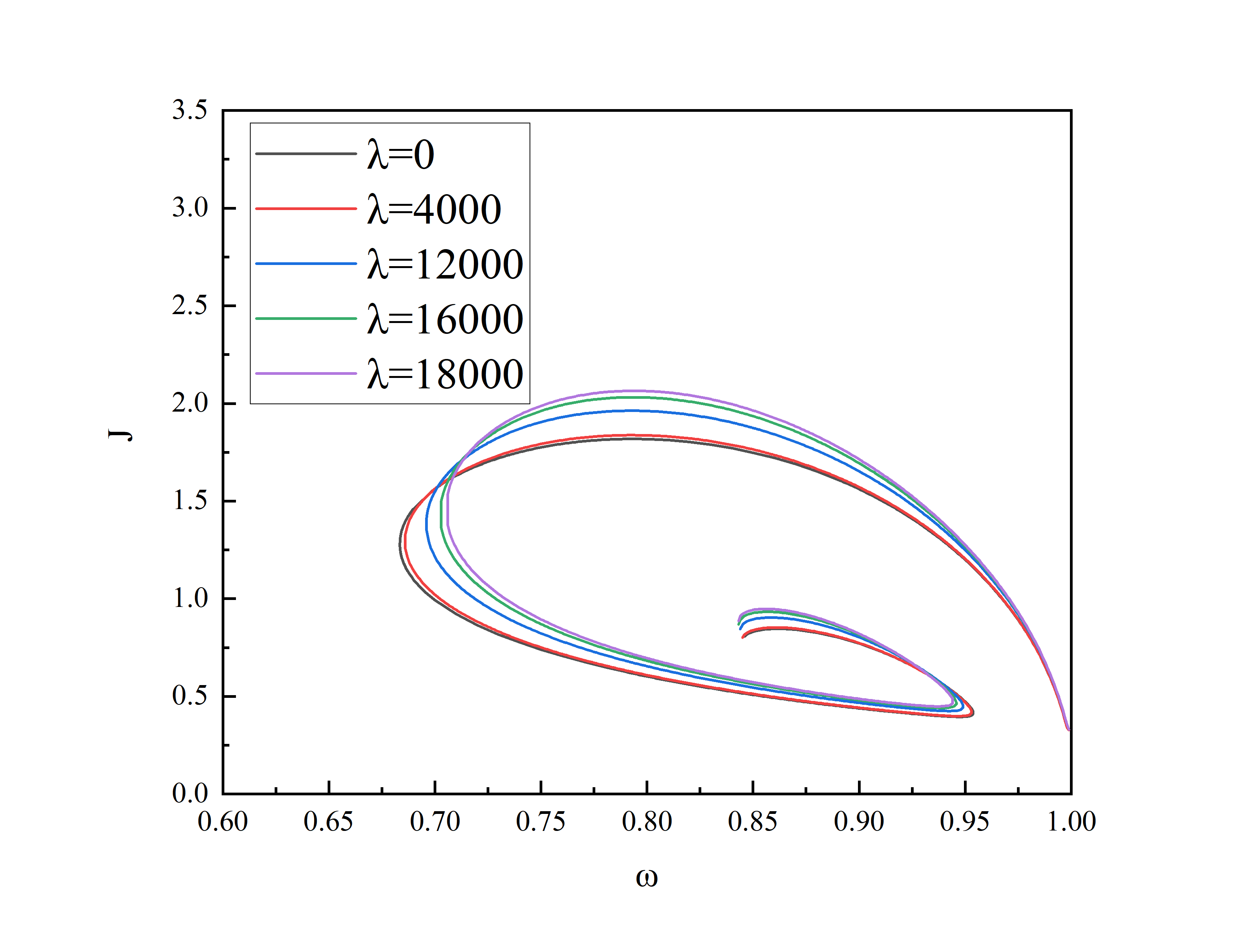"}
		\end{minipage}
		\begin{minipage}[b]{.3\linewidth}
			\centering
			\includegraphics[scale=0.2]{"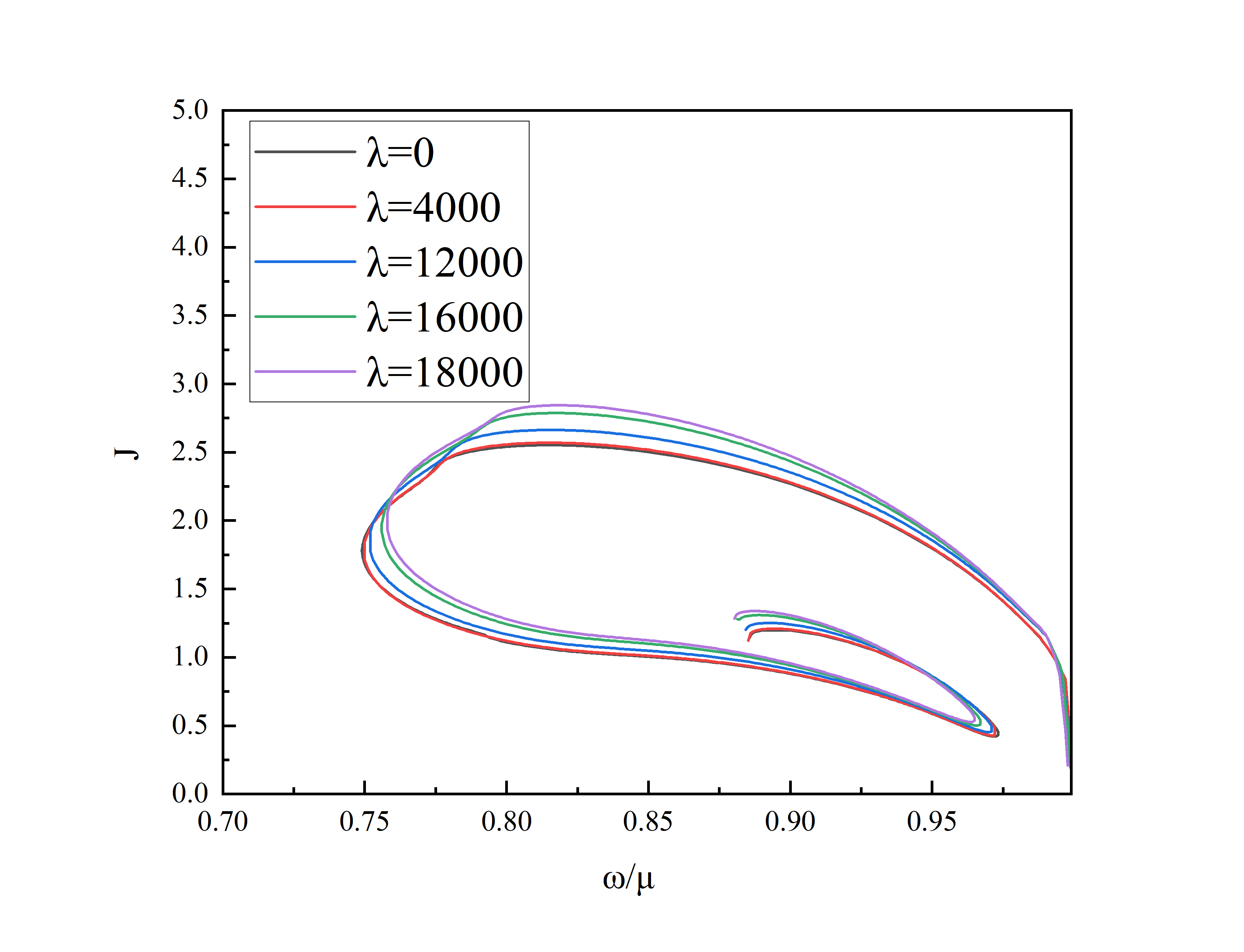"}
		\end{minipage}
	}
	\caption{The total angular momentum $J$ as a function of the frequency $\omega$. Left: The rotating BSs with sextic self-interaction of one constituent. Middle: The rotating BSs with sextic self-interaction of two constituents. Right: The rotating BSs with sextic self-interaction of four constituents.}
	\label{Fig.22}
\end{figure}

\begin{figure}[htbp]
	\centering
	\subfigure
	{
		\begin{minipage}[b]{0.45\textwidth}
			\centering
			\includegraphics[scale=0.3]{"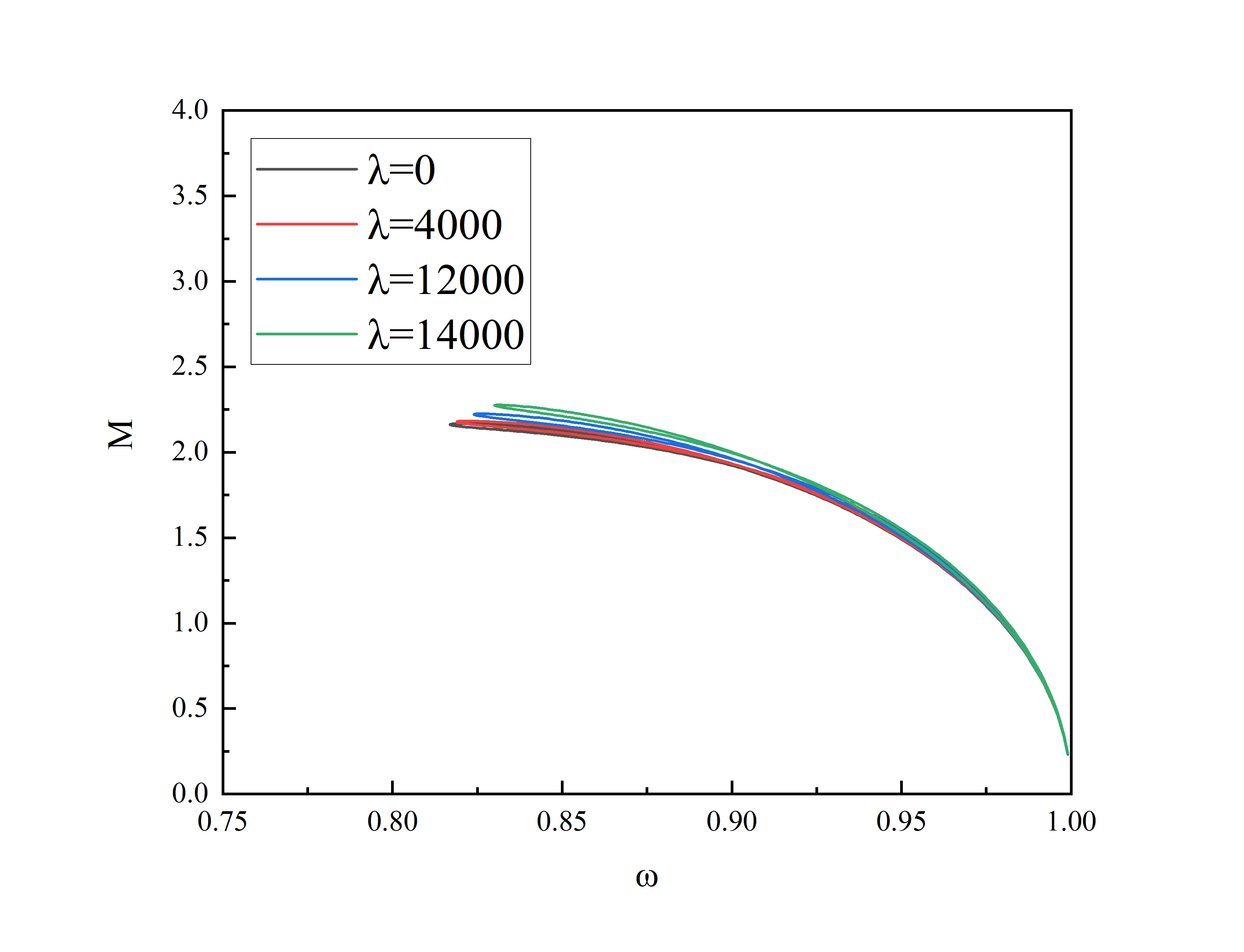"}
		\end{minipage}
	}
	\subfigure
	{
		\begin{minipage}[b]{0.45\textwidth}
			\centering
			\includegraphics[scale=0.3]{"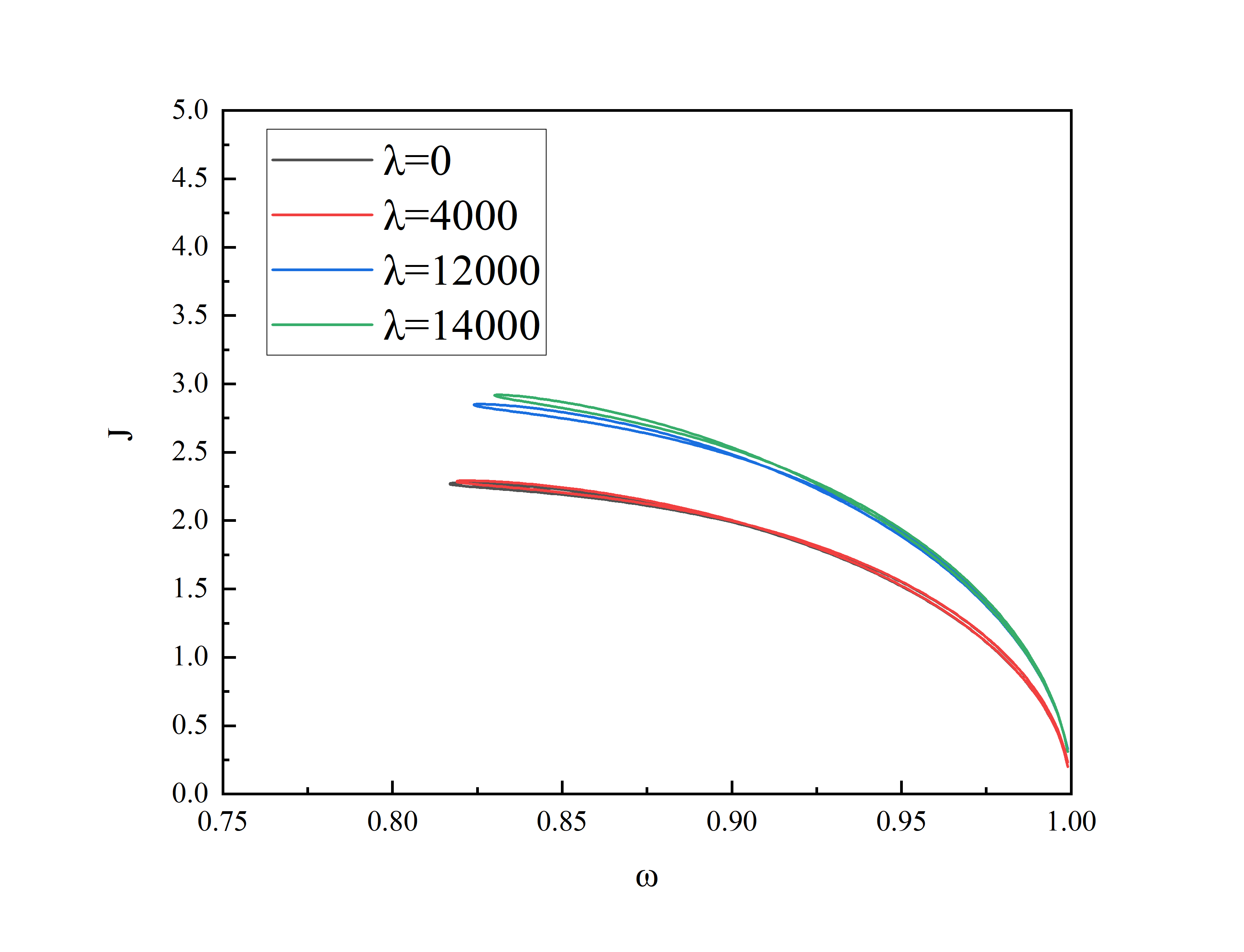"}
		\end{minipage}
	}
	\caption{The curves of $(\omega, M)$ and $(\omega, J)$ for three-constituent case. Left: The ADM mass $M$ with different frequency $\omega$ of the rotating BSs with sextic self-interaction of three constituents. Right: The total angular momentum $J$ with different frequency $\omega$ of the rotating BSs with sextic self-interaction of three constituents.}
	\label{Fig.33}
\end{figure}

Fig. \ref{Fig.33} depicts the dependence of $M$ and $J$ on $\omega$ for three-constituent rotating BSs chains with sextic self-interaction. Similarly to the quartic self-interaction case, the figures also exhibit a loop structure in $(\omega, M)$ and $(\omega, J)$. Although the sextic self-interaction avoids the large-coupling constraint of the quartic case, the maximum attainable $M$ and $J$ values, even at significantly larger couplings, do not exceed the maximum values achieved in the quartic case, even at its maximum stable coupling of $\lambda = 250$. This implies that no stable solutions exist at higher coupling strengths for the sextic self-interaction either.

\begin{figure}[htbp]
	\centering
	{
		\begin{minipage}[b]{.3\linewidth}
			\centering
			\includegraphics[scale=0.133]{"FIG1/1k0.658ergo.png"}
		\end{minipage}
		\begin{minipage}[b]{.3\linewidth}
			\centering
			\includegraphics[scale=0.133]{"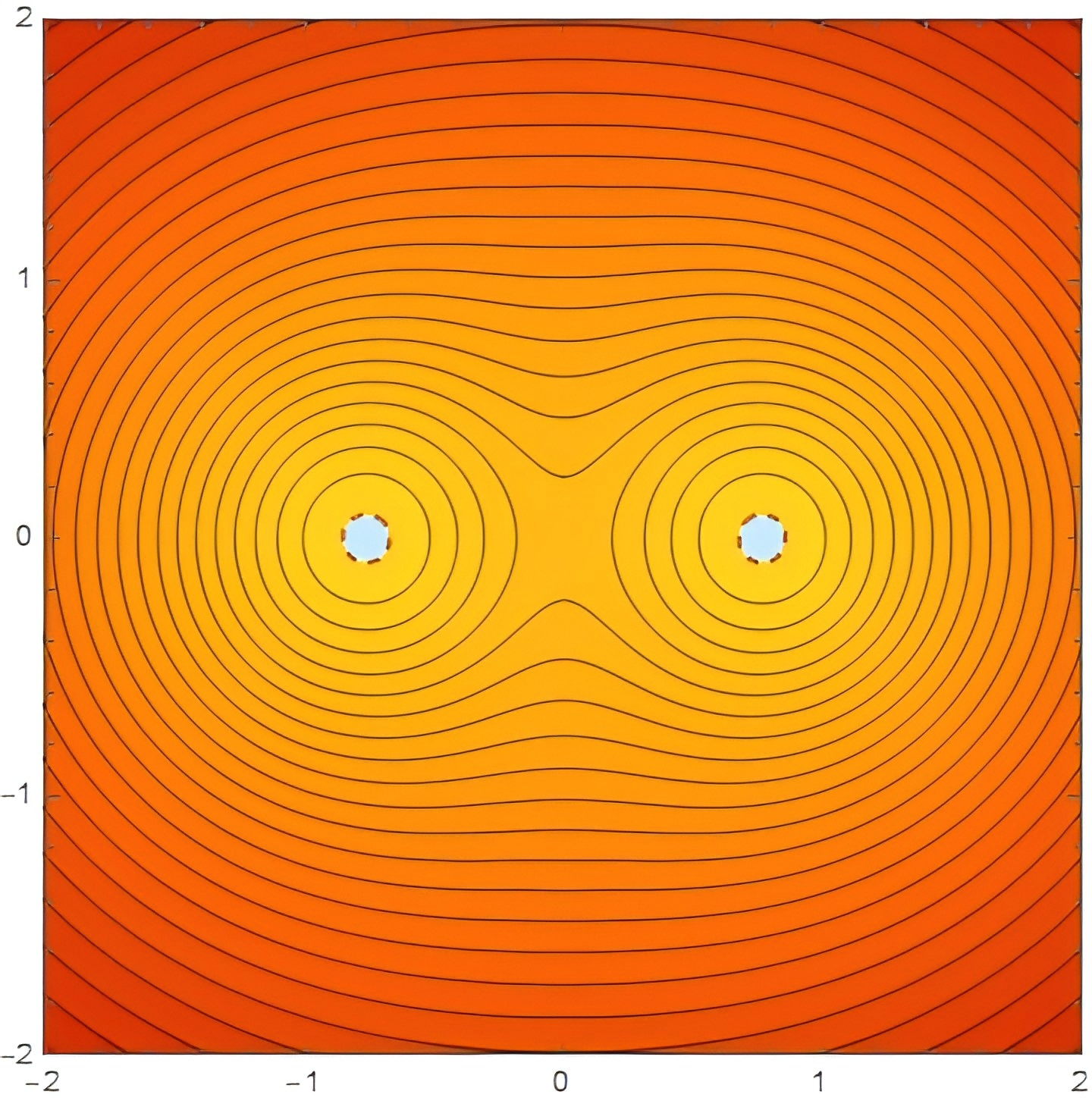"}
		\end{minipage}
		\begin{minipage}[b]{.3\linewidth}
			\centering
			\includegraphics[scale=0.1]{"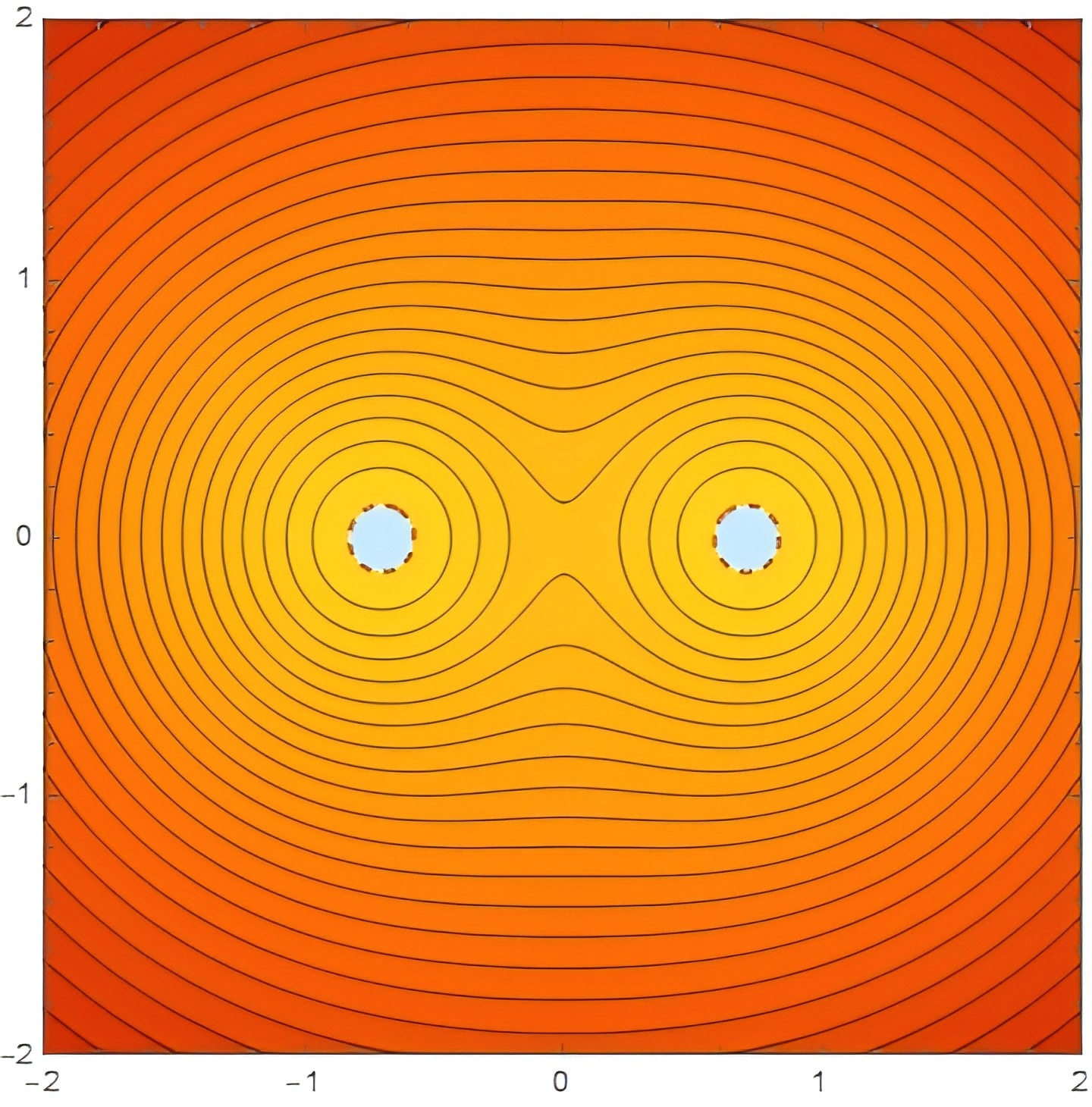"}
		\end{minipage}
	}
	\caption{Contours of $g_{tt}$ for chains of rotating BSs with one constituent. Left: When $\lambda = 0$, the ergosphere emerges at $\omega = 0.658$ on the first branch. Middle: When $\lambda = 4000$, the ergosphere emerges at $\omega = 0.661$ on the second branch. Right: When $\lambda = 16000$, the ergosphere emerges at $\omega = 0.6875$ on the second branch. Warm and cold color schemes denote negative and positive $g_{tt}$ values, respectively. The red dashed line represents ergospheres ($g_{tt}$ = 0), and the black solid lines represent the contours of $g_{tt}$.}
	\label{Fig.44}
\end{figure}

Similar to the quartic case, rotating BSs chains with sextic self-interaction also exhibit ergospheres for one, two, and four constituents. Furthermore, the ergospheres in both two- and four-constituent systems merge as $\omega$ increases, following a pattern similar to that observed in the quartic case.

\begin{figure}[htbp]
	\centering
	{
		\begin{minipage}[b]{.3\linewidth}
			\centering
			\includegraphics[scale=0.133]{"FIG1/2_2k0.719ergo.png"}
		\end{minipage}
		\begin{minipage}[b]{.3\linewidth}
			\centering
			\includegraphics[scale=0.1]{"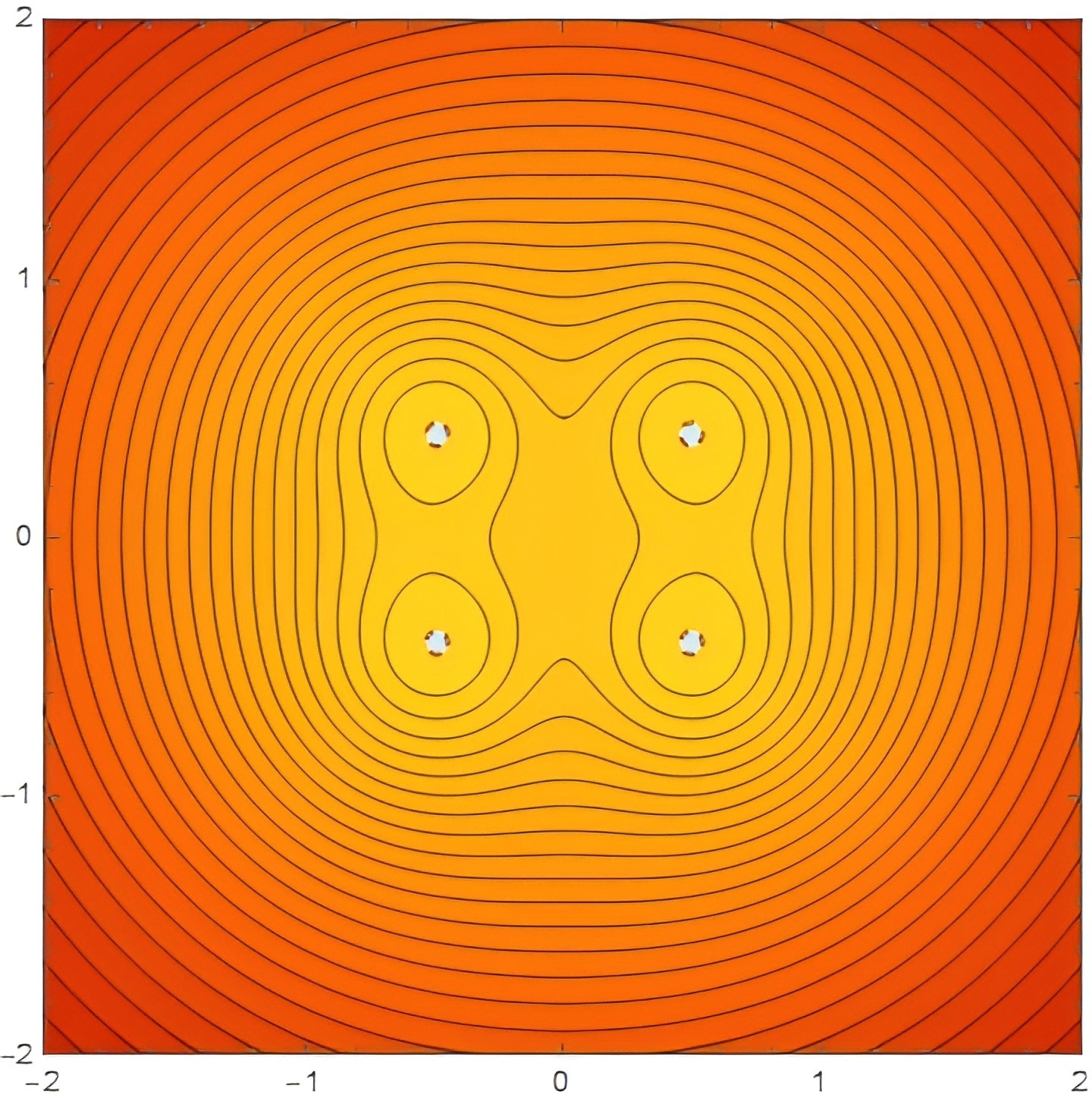"}
		\end{minipage}
		\begin{minipage}[b]{.3\linewidth}
			\centering
			\includegraphics[scale=0.1]{"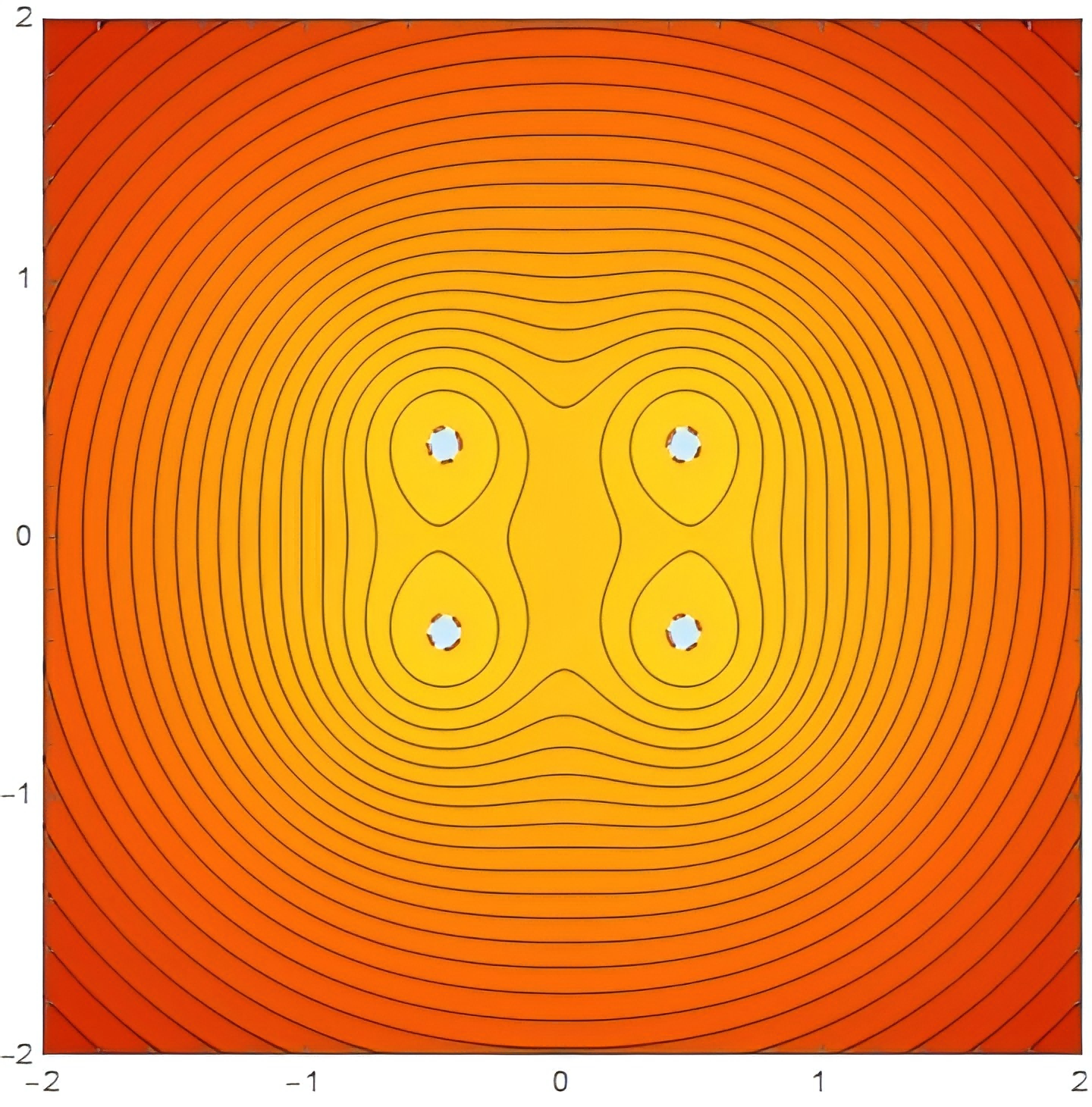"}
		\end{minipage}
	}
	\caption{Contours of $g_{tt}$ for chains of rotating BSs with two constituents. Left: When $\lambda = 0$, the ergospheres emerge at $\omega=0.719$ on the first branch. Middle: When $\lambda = 4000$, the ergospheres emerge at $\omega = 0.725$ on the second branch. Right: When $\lambda = 16000$, the ergospheres emerge at $\omega = 0.76$ on the second branch. Warm and cold color schemes denote negative and positive $g_{tt}$ values, respectively. The red dashed line represents ergospheres ($g_{tt}$ = 0), and the black solid lines represent the contours of $g_{tt}$.}
	\label{Fig.55}
\end{figure}

\begin{figure}[htbp]
	\centering
	{
		\begin{minipage}[b]{.3\linewidth}
			\centering
			\includegraphics[scale=0.133]{"FIG1/3_2k0.81ergobegin.png"}
		\end{minipage}
		\begin{minipage}[b]{.3\linewidth}
			\centering
			\includegraphics[scale=0.1]{"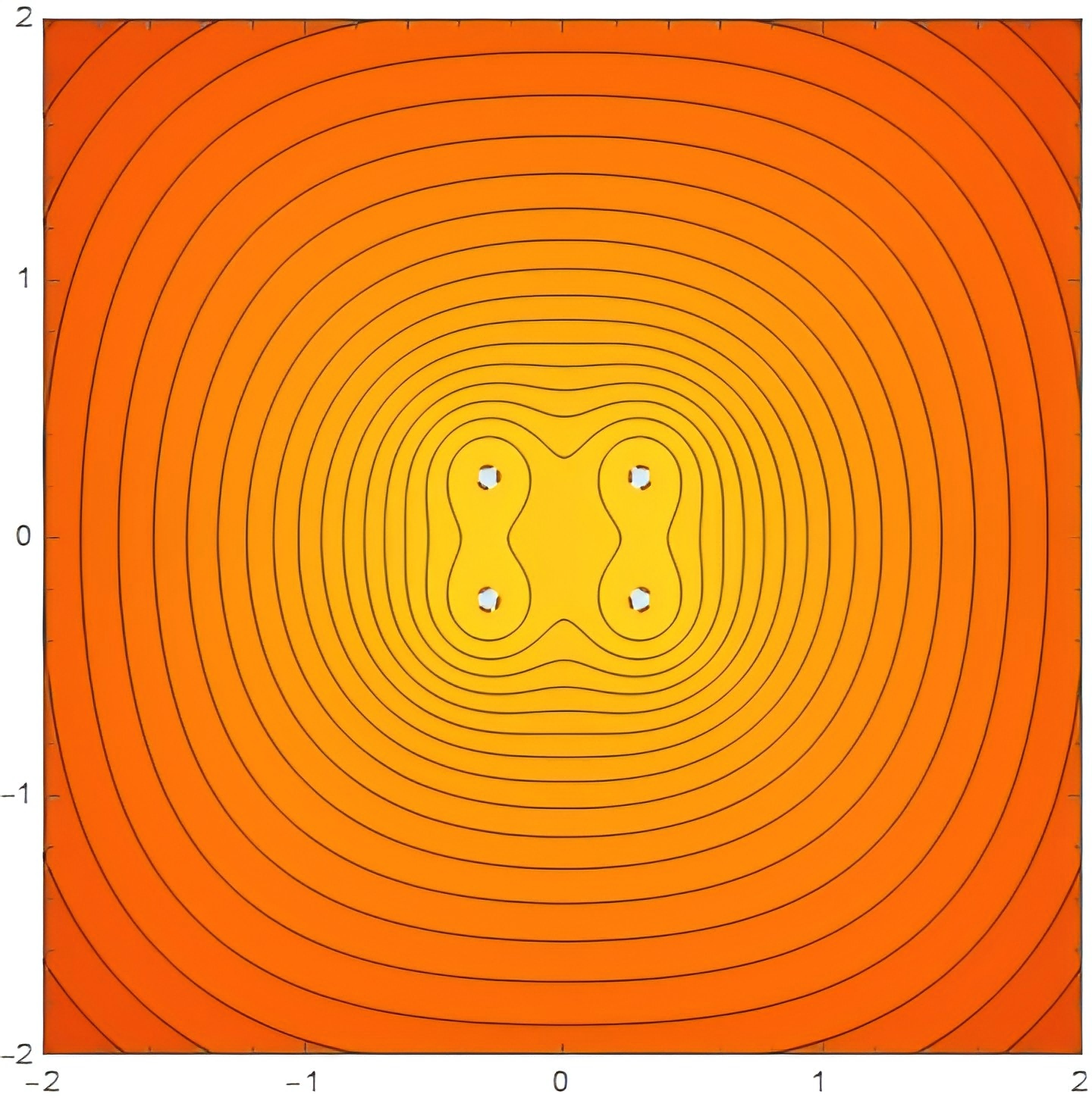"}
		\end{minipage}
		\begin{minipage}[b]{.3\linewidth}
			\centering
			\includegraphics[scale=0.1]{"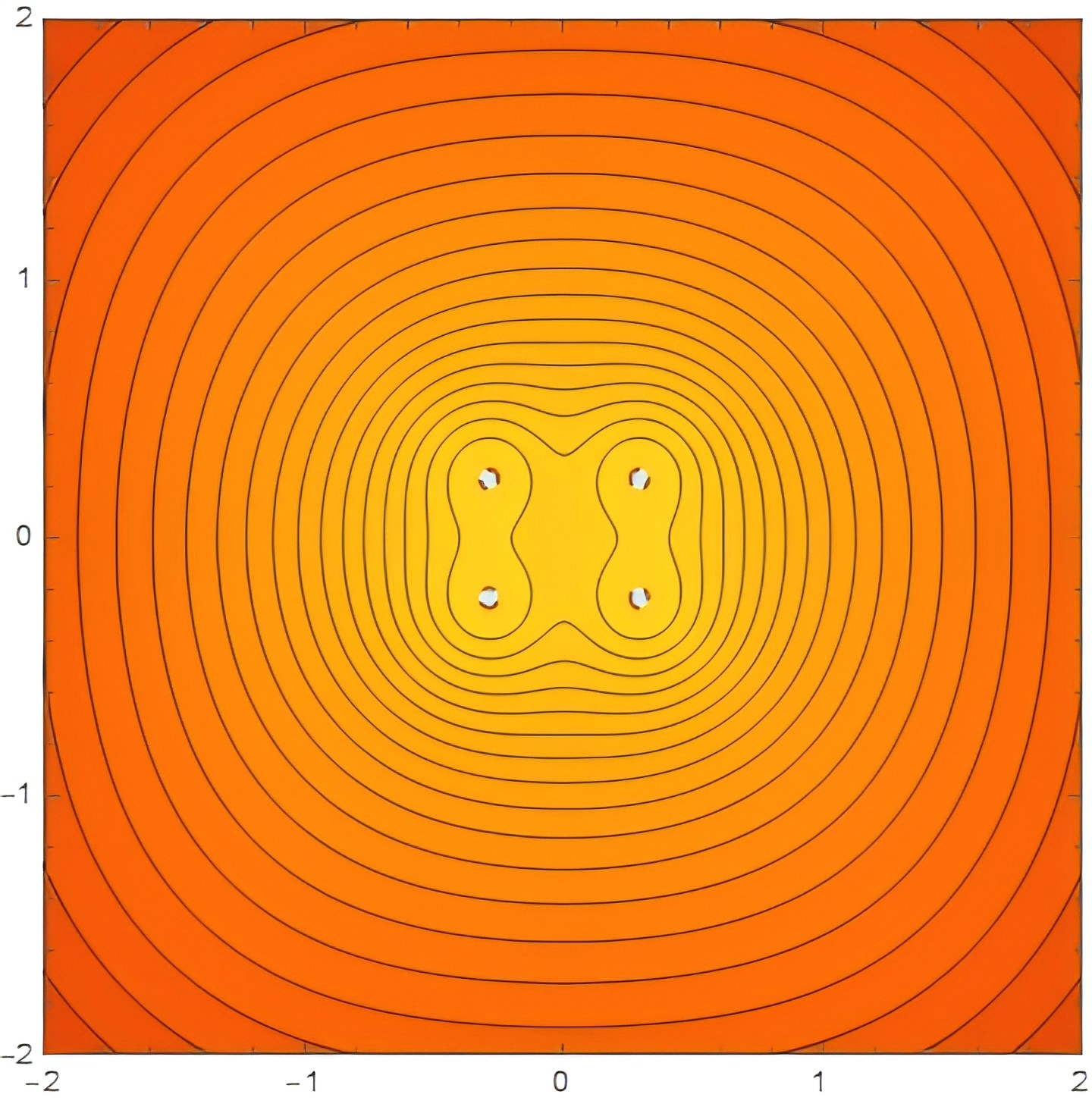"}
		\end{minipage}
	}
	\caption{Contours of $g_{tt}$ for chains of rotating BSs with four constituents. Left: When $\lambda = 0$, the ergospheres emerge at $\omega = 0.801$ on the first branch. Middle: When $\lambda = 4000$, the ergospheres emerge at $\omega = 0.807$ on the second branch. Right: When $\lambda = 16000$, the ergospheres emerge at $\omega = 0.817$ on the second branch. Warm and cold color schemes denote negative and positive $g_{tt}$ values, respectively. The red dashed line represents ergospheres ($g_{tt}$ = 0), and the black solid lines represent the contours of $g_{tt}$.}
	\label{Fig.66}
\end{figure}

\begin{figure}[htbp]
	\centering
	{
		\begin{minipage}[b]{.3\linewidth}
			\centering
			\includegraphics[scale=0.1]{"FIG2/2_0.76_1_.png"}
		\end{minipage}
		\begin{minipage}[b]{.3\linewidth}
			\centering
			\includegraphics[scale=0.1]{"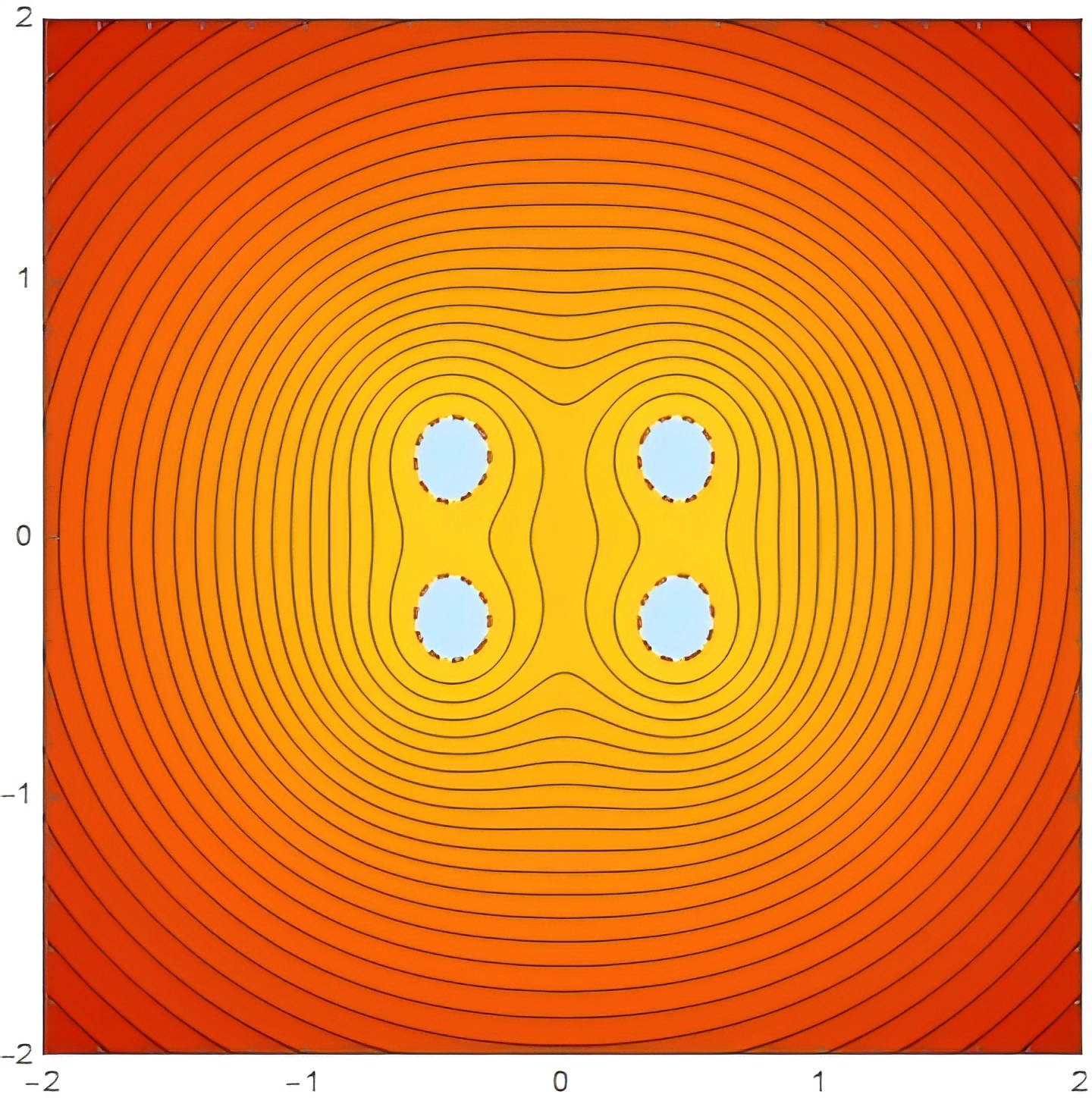"}
		\end{minipage}
		\begin{minipage}[b]{.3\linewidth}
			\centering
			\includegraphics[scale=0.1]{"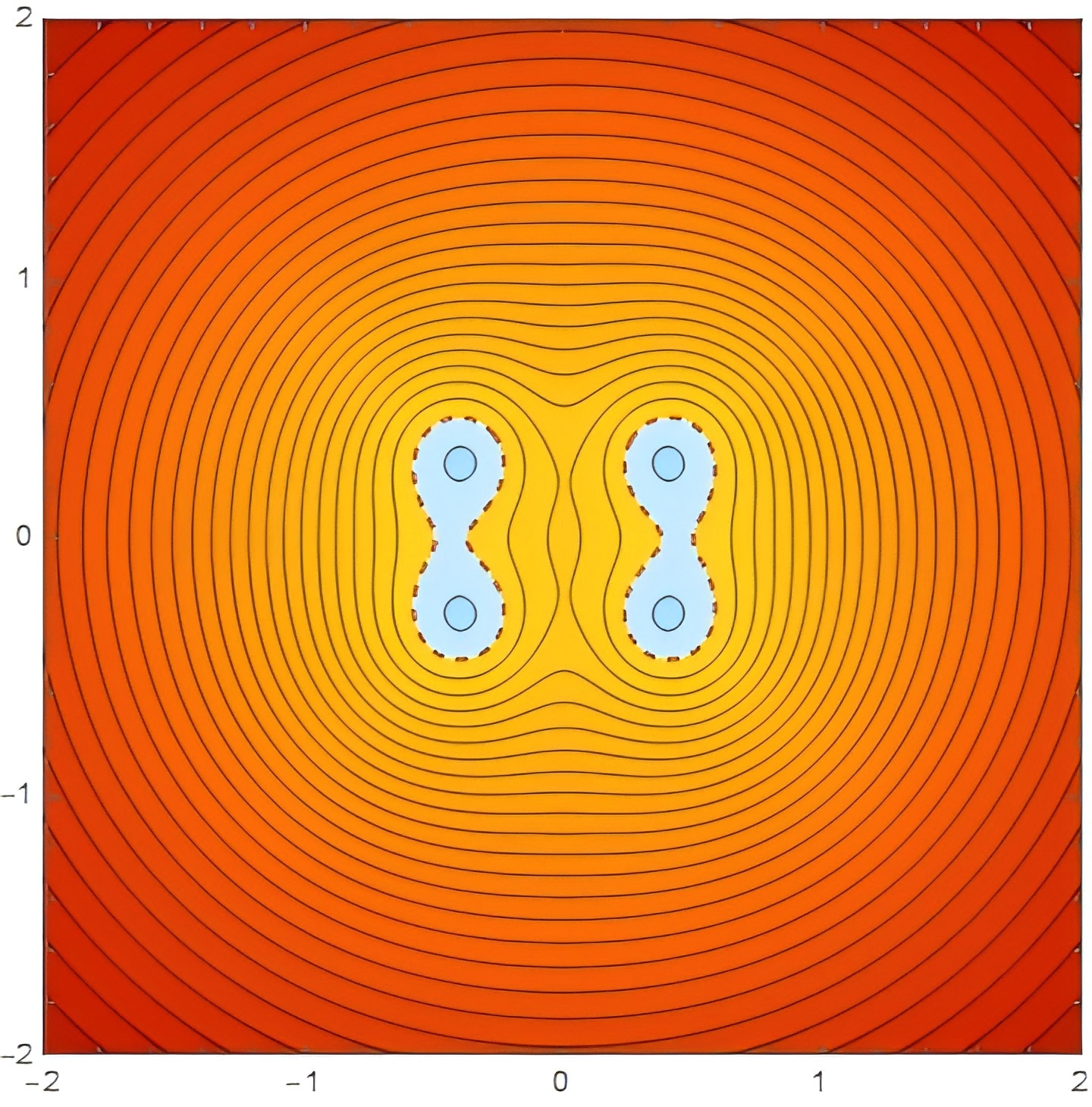"}
		\end{minipage}
	}
	\caption{Contours of $g_{tt}$ for chains of rotating BSs with two constituents at $\lambda = 16000$. Left: The ergospheres emerge at $\omega = 0.76$ on the second branch. Middle: With the increase of $\omega$, the ergospheres begin to merge at $\omega = 0.78$ on the second branch. Right: The two ergospheres merge into one at $\omega=0.8$ on the second branch. Warm and cold color schemes denote negative and positive $g_{tt}$ values, respectively. The red dashed line represents ergospheres ($g_{tt}$ = 0), and the black solid lines represent the contours of $g_{tt}$.}
	\label{Fig.77}
\end{figure}

\begin{figure}[htbp]
	\centering
	{
		\begin{minipage}[b]{.3\linewidth}
			\centering
			\includegraphics[scale=0.1]{"FIG2/4_0.817_1_.png"}
		\end{minipage}
		\begin{minipage}[b]{.3\linewidth}
			\centering
			\includegraphics[scale=0.1]{"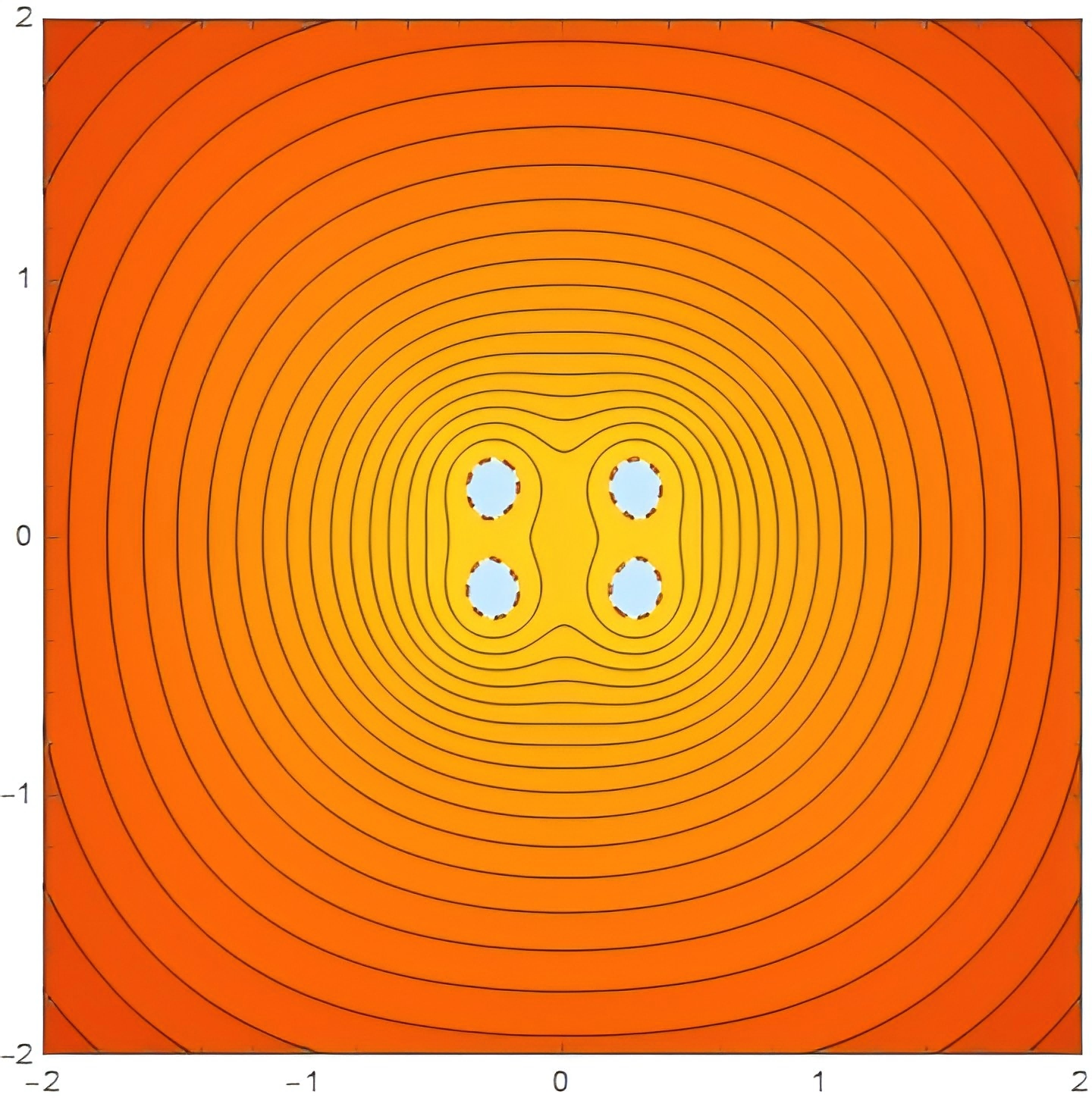"}
		\end{minipage}
		\begin{minipage}[b]{.3\linewidth}
			\centering
			\includegraphics[scale=0.1]{"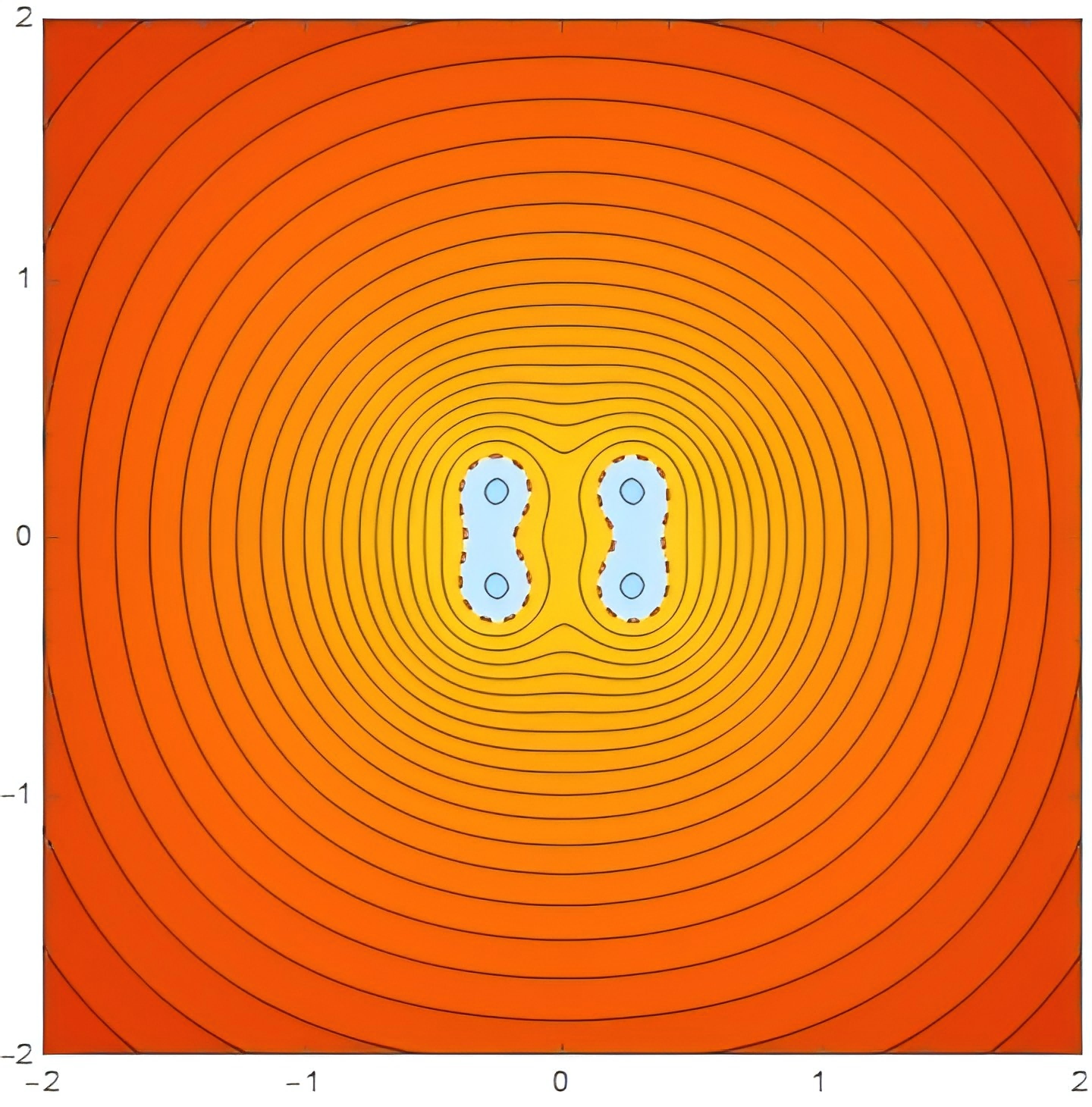"}
		\end{minipage}
	}
	\caption{Contours of $g_{tt}$ for chains of rotating BSs with four constituents at $\lambda = 16000$. Left: The ergospheres emerge at $\omega = 0.817$ on the second branch. Middle: With the increase of $\omega$, the ergospheres begin to merge at $\omega = 0.83$ on the second branch. Right: The two ergospheres merge into one at $\omega = 0.85$ on the second branch. Warm and cold color schemes denote negative and positive $g_{tt}$ values, respectively. The red dashed line represents ergospheres ($g_{tt}$ = 0), and the black solid lines represent the contours of $g_{tt}$.}
	\label{Fig.88}
\end{figure}

Figs. \ref{Fig.44}, \ref{Fig.55}, and \ref{Fig.66} display the incipient ergospheres for the one-, two-, and four-constituent cases with sextic self-interaction. With increasing coupling strength $\lambda$, the emergence point of the ergospheres shifts to the lower $\omega$ along the $(\omega, M)$ and $(\omega, J)$ curves in all three cases. Similarly, as the number of constituents increases, the ergosphere emergence point shifts toward the lower $\omega$ on the $(\omega, M)$ and $(\omega, J)$ curves. As with the quartic interaction, the ergospheres merge with increasing $\omega$. Figs. \ref{Fig.77} and \ref{Fig.88} demonstrate this merging behavior for the two- and four-constituent cases, respectively, both at $\lambda=16000$, confirming the behavior's generality across different chain configurations.

\begin{figure}[htbp]
	\centering	
	{
		\begin{minipage}[b]{.3\linewidth}
			\centering
			\includegraphics[scale=0.4]{"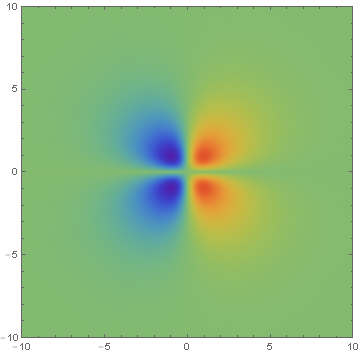"}
		\end{minipage}
		\begin{minipage}[b]{.3\linewidth}
			\centering
			\includegraphics[scale=0.4]{"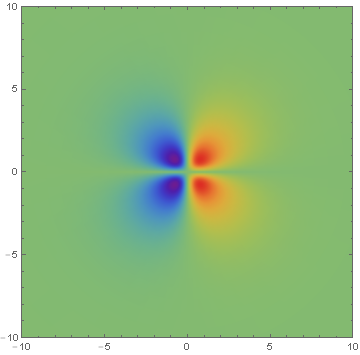"}
		\end{minipage}
		\begin{minipage}[b]{.3\linewidth}
			\centering
			\includegraphics[scale=0.4]{"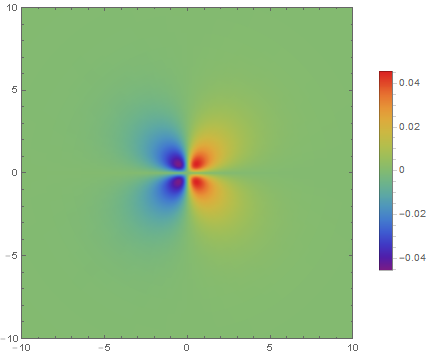"}
		\end{minipage}
	}
	\caption{Distributions of the scalar field function $\phi$ for chains of rotating BSs with two constituents with $\lambda = 12000$ case in the $z-\rho$ plane. The three panels from left to right correspond to solutions with  $\omega = 0.75$, $\omega = 0.76$, and $\omega = 0.79$ in the second branch.}
	\label{Fig.20}
\end{figure}

\begin{figure}[htbp]
	\centering	
	{
		\begin{minipage}[b]{.3\linewidth}
			\centering
			\includegraphics[scale=0.4]{"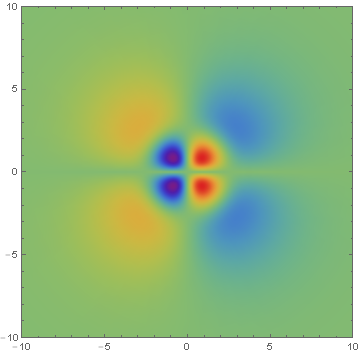"}
		\end{minipage}
		\begin{minipage}[b]{.3\linewidth}
			\centering
			\includegraphics[scale=0.4]{"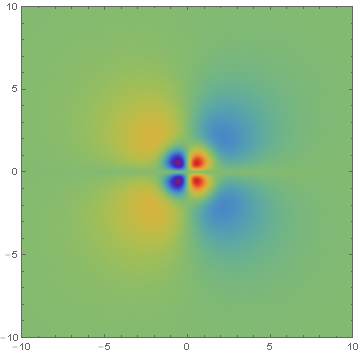"}
		\end{minipage}
		\begin{minipage}[b]{.3\linewidth}
			\centering
			\includegraphics[scale=0.4]{"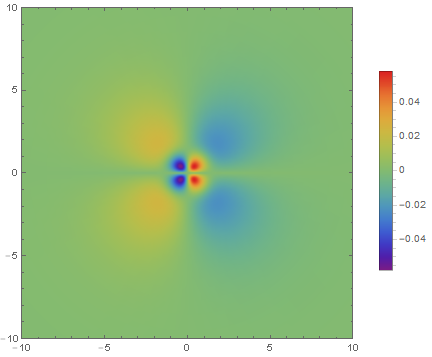"}
		\end{minipage}
	}
	\caption{Distributions of the scalar field function $\phi$ for chains of rotating BSs with four constituents with $\lambda = 12000$ case in the $z-\rho$ plane. The three panels from left to right correspond to solutions with  $\omega = 0.76$, $\omega = 0.78$, and $\omega = 0.8$ on the second branch.}
	\label{Fig.21}
\end{figure}

Figs. \ref{Fig.20} and \ref{Fig.21} show the $\phi$-field distributions for the sextic interaction to clarify the ergosphere merger mechanism. We observe a frequency-induced amplitude growth and spatial contraction of the field, indicating its intensified central concentration. The dominance of the central over the outer extremum in the four-constituent case (Fig. \ref{Fig.21}) explains the formation of merely two ergospheres. Moreover, the fact that the central peak scales with frequency in the same way as in the two-constituent system demonstrates that the ergospheres expand and eventually merge through a similar mechanism.

The graphical analysis in this section reveals a behavioral pattern in the sextic self-interaction case similar to the quartic self-interaction case. For sextic self-interaction, the emergence point of the ergospheres shifts along the evolutionary trajectory of the chains of rotating BSs as the coupling strength parameter $\lambda$ increases. Our results indicate that for sufficiently large $\lambda$ or a high number of constituents, the ergospheres will emerge near the end of the third branch, making their merger unobservable. All numerical results are compiled in Table \ref{tab1}. Each configuration is classified by its self-interaction type and constituent number parity, along with the corresponding curve type, number of ergospheres, and merger status.

\begin{table}[!htbp] 
	\centering 
	\caption{Solutions and their properties for two distinct types of self-interaction.} 
	\label{tab1} 
	\begin{tabular}{ccccc} 
		\hline 
		 Configuration &\quad Constituents number \quad&\quad Curve Type \quad&\quad Ergospheres number \quad& Merger occurrence \\ 
		\midrule\hline 
		\multirow{2}{*}{Quartic Self-interaction} &\quad Even &\quad Spiral & two & Yes \\ 
		&\quad Odd &\quad Loop & Null & Null \\ 
		\midrule\hline 
		\multirow{2}{*}{Sextic Self-interaction} &\quad Even &\quad Spiral & two & Yes \\ 
		&\quad Odd &\quad Loop & Null & Null \\ 
		\bottomrule\hline 
	\end{tabular}
\end{table}

\section{Conclusion}\label{sec5}
This paper investigates rotating chains of BSs with either quartic or sextic self-interactions. These solitonic solutions, composed of complex scalar fields minimally coupled to Einstein gravity, are constructed numerically for configurations containing one to four constituents.



In the quartic self-interaction case, the behavior of ADM mass $M$ and total angular momentum $J$ as functions of the frequency $\omega$ reveals distinct morphological patterns across different constituent numbers. While one-, two-, and four-constituent configurations follow characteristic spiral curves, the three-constituent system forms a closed loop—reflecting a unique gravitational-scalar equilibrium specific to this configuration. The spiral curves indicate a phase transition from stable to metastable and unstable states, whereas the loop implies a delicate balance that is maintained only in the three-constituent case. Moreover, an increase in the coupling strength $\lambda$ raises both $M$ and $J$, though stability is lost once $\lambda$ reaches approximately 250 in the three-constituent system.

Nevertheless, in the sextic self-interaction case, although higher coupling strengths are permitted, the resulting values of $M$ and $J$ do not surpass those achieved in the quartic case at $\lambda=250$. Significant deviations from the non-interacting scenario occur only at large $\lambda$, consistent with the sextic term acting as a higher-order correction. Nevertheless, the overall morphology of the curves remains similar to that observed in the quartic case. These results suggest that stable solutions with sextic self-interactions are not only bounded in magnitude but also restricted to a finite interval of $\lambda$, underscoring a fundamental constraint shared across both interaction types.

This study further examines how self-interactions and frequency influence the structure of the ergospheres in chains of rotating BSs, including the conditions under which two distinct ergospheres merge into one. We find that ergospheres form in chains with one, two, and four constituents. In both the two- and four-constituent configurations, the two ergospheres coalesce into a single region as the frequency increases. In contrast, the absence of ergospheres in the three-constituent case is explained by its insufficient compactness and the associated lack of required spacetime curvature \cite{Sun:2023ord}. Additionally, the frequency threshold for ergosphere emergence rises with coupling strength. The merger of ergospheres could amplify modulations in the gravitational wave emission, offering a potential observational signature for identifying chains of BSs with detectors such as LISA, LIGO, and Virgo.

Based on this study, we propose three key directions for future research. Firstly, the dynamical stability of chains of BSs with self-interaction under perturbation remains an open problem, not addressed in related evolutionary studies such as \cite{Liang:2023ywv}, making it a meaningful research topic. Secondly, the observational signatures of such chains, including photon rings, shadows and gravitational lensing effects, present significant research value. Finally, to the best of our knowledge, no systematic studies have yet explored BSs chains in the context of modified gravity theories; this gap in the literature represents a promising direction for further investigation.


\section*{Acknowledgements}
 We thank Prof. Yongqiang Wang and Dr. Shixian Sun for the important discussions. This work is supported by National Key Research and Development Program of China (Grant No. 2020YFC2201503 and No. 2021YFC2203003) and  the National Natural Science Foundation of China (Grants No.~12275110 and No.~12247101). Parts of computations were performed on the shared memory system at institute of computational physics and complex systems in Lanzhou university.


\begin{thebibliography}{10}
\bibitem{LIGOScientific:2016aoc}
B.~P.~Abbott \textit{et al.} [LIGO Scientific and Virgo],
Phys. Rev. Lett. \textbf{116}, no.6, 061102 (2016)
doi:10.1103/PhysRevLett.116.061102
[arXiv:1602.03837 [gr-qc]].

\bibitem{LIGOScientific:2017vwq}
B.~P.~Abbott \textit{et al.} [LIGO Scientific and Virgo],
Phys. Rev. Lett. \textbf{119}, no.16, 161101 (2017)
doi:10.1103/PhysRevLett.119.161101
[arXiv:1710.05832 [gr-qc]].

\bibitem{Bondi:1962px}
H.~Bondi, M.~G.~J.~van der Burg and A.~W.~K.~Metzner,
Proc. Roy. Soc. Lond. A \textbf{269}, 21-52 (1962)
doi:10.1098/rspa.1962.0161.

\bibitem{LIGOScientific:2017ync}
B.~P.~Abbott \textit{et al.} [LIGO Scientific, Virgo, Fermi GBM, INTEGRAL, IceCube, AstroSat Cadmium Zinc Telluride Imager Team, IPN, Insight-Hxmt, ANTARES, Swift, AGILE Team, 1M2H Team, Dark Energy Camera GW-EM, DES, DLT40, GRAWITA, Fermi-LAT, ATCA, ASKAP, Las Cumbres Observatory Group, OzGrav, DWF (Deeper Wider Faster Program), AST3, CAASTRO, VINROUGE, MASTER, J-GEM, GROWTH, JAGWAR, CaltechNRAO, TTU-NRAO, NuSTAR, Pan-STARRS, MAXI Team, TZAC Consortium, KU, Nordic Optical Telescope, ePESSTO, GROND, Texas Tech University, SALT Group, TOROS, BOOTES, MWA, CALET, IKI-GW Follow-up, H.E.S.S., LOFAR, LWA, HAWC, Pierre Auger, ALMA, Euro VLBI Team, Pi of Sky, Chandra Team at McGill University, DFN, ATLAS Telescopes, High Time Resolution Universe Survey, RIMAS, RATIR and SKA South Africa/MeerKAT],
Astrophys. J. Lett. \textbf{848}, no.2, L12 (2017)
doi:10.3847/2041-8213/aa91c9
[arXiv:1710.05833 [astro-ph.HE]].

\bibitem{Kapadia:2019uut}
S.~J.~Kapadia, S.~Caudill, J.~D.~E.~Creighton, W.~M.~Farr, G.~Mendell, A.~Weinstein, K.~Cannon, H.~Fong, P.~Godwin and R.~K.~L.~Lo, \textit{et al.}
Class. Quant. Grav. \textbf{37}, no.4, 045007 (2020)
doi:10.1088/1361-6382/ab5f2d
[arXiv:1903.06881 [astro-ph.HE]].

\bibitem{Rosen:1968mfz}
G.~Rosen,
J. Math. Phys. \textbf{9}, 996 (1968)
doi:10.1063/1.1664693.

\bibitem{Friedberg:1976me}
R.~Friedberg, T.~D.~Lee and A.~Sirlin,
Phys. Rev. D \textbf{13}, 2739-2761 (1976)
doi:10.1103/PhysRevD.13.2739.

\bibitem{Liebling:2012fv}
S.~L.~Liebling and C.~Palenzuela,
Living Rev. Rel. \textbf{26}, no.1, 1 (2023)
doi:10.1007/s41114-023-00043-4
[arXiv:1202.5809 [gr-qc]].

\bibitem{Kaup:1968zz}
D.~J.~Kaup,
Phys. Rev. \textbf{172}, 1331-1342 (1968)
doi:10.1103/PhysRev.172.1331.

\bibitem{Wheeler:1955zz}
J.~A.~Wheeler,
Phys. Rev. \textbf{97}, 511-536 (1955)
doi:10.1103/PhysRev.97.511.

\bibitem{Ruffini:1969qy}
R.~Ruffini and S.~Bonazzola,
Phys. Rev. \textbf{187}, 1767-1783 (1969)
doi:10.1103/PhysRev.187.1767.

\bibitem{DellaMonica:2022kow}
R.~Della Monica and I.~de Martino,
Astron. Astrophys. \textbf{670}, L4 (2023)
doi:10.1051/0004-6361/202245150
[arXiv:2206.03980 [gr-qc]].

\bibitem{Olivares:2018abq}
H.~Olivares, Z.~Younsi, C.~M.~Fromm, M.~De Laurentis, O.~Porth, Y.~Mizuno, H.~Falcke, M.~Kramer and L.~Rezzolla,
Mon. Not. Roy. Astron. Soc. \textbf{497}, no.1, 521-535 (2020)
doi:10.1093/mnras/staa1878
[arXiv:1809.08682 [gr-qc]].

\bibitem{Hui:2016ltb}
L.~Hui, J.~P.~Ostriker, S.~Tremaine and E.~Witten,
Phys. Rev. D \textbf{95}, no.4, 043541 (2017)
doi:10.1103/PhysRevD.95.043541
[arXiv:1610.08297 [astro-ph.CO]].

\bibitem{Colpi:1986ye}
M.~Colpi, S.~L.~Shapiro and I.~Wasserman,
Phys. Rev. Lett. \textbf{57}, 2485-2488 (1986)
doi:10.1103/PhysRevLett.57.2485.

\bibitem{Guzman:2009zz}
F.~S.~Guzman and J.~M.~Rueda-Becerril,
Phys. Rev. D \textbf{80}, 084023 (2009)
doi:10.1103/PhysRevD.80.084023
[arXiv:1009.1250 [astro-ph.HE]].

\bibitem{Guzman:2009xre}
F.~S.~Guzm{\'a}n,
Rev. Mex. Fis. \textbf{55}, 321-326 (2009)
[arXiv:1907.08193 [gr-qc]].



\bibitem{Palenzuela:2006wp}
C.~Palenzuela, I.~Olabarrieta, L.~Lehner and S.~L.~Liebling,
Phys. Rev. D \textbf{75}, 064005 (2007)
doi:10.1103/PhysRevD.75.064005
[arXiv:gr-qc/0612067 [gr-qc]].


\bibitem{Cardoso:2016oxy}
V.~Cardoso, S.~Hopper, C.~F.~B.~Macedo, C.~Palenzuela and P.~Pani,
Phys. Rev. D \textbf{94}, no.8, 084031 (2016)
doi:10.1103/PhysRevD.94.084031
[arXiv:1608.08637 [gr-qc]].

\bibitem{CalderonBustillo:2020fyi}
J.~Calder{\'o}n Bustillo, N.~Sanchis-Gual, A.~Torres-Forn{\'e}, J.~A.~Font, A.~Vajpeyi, R.~Smith, C.~Herdeiro, E.~Radu and S.~H.~W.~Leong,
Phys. Rev. Lett. \textbf{126}, no.8, 081101 (2021)
doi:10.1103/PhysRevLett.126.081101
[arXiv:2009.05376 [gr-qc]].

\bibitem{Bezares:2018qwa}
M.~Bezares and C.~Palenzuela,
Class. Quant. Grav. \textbf{35}, no.23, 234002 (2018)
doi:10.1088/1361-6382/aae87c
[arXiv:1808.10732 [gr-qc]].

\bibitem{Lee:2008jp}
J.~W.~Lee and S.~Lim,
JCAP \textbf{01}, 007 (2010)
doi:10.1088/1475-7516/2010/01/007
[arXiv:0812.1342 [astro-ph]].


\bibitem{Feng:2010gw}
J.~L.~Feng,
Ann. Rev. Astron. Astrophys. \textbf{48}, 495-545 (2010)
doi:10.1146/annurev-astro-082708-101659
[arXiv:1003.0904 [astro-ph.CO]].

\bibitem{Palenzuela:2017kcg}
C.~Palenzuela, P.~Pani, M.~Bezares, V.~Cardoso, L.~Lehner and S.~Liebling,
Phys. Rev. D \textbf{96}, no.10, 104058 (2017)
doi:10.1103/PhysRevD.96.104058
[arXiv:1710.09432 [gr-qc]].

\bibitem{Lee:1988av}
T.~D.~Lee and Y.~Pang,
Nucl. Phys. B \textbf{315}, 477 (1989)
doi:10.1016/0550-3213(89)90365-9.

\bibitem{Friedberg:1986tq}
R.~Friedberg, T.~D.~Lee and Y.~Pang,
Phys. Rev. D \textbf{35}, 3658 (1987)
doi:10.1103/PhysRevD.35.3658.

\bibitem{Herdeiro:2022gzp}
C.~A.~R.~Herdeiro and E.~Radu,
Int. J. Mod. Phys. D \textbf{31}, no.14, 2242022 (2022)
doi:10.1142/S0218271822420226
[arXiv:2205.05395 [gr-qc]].

\bibitem{Mielke:1980sa}
E.~W.~Mielke and R.~Scherzer,
Phys. Rev. D \textbf{24}, 2111 (1981)
doi:10.1103/PhysRevD.24.2111.


\bibitem{Eby:2015hsq}
J.~Eby, C.~Kouvaris, N.~G.~Nielsen and L.~C.~R.~Wijewardhana,
JHEP \textbf{02}, 028 (2016)
doi:10.1007/JHEP02(2016)028
[arXiv:1511.04474 [hep-ph]].

\bibitem{Sanchis-Gual:2021phr}
N.~Sanchis-Gual, C.~Herdeiro and E.~Radu,
Class. Quant. Grav. \textbf{39}, no.6, 064001 (2022)
doi:10.1088/1361-6382/ac4b9b
[arXiv:2110.03000 [gr-qc]].

\bibitem{Herdeiro:2020kvf}
C.~A.~R.~Herdeiro, J.~Kunz, I.~Perapechka, E.~Radu and Y.~Shnir,
Phys. Lett. B \textbf{812}, 136027 (2021)
doi:10.1016/j.physletb.2020.136027
[arXiv:2008.10608 [gr-qc]].


\bibitem{Jaramillo:2024cus}
V.~Jaramillo and S.~Y.~Zhou,
Phys. Rev. D \textbf{111}, no.2, 024027 (2025)
doi:10.1103/PhysRevD.111.024027
[arXiv:2411.08985 [gr-qc]].

\bibitem{Herdeiro:2016gxs}
C.~A.~R.~Herdeiro, E.~Radu and H.~F.~R{\'u}narsson,
Int. J. Mod. Phys. D \textbf{25}, no.09, 1641014 (2016)
doi:10.1142/S0218271816410145
[arXiv:1604.06202 [gr-qc]].

\bibitem{Ryan:1996nk}
F.~D.~Ryan,
Phys. Rev. D \textbf{55}, 6081-6091 (1997)
doi:10.1103/PhysRevD.55.6081.

\bibitem{Zhang:2023rwc}
R.~Zhang, S.~X.~Sun, L.~X.~Huang and Y.~Q.~Wang,
Phys. Rev. D \textbf{111}, no.2, 024076 (2025)
doi:10.1103/PhysRevD.111.024076
[arXiv:2312.15755 [gr-qc]].

\bibitem{Liang:2023ywv}
C.~Liang, J.~R.~Rena, S.~X.~Sun and Y.~Q.~Wang,
Eur. Phys. J. C \textbf{84}, no.1, 14 (2024)
doi:10.1140/epjc/s10052-023-12345-6
[arXiv:2306.11437 [hep-th]].

\bibitem{Sun:2022duv}
S.~X.~Sun, L.~Zhao and Y.~Q.~Wang,
JHEP \textbf{08}, 152 (2023)
doi:10.1007/JHEP08(2023)152
[arXiv:2210.09265 [gr-qc]].

\bibitem{Herdeiro:2014goa}
C.~A.~R.~Herdeiro and E.~Radu,
Phys. Rev. Lett. \textbf{112}, 221101 (2014)
doi:10.1103/PhysRevLett.112.221101
[arXiv:1403.2757 [gr-qc]].

\bibitem{Mourelle:2024qgo}
J.~C.~Mourelle, C.~Adam, J.~Calder{\'o}n Bustillo and N.~Sanchis-Gual,
Phys. Rev. D \textbf{110}, no.12, 123019 (2024)
doi:10.1103/PhysRevD.110.123019
[arXiv:2403.13052 [gr-qc]].

\bibitem{Kleihaus:2005me}
B.~Kleihaus, J.~Kunz and M.~List,
Phys. Rev. D \textbf{72}, 064002 (2005)
doi:10.1103/PhysRevD.72.064002
[arXiv:gr-qc/0505143 [gr-qc]].

\bibitem{Astefanesei:2003rw}
D.~Astefanesei and E.~Radu,
Phys. Lett. B \textbf{587}, 7-15 (2004)
doi:10.1016/j.physletb.2004.03.006
[arXiv:gr-qc/0310135 [gr-qc]].

\bibitem{Vaglio:2022flq}
M.~Vaglio, C.~Pacilio, A.~Maselli and P.~Pani,
Phys. Rev. D \textbf{105}, no.12, 124020 (2022)
doi:10.1103/PhysRevD.105.124020
[arXiv:2203.07442 [gr-qc]].

\bibitem{Sun:2023ord}
S.~X.~Sun and Y.~Q.~Wang,
[arXiv:2312.16921 [gr-qc]].

\bibitem{Gervalle:2022fze}
R.~Gervalle,
Phys. Rev. D \textbf{105}, no.12, 124052 (2022)
doi:10.1103/PhysRevD.105.124052
[arXiv:2206.03982 [gr-qc]].

\bibitem{Liang:2025myf}
C.~Liang, C.~A.~R.~Herdeiro and E.~Radu,
JHEP \textbf{03}, 119 (2025)
doi:10.1007/JHEP03(2025)119
[arXiv:2501.05342 [gr-qc]].

\bibitem{Herdeiro:2021mol}
C.~A.~R.~Herdeiro, J.~Kunz, I.~Perapechka, E.~Radu and Y.~Shnir,
Phys. Rev. D \textbf{103}, no.6, 065009 (2021)
doi:10.1103/PhysRevD.103.065009
[arXiv:2101.06442 [gr-qc]].

\bibitem{Wang:2018xhw}
Y.~Q.~Wang, Y.~X.~Liu and S.~W.~Wei,
Phys. Rev. D \textbf{99}, no.6, 064036 (2019)
doi:10.1103/PhysRevD.99.064036
[arXiv:1811.08795 [gr-qc]].

\bibitem{Herdeiro:2014jaa}
C.~Herdeiro and E.~Radu,
Phys. Rev. D \textbf{89}, no.12, 124018 (2014)
doi:10.1103/PhysRevD.89.124018
[arXiv:1406.1225 [gr-qc]].

\bibitem{Herdeiro:2014goa}
C.~A.~R.~Herdeiro and E.~Radu,
Phys. Rev. Lett. \textbf{112}, 221101 (2014)
doi:10.1103/PhysRevLett.112.221101
[arXiv:1403.2757 [gr-qc]].

\bibitem{Kunz:2019sgn}
J.~Kunz, I.~Perapechka and Y.~Shnir,
JHEP \textbf{07}, 109 (2019)
doi:10.1007/JHEP07(2019)109
[arXiv:1904.13379 [gr-qc]].


\bibitem{Cunha:2018acu}
P.~V.~P.~Cunha and C.~A.~R.~Herdeiro,
Gen. Rel. Grav. \textbf{50}, no.4, 42 (2018)
doi:10.1007/s10714-018-2361-9
[arXiv:1801.00860 [gr-qc]].

\bibitem{deSa:2024dhj}
P.~L.~B.~de S{\'a}, H.~C.~D.~Lima, Jr., C.~A.~R.~Herdeiro and L.~C.~B.~Crispino,
Phys. Rev. D \textbf{110}, no.10, 104047 (2024)
doi:10.1103/PhysRevD.110.104047
[arXiv:2406.02695 [gr-qc]].

\bibitem{Rosa:2022tfv}
J.~L.~Rosa and D.~Rubiera-Garcia,
Phys. Rev. D \textbf{106}, no.8, 084004 (2022)
doi:10.1103/PhysRevD.106.084004
[arXiv:2204.12949 [gr-qc]].


\end{thebibliography}
\end{document}